\documentclass[acmsmall]{acmart}
\AtBeginDocument{%
  \providecommand\BibTeX{{%
    \normalfont B\kern-0.5em{\scshape i\kern-0.25em b}\kern-0.8em\TeX}}}

\setcopyright{acmlicensed}
\copyrightyear{2018}
\acmYear{2018}
\acmDOI{XXXXXXX.XXXXXXX}

\acmJournal{JACM}
\acmVolume{37}
\acmNumber{4}
\acmArticle{111}
\acmMonth{8}

\usepackage{float}
\usepackage{graphicx}
\usepackage{enumerate}
\usepackage{enumitem}
\usepackage{subfigure}
\usepackage{color}
\usepackage{mathtools}
\usepackage{epstopdf}
\usepackage{tabularx}

\usepackage[linesnumbered,vlined,ruled,noend]{algorithm2e}

\newcommand{\cut}[1]{}

\newcommand{\todo}[1]{\textcolor{red}{TODO: #1}}

\newcommand{\xy}[1]{\textcolor{cyan}{#1}}

\newtheorem{theorem}{Theorem}
\newtheorem{definition}{Definition}

\newtheorem{example}{Example}
\newtheorem{lemma}{Lemma}

\newtheorem{assumption}{Assumption}

\newtheorem{procedure}{Procedure}

\DeclareMathOperator{\cost}{cost}

\DeclareMathOperator{\Sim}{Sim}

\usepackage{xspace}
\newcommand{\sysname}{Wii\xspace}

\begin{document}

\title{Wii: Dynamic Budget Reallocation In Index Tuning}

% \affiliation{%

\author{Xiaoying Wang}
\email{xiaoying_wang@sfu.ca}
\orcid{0009-0004-7817-7196}
\affiliation{%
  \institution{Simon Fraser University}
  \city{Burnaby}
  \state{British Columbia}
  \country{Canada}
}

\author{Wentao Wu}
\affiliation{
  \institution{Microsoft Research}
  \city{Redmond}
  \country{USA}}
\email{wentao.wu@microsoft.com}

\author{Chi Wang}
\affiliation{
  \institution{Microsoft Research}
  \city{Redmond}
  \country{USA}}
\email{wang.chi@microsoft.com}

\author{Vivek Narasayya}
\affiliation{
  \institution{Microsoft Research}
  \city{Redmond}
  \country{USA}}
\email{viveknar@microsoft.com}

\author{Surajit Chaudhuri}
\affiliation{
  \institution{Microsoft Research}
  \city{Redmond}
  \country{USA}}
\email{surajitc@microsoft.com}
%\renewcommand{\shortauthors}{Xiaoying Wang, Wentao Wu, Chi Wang, Vivek Narasayya, and Surajit Chaudhuri}

\begin{abstract}
%\todo{Revise, copied from the SIGMOD paper.}

Index tuning aims to find the optimal index configuration for an input workload.
It is often a time-consuming and resource-intensive process, largely attributed to the huge amount of ``what-if'' calls made to the query optimizer during configuration enumeration.
Therefore, in practice it is desirable to set a \emph{budget constraint} that limits the number of what-if calls allowed.
%A dominant factor of 
%which dominate the overhead (e.g., time and resource consumption) during configuration enumeration.
This yields a new problem of \emph{budget allocation}, namely, deciding on which query-configuration pairs (QCP's) to issue what-if calls.
Unfortunately, optimal budget allocation is NP-hard, and budget allocation decisions made by existing solutions can be inferior.
%rely on various heuristics.
%this \emph{budget-aware index tuning} problem remains NP-hard and existing budget-aware configuration enumeration algorithms rely on various heuristics for 
%Such budget allocation decisions can be inferior. 
In particular, many of the what-if calls allocated by using existing solutions are devoted to QCP's whose what-if costs can be approximated by using \emph{cost derivation}, a well-known technique that is computationally much more efficient and has been adopted by commercial index tuning software.
This results in considerable waste of the budget, as these what-if calls are unnecessary.
In this paper, we propose ``\sysname,'' a lightweight mechanism that aims to avoid such spurious what-if calls. 
It can be seamlessly integrated with existing configuration enumeration algorithms.
Experimental evaluation on top of both standard industrial benchmarks and real workloads demonstrates that \sysname can eliminate significant number of spurious what-if calls.
Moreover, by \emph{reallocating} the saved budget to QCP's where cost derivation is less accurate, existing algorithms can be significantly improved in terms of the final configuration found.
\end{abstract}




\received{October 2023}
\received[revised]{January 2024}
\received[accepted]{February 2024}

\setcopyright{acmlicensed}
\acmJournal{PACMMOD}
\acmYear{2024} \acmVolume{2} \acmNumber{3 (SIGMOD)} \acmArticle{182} \acmMonth{6}\acmDOI{10.1145/3654985}

\maketitle

\section{Introduction}

\begin{figure}
\centering
    \includegraphics[width=.75\columnwidth]{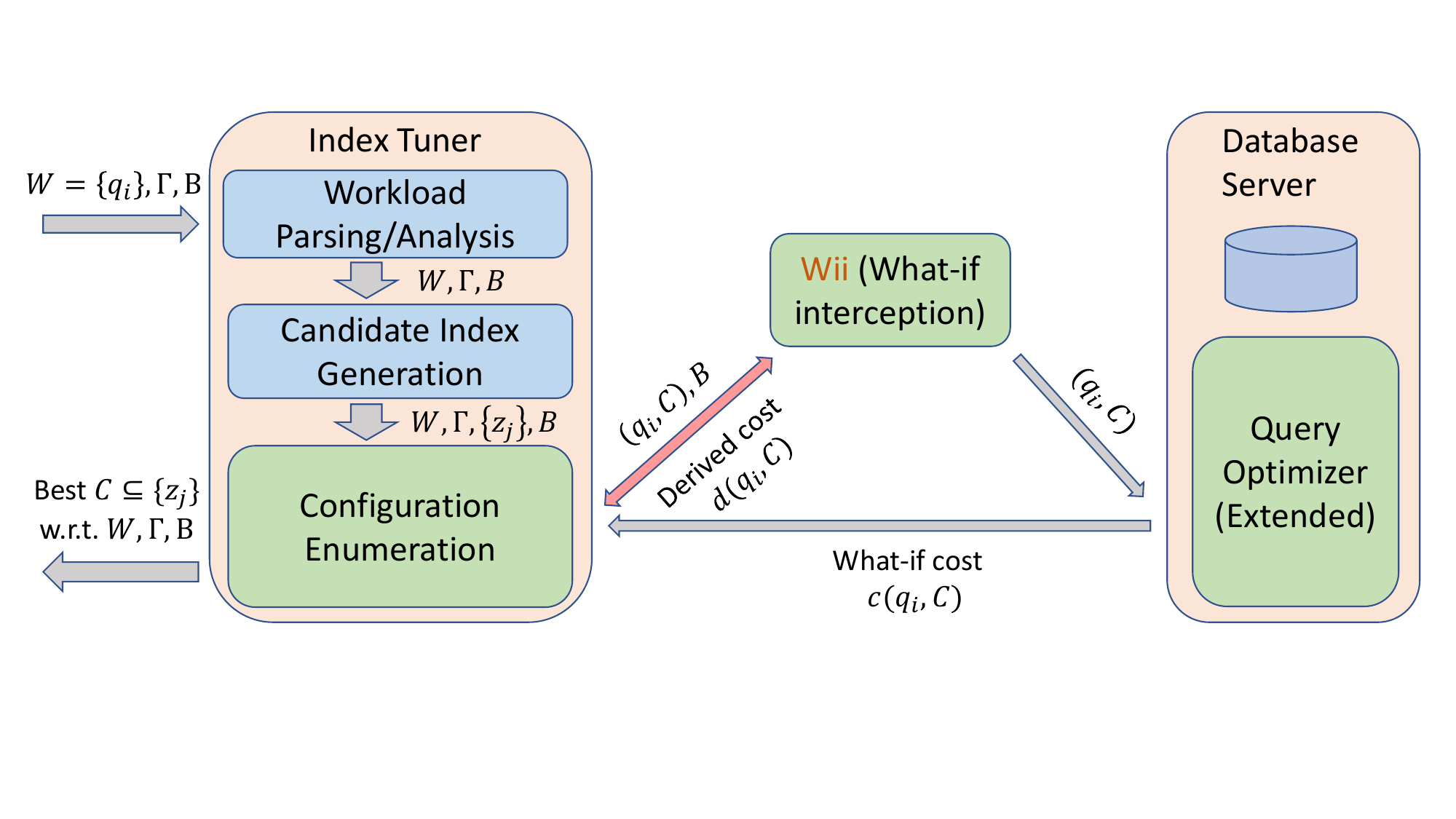}
\vspace{-4em}
\caption{The architecture of budget-aware index tuning with ``\sysname'', i.e., what-if (call) interception, where $W$ %$W=\{q_i\}_{i=1}^m$ 
represents the input workload, $q_i\in W$ represents an individual SQL query in the workload, $\Gamma$ represents a set of tuning constraints, $B$ represents the budget on the number of what-if calls allowed. Moreover, $\{z_j\}$ represents the set of candidate indexes generated for $W$, and $C\subseteq\{z_j\}$ represents an index configuration proposed during configuration enumeration.}
%\todo{Can we give a name to the booster?}}
\label{fig:what-if-architecture}
\vspace{0.5em}
\end{figure}

Index tuning aims to find the optimal index \emph{configuration} (i.e., a set of indexes) for an input workload of SQL queries.
It is often a time-consuming and resource-intensive process for large and complex workloads in practice.
From user's perspective, it is therefore desirable to constrain the index tuner/advisor by limiting its execution time and resource, with the compromise that the goal of index tuning shifts to seeking the best configuration within the given time and resource constraints.
Indeed, commercial index tuners, such as the Database Tuning Advisor (DTA) developed for Microsoft SQL Server, have been offering a \emph{timeout} option that allows user to explicitly control the execution time of index tuning to prevent it from running indefinitely~\cite{dta,dta-utility}.
More recently, there has been a proposal of \emph{budget-aware index tuning} that puts a \emph{budget constraint} on the number of ``what-if'' (optimizer) calls~\cite{WuWSWNCB22}, motivated by the observation that most of the time and resource in index tuning is spent on what-if calls~\cite{KossmannHJS20,PapadomanolakisDA07} made to the query optimizer during configuration enumeration (see Figure~\ref{fig:what-if-architecture}).
%within a given budget on the number of ``what-if'' (optimizer) calls~\cite{WuWSWNCB22}.
%It is important and beneficial in practice  since index tuning is often resource-intensive and time-consuming, 

\begin{figure}
\centering
\subfigure[\emph{two-phase greedy}]{ \label{fig:closeness:two-phase}
    \includegraphics[width=0.46\columnwidth]{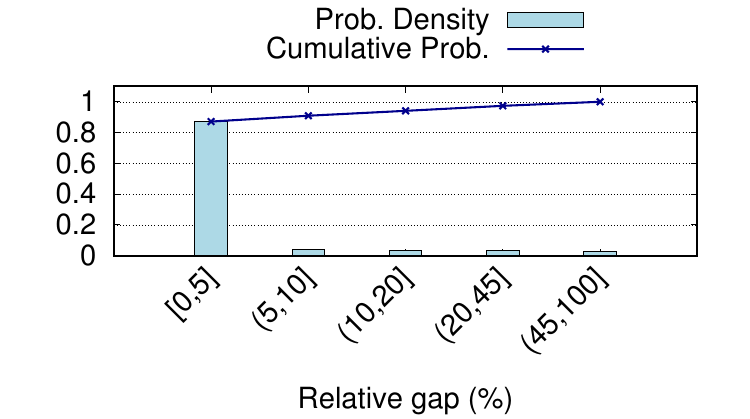}}
%\hspace{0.1\columnwidth}
\subfigure[\emph{MCTS}]{ \label{fig:closeness:mcts}
    \includegraphics[width=0.46\columnwidth]{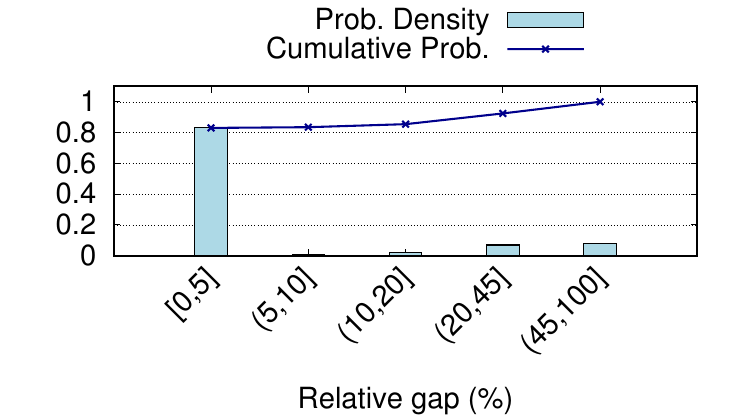}}
\vspace{-1em}
\caption{Distribution of the relative gap between what-if cost and derived cost when tuning TPC-DS under a budget of 5,000 what-if calls. Here the relative gap is defined as $\frac{\text{derived cost}-\text{what-if cost}}{\text{derived cost}}\times 100\%$, as derived cost is an upper bound of the what-if cost under monotonicity assumption.} %\xy{max value 100}}
\label{fig:motivation}
\vspace{0.5em}
\end{figure}

A what-if call takes as input a query-configuration pair (QCP) and returns the estimated cost of the query by utilizing the indexes in the configuration.
It is the same as a regular query optimizer call except for that it also takes \emph{hypothetical} indexes, i.e., indexes that are proposed by the index tuner but have not been materialized, into consideration~\cite{ChaudhuriN98,ValentinZZLS00}.
There can be thousands or even millions of potential what-if calls when tuning large and complex workloads~\cite{ml-index-tuning-overview}.
Therefore, it is not feasible to make a what-if call for \emph{every} QCP encountered in configuration enumeration/search.
As a result, one key problem in budget-aware index tuning is \emph{budget allocation}, where one needs to determine which QCP's to make what-if calls for so that the index tuner can find the best index configuration.
Unfortunately, optimal budget allocation is NP-hard~\cite{Comer78,ChaudhuriDN04,WuWSWNCB22}.
Existing budget-aware configuration search algorithms~\cite{WuWSWNCB22} range from adaptations of the classic greedy search algorithm~\cite{ChaudhuriN97} to more sophisticated approaches with Monte Carlo tree search (MCTS)~\cite{uct}, which allocate budget by leveraging various heuristics.
For example, the greedy-search variants adopt a simple ``first come first serve'' (FCFS) strategy where what-if calls are allocated on demand, and the MCTS-based approach considers the \emph{rewards} observed in previous budget allocation steps to decide the next allocation step.
These budget allocation strategies can be inferior.
In particular, we find in practice that many of the what-if calls made are unnecessary, as their corresponding what-if costs are close to the approximations given by a well-known technique called \emph{cost derivation}~\cite{ChaudhuriN97}.
Compared to making a what-if call, cost derivation is computationally much more efficient and has been integrated into commercial index tuning software such as DTA~\cite{dta,dta-utility}.
In the rest of this paper, we refer to the approximation given by cost derivation as the \emph{derived cost}.
Figure~\ref{fig:motivation} presents the distribution of the \emph{relative gap} between what-if cost and derived cost when tuning the TPC-DS benchmark workload with 99 complex queries.
We observe that 80\% to 90\% of the what-if calls were made for QCP's with relative gap below 5\%, for two state-of-the-art budget-aware configuration search algorithms \emph{two-phase greedy} and \emph{MCTS} (Section~\ref{sec:preliminaries:budget-aware-search}).
If we know that the derived cost is indeed a good approximation, we can avoid such a \emph{spurious} what-if call.
The challenge, however, is that we need to learn this fact \emph{before} the what-if call is made.

The best knowledge we have so far is that, under mild assumption on the \emph{monotonicity} of query optimizer's cost function (i.e., a larger configuration with more indexes should not increase the query execution cost), the derived cost acts as an \emph{upper bound} of the what-if cost (Section~\ref{sec:framework:upper-bound}).
However, the what-if cost can still lie \emph{anywhere} between zero and the derived cost.
In this paper, we take one step further by proposing a generic framework that develops a \emph{lower bound} for the what-if cost.
The gap between the lower bound and the upper bound (i.e., the derived cost) therefore measures the closeness between the what-if cost and the derived cost.
As a result, it is safe to avoid a what-if call when this gap is small and use the derived cost as a surrogate. %(for the what-if cost).

Albeit a natural idea, there are a couple of key requirements to make it relevant in practice.
First, the lower bound needs to be \emph{nontrivial}, i.e., it needs to be as close to the what-if cost as possible---an example of a trivial but perhaps useless lower bound would be always setting it to zero.
Second, the lower bound needs to be computationally \emph{efficient} compared to making a what-if call.
Third, the lower bound needs to be \emph{integratable} with existing budget-aware configuration enumeration algorithms.
In this paper, we address these requirements as follows.

\vspace{-0.5em}
\paragraph*{Nontriviality}

We develop a lower bound that depends only on common properties of the cost functions used by the query optimizer, such as \emph{monotonicity} and \emph{submodularity}, which have been widely assumed by previous work~\cite{GuptaHRU97,SchnaitterPG09,ChoenniBC93,WuCZTHN13,qo-bench} and independently verified in our own experiments~\cite{full-version}.
In a nutshell, it looks into the \emph{marginal cost improvement} (MCI) that each individual index in the given configuration can achieve, and then establishes an upper bound on the \emph{cost improvement} (and therefore a lower bound on the what-if cost) of the given configuration by summing up the upper bounds on the MCI's of individual indexes (Section~\ref{sec:framework:lower-bound}).
We further propose optimization techniques to refine the lower bound for budget-aware greedy search algorithms (Section~\ref{sec:bounds:mci:greedy}) and MCTS-based algorithms (Section~\ref{sec:optimization:coverage}).

\vspace{-0.5em}
\paragraph*{Efficiency} We demonstrate that the computation time of our lower bound is \emph{orders of magnitude less} compared to a what-if call, though it is in general more expensive than computing the upper bound, i.e., the derived cost (Section~\ref{sec:evaluation:computation:overhead}). 
For example, as shown in Figure~\ref{fig:bound-computation:tpcds}, when running the \emph{MCTS} configuration enumeration algorithm on top of the TPC-DS benchmark, on average it takes 0.02 ms and 0.04 ms to compute the derived cost and our lower bound, respectively; in contrast, the average time of making a what-if call to the query optimizer is around 800 ms.
%\todo{Highlight some results. Also compare to ML inference time and time of computing derived cost here.}

\vspace{-0.5em}
\paragraph*{Integratability} We demonstrate that our lower bound can be seamlessly integrated with existing budget-aware index tuning algorithms (Section~\ref{sec:integration}). From a software engineering perspective, the integration is \emph{non-intrusive}, meaning that there is no need to change the architecture of the current cost-based index tuning software stack.
As illustrated in Figure~\ref{fig:what-if-architecture}, we encapsulate the lower-bound computation inside a component called ``\sysname,'' which is shorthand for ``what-if (call) interception.''
During configuration enumeration, \sysname intercepts every what-if call made to the query optimizer, computes the lower bound of the what-if cost, and then checks the closeness between the lower bound and the derived cost (i.e., the upper bound) with a \emph{confidence}-based mechanism (Section~\ref{sec:integration:confidence}).
If \sysname feels confident enough, it will skip the what-if call and instead send the derived cost back to the configuration enumerator.

%\todo{More description. In particular, talk about the interception mechanism, bring up the \sysname, and its position in Figure 1. Moreover, mention how easy it is to integrate with existing algorithms.}

%\vspace{0.5em}
More importantly, we demonstrate the efficacy of \sysname in terms of (1) the number of what-if calls it allows to skip (Section~\ref{sec:evaluation:skipping}) and (2) the end-to-end improvement on the final index configuration found (Section~\ref{sec:evaluation:e2e}).
The latter is perhaps the most valuable benefit of \sysname in practice, and we show that, by \emph{reallocating} the saved budget to what-if calls where \sysname is less confident, it can yield significant improvement on both standard industrial benchmarks and real customer workloads (Section~\ref{sec:evaluation:e2e}).
For example, as showcased in Figure~\ref{fig:two-phase:call-level:tpcds:K20}, with 5,000 what-if calls as budget and 20 as the maximum configuration size allowed, on TPC-DS \sysname improves the baseline \emph{two-phase greedy} configuration enumeration algorithm by increasing the \emph{percentage improvement} of the final configuration found from 50\% to 65\%; this is achieved by skipping around 18,000 unnecessary what-if calls, as shown in Figure~\ref{fig:two-phase:call-level:tpcds:skip}.

%\todo{Add some result highlights.}

Last but not least, while we focus on budget-aware index tuning in this paper, \sysname can also be used in a special situation where one does not enforce a budget on the index tuner, namely, the tuner has \emph{unlimited budget} on the number of what-if calls.
This special situation may make sense if, for example, one has a relatively small workload.
\sysname plays a different role here.
Since there is no budget constraint, \sysname cannot improve the quality of the final configuration found, as the best quality can anyways be achieved by keeping on issuing what-if calls to the query optimizer. 
Instead, by skipping spurious what-if calls, \sysname can significantly improve the overall efficiency of index tuning.
For example, without a budget constraint, when tuning the standard TPC-H benchmark with 22 queries, \sysname can reduce index tuning time by 4$\times$ while achieving the same quality on the best configuration found (Section~\ref{sec:eval:no-budget-limit}).
%~\cite{full-version}.

\iffalse
%\vspace{-0.5em}
%\paragraph*{Paper organization} 
\noindent
\textbf{(Paper Organization)} The rest of the paper is organized as follows. Section~\ref{sec:preliminaries} presents a brief overview of the budget-aware configuration enumeration problem and the existing solutions to budget allocation that motivate the development of \sysname.
Section~\ref{sec:framework} presents the lower and upper bounds used by \sysname to skip unnecessary what-if calls to improve budget allocation, and Section~\ref{sec:refinements} presents the coverage-based refinements.
Section~\ref{sec:integration} further presents the integration of \sysname with existing budget-aware configuration enumeration algorithms.
Section~\ref{sec:evaluation} reports experimental evaluation results.
Section~\ref{sec:related-work} discusses related work in detail, and Section~\ref{sec:conclusion} concludes the paper. \todo{Revise.}
\fi

\section{Preliminaries}
\label{sec:preliminaries}

In this section, we present a brief overview of the budget-aware index configuration search problem.

%\todo{Add the overview here. Points to cover: (0) basic notions in cost-based index tuning, such as configuration, hypothetical index, what-if call, indexable column, candidate generation, configuration enumeration, etc; (1) the problem definition of budget-aware configuration search; (2) existing solutions to budget-aware configuration search and their limitations; (3) highlight the challenges on the budget allocation problem and reiterate our motivation on developing a uniform performance booster.}

\vspace{-0.5em}
\subsection{Cost-based Index Tuning}
\label{sec:preliminaries:index-tuning}

As Figure~\ref{fig:what-if-architecture} shows, cost-based index tuning consists of two stages: 
\begin{itemize}%[leftmargin=*]
    \item \textbf{Candidate index generation.}  We generate a set of \emph{candidate indexes} for each query in the workload based on the \emph{indexable columns}~\cite{ChaudhuriN97}.
    Indexable columns are those that appear in the \emph{selection}, \emph{join}, \emph{group-by}, and \emph{order-by} expressions of a SQL query, which are used as \emph{key} columns for fast seek-based index look-ups.
    We then take the union of the candidate indexes from individual queries as the candidate indexes for the entire workload.
    %For each query in the input workload,  by looking for the ; we take the union as the candidate indexes for the entire workload.
    \item \textbf{Configuration enumeration.} 
    %Based on the candidate indexes generated for the input workload, 
    We search for a subset (i.e., a \emph{configuration}) of the candidate indexes that can minimize the what-if cost of the workload, with respect to constraints such as the maximum number of indexes allowed or the total amount of storage taken by the index configuration.
    %For a given set of candidate indexes, we search for a subset (i.e., \emph{configuration}) of indexes that minimizes the what-if cost of the input workload.
\end{itemize}

Index tuning is time-consuming and resource-intensive, due to the large amount of what-if calls issued to the query optimizer during configuration enumeration/search.
%\footnote{A what-if call is similar to a regular optimizer call, with the exception that the \emph{hypothetical} indexes are also made visible to the query optimizer. An hypothetical index is a candidate index considered by the index tuner that has not yet been built.}
Therefore, previous work proposes putting a \emph{budget} on the amount of what-if calls that can be issued during configuration search~\cite{WuWSWNCB22}.
We next present this \emph{budget-aware} configuration search problem in more detail.

\subsection{Budget-aware Configuration Search}
\label{sec:preliminaries:budget-aware-search}

%\vspace{-0.5em}
\subsubsection{Problem Statement}

Given an input workload $W$ with a set of candidate indexes $I$~\cite{ChaudhuriN97}, a set of constraints $\Gamma$, and a budget $B$ on the number of what-if calls allowed during configuration enumeration, our goal is to find a configuration $C^*\subseteq I$ whose what-if cost $c(W, C^*)$ is minimized under the constraints given by $\Gamma$ and $B$.

%\vspace{-0.5em}
%\paragraph*{Remarks}
In this paper, we focus on index tuning for data analytic workloads $W$ (e.g., the TPC-H
and TPC-DS benchmark workloads).
Although the constraints in $\Gamma$ can be arbitrary, we focus on the \emph{cardinality constraint} $K$ that specifies the maximum \emph{configuration size} (i.e., the number of indexes contained by the configuration) allowed.
%\footnote{Based on~\cite{WuWSWNCB22}, the tuning results by using the \emph{storage constraint} are correlated.}
Moreover, under a limited budget $B$, it is often impossible to know the what-if cost of \emph{every} query-configuration pair (QCP) encountered during configuration enumeration.
Therefore, to estimate the costs for QCP's where what-if calls are not allocated, one has to rely on approximation of the what-if cost without invoking the query optimizer.
One common approximation technique is \emph{cost derivation}~\cite{ChaudhuriN97,dta}, as we discuss below. %\xy{all existing budget-aware tuning algorithm uses.}

%and we call this approximated cost the \emph{derived cost}.

%\vspace{-0.5em}
\subsubsection{Cost Derivation}
\label{sec:framework:upper-bound}

%\todo{Move this section to preliminaries.}
Given a QCP $(q, C)$, its \emph{derived cost} %~\cite{ChaudhuriN97,WuCZTHN13} 
$d(q, C)$ is the minimum cost over all subset configurations of $C$ with \emph{known} what-if costs. Formally,
\begin{definition}[Derived Cost]
The derived cost of $q$ over $C$ is
\begin{equation}\label{eq:derived-cost}
  d(q, C)=\min_{S\subseteq C}c(q, S).
\end{equation}
Here, $c(q, S)$ is the what-if cost of $q$ using only a subset $S$ of indexes from the configuration $C$.
\end{definition}

We assume the following \emph{monotone} property~\cite{GuptaHRU97,SchnaitterPG09} of index configuration costs w.r.t. to an arbitrary query $q$:

\begin{assumption}[Monotonicity]\label{assumption:monotone}
Let $C_1$ and $C_2$ be two index configurations where $C_1\subseteq C_2$. Then $c(q, C_2)\leq c(q, C_1)$.
\end{assumption}
That is, including more indexes into a configuration does not increase the what-if cost.
Our validation results using Microsoft SQL Server show that monotonicity holds with probability between 0.95 and 0.99, on a variety of benchmark and real workloads (see~\cite{full-version} for details).
%\todo{Add pointers to the full version on the validation results.}
\iffalse
Assumption~\ref{assumption:monotone} may not always hold in practice, as it depends on the specific implementation of a query optimizer's cost model.
\fi
Under Assumption~\ref{assumption:monotone}, we have 
$$d(q, C)\geq c(q, C),$$
i.e., derived cost is an \emph{upper bound} $U(q, C)$ of what-if cost:
%Therefore, we can set the upper bound $U(q, C)$ as
$$U(q, C) = d(q, C)=\min_{S\subseteq C}c(q, S).$$

%for an arbitrary query-configuration pair $(q, C)$.

%For a given query $q$, assume that the cost function $c(q, C)$ is non-increasing with respect to $C$.
%That is, $c(q, C)\leq c(q, S)$ if $S\subseteq C$.
%We then have $d(q, C)\geq c(q, C)$.
%Therefore, we can set the upper bound $U(q, C)$ as
%$$U(q, C) = d(q, C).$$
%\todo{Formally state this monotonicity property.}

%\xy{To align with the MCI used for lower bound, here: $c(q, C)\leq\min_{S\subseteq C}\Big(c(q, S)-\sum\nolimits_{i=1}^k l(q, z_i)\Big)$, and we directly use $l(q,z_i)=0$ as the lower bound of the MCI.}

%The derived cost is actually an upper-bound of the actual what-if cost (see Section~\ref{sec:framework:upper-bound} for more details).

%For budget-aware configuration search algorithms that work on large/complex workloads with limited budget, they therefore try to minimize the derived cost instead, with the hypothesis that the derived cost of the final configuration found will be close to its actual what-if cost.

\begin{figure}[t]
\centering
    \includegraphics[clip, trim=5cm 7.5cm 7cm 3.5cm, width=.7\columnwidth]{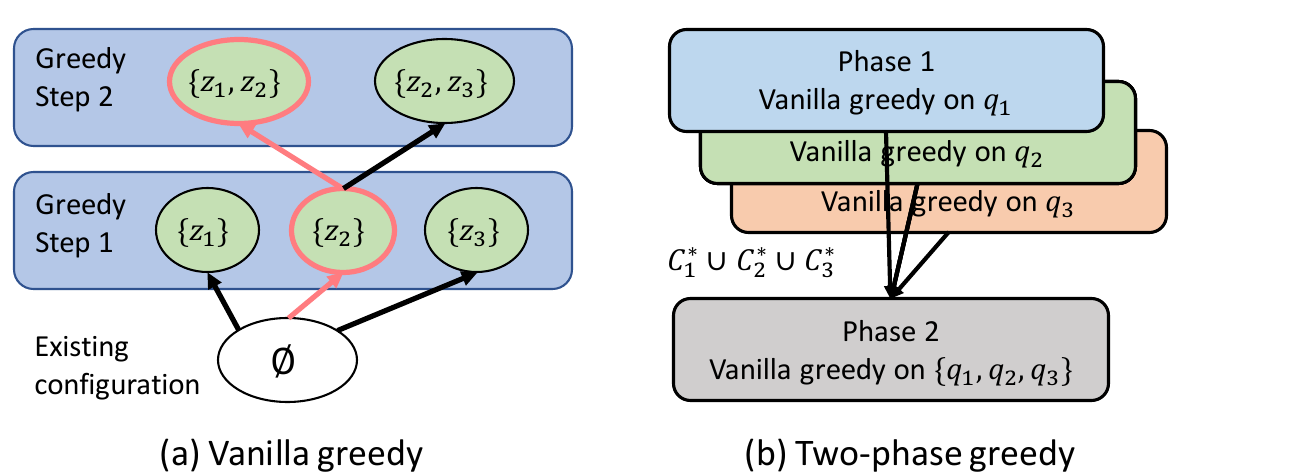}  
\vspace{-1em}
\caption{Example of budget-aware greedy search.}
\label{fig:greedy}
\vspace{0.5em}
\end{figure}

\subsubsection{Existing Solutions}
\label{sec:framework:existing}

The budget-aware configuration search problem is NP-hard.
At the core of this problem is \emph{budget allocation}, namely, to decide on which QCP's to make what-if calls.
Existing heuristic solutions to the problem include: (1) \emph{vanilla greedy}, (2) \emph{two-phase greedy}, (3) \emph{AutoAdmin greedy}, and (4) \emph{MCTS}. Since (2) and (3) are similar, we omit (3) in this paper.

\vspace{-0.5em}
\paragraph*{Vanilla greedy}

Figure~\ref{fig:greedy}(a) illustrates the \emph{vanilla greedy} algorithm with an example of three candidate indexes $\{z_1, z_2, z_3\}$ and the cardinality constraint $K=2$. Throughout this paper, we use $\emptyset$ to represent the \emph{existing configuration}. 
\emph{Vanilla greedy} works step-by-step, where each step adopts a \emph{greedy policy} to choose the next index to be included that can minimize the workload cost on the chosen configuration.
In this example, we have two greedy steps.
The first step examines the three \emph{singleton configurations} $\{z_1\}$, $\{z_2\}$, and $\{z_3\}$. 
Suppose that $\{z_2\}$ results in the lowest workload cost.
The second step tries to expand $\{z_2\}$ by adding one more index, which leads to two candidate configurations $\{z_1, z_2\}$ and $\{z_2, z_3\}$. 
Suppose that $\{z_1, z_2\}$ is better and therefore returned by \emph{vanilla greedy}.
Note that the configuration $\{z_1, z_3\}$ is never visited in this example.
\emph{Vanilla greedy} adopts a simple ``first come first serve (FCFS)'' budget allocation policy to make what-if calls.

\vspace{-0.5em}
\paragraph*{Two-phase greedy}

Figure~\ref{fig:greedy}(b) illustrates the \emph{two-phase greedy} algorithm that can be viewed as an optimization on top of \emph{vanilla greedy}.
Specifically, there are two phases of greedy search in \emph{two-phase greedy}.
In the first phase, we view each query as a workload by itself and run \emph{vanilla greedy} on top of it to obtain the best configuration for that query.
In this particular example, we have three queries $q_1$, $q_2$, and $q_3$ in the workload.
After running \emph{vanilla greedy}, we obtain their best configurations $C_1^*$, $C_2^*$, and $C_3^*$, respectively.
In the second phase, we take the union of the best configurations found for individual queries and use that as the refined set of candidate indexes for the entire workload.
We then run \emph{vanilla greedy} again for the workload with this refined set of candidate indexes, as depicted in Figure~\ref{fig:greedy}(b) for the given example.
\emph{Two-phase greedy} has particular importance in practice as it has been adopted by commercial index tuning software such as Microsoft's Database Tuning Advisor (DTA)~\cite{dta,dta-utility}. Again, budget is allocated with the simple FCFS policy---the same as in \emph{vanilla greedy}.

\vspace{-0.5em}
\paragraph*{MCTS}

\begin{figure}[t]
\centering
    \includegraphics[clip, trim=6cm 6.5cm 7cm 3cm, width=.7\columnwidth]{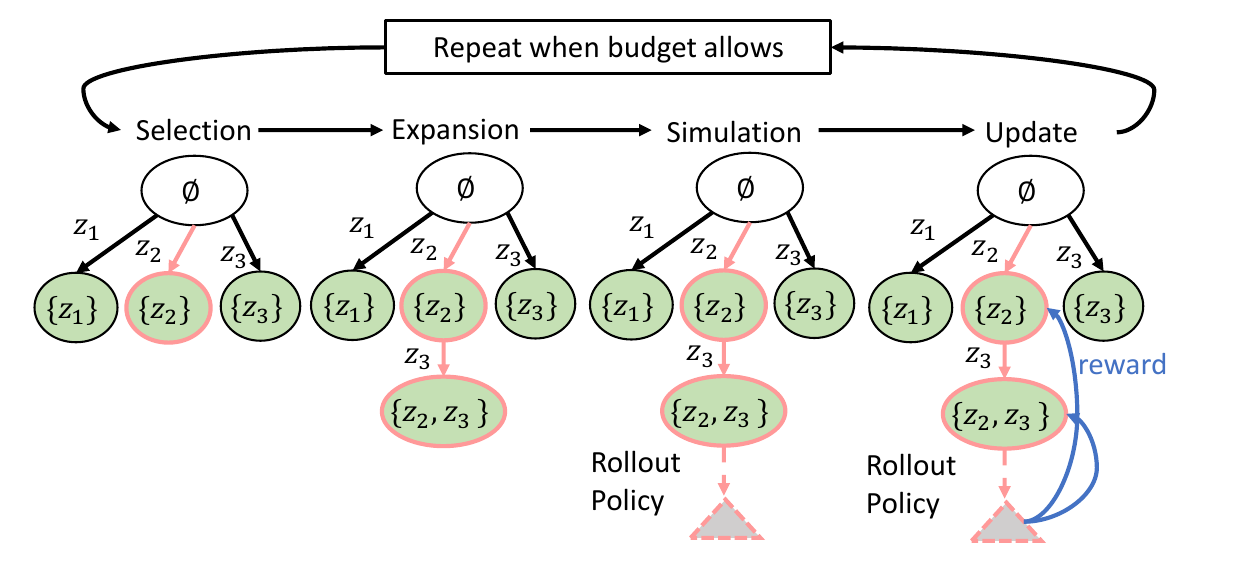}  
\vspace{-1em}
\caption{Example of budget allocation in MCTS.}
\label{fig:mcts-greedy}
\vspace{0.5em}
\end{figure}

Figure~\ref{fig:mcts-greedy} illustrates the \emph{MCTS} algorithm with the same example used in Figure~\ref{fig:greedy}.
It is an iterative procedure that allocates one what-if call in each iteration until the budget runs out.
The decision procedure in each iteration on which query and which configuration to issue the what-if call is an application of the classic \emph{Monte Carlo tree search} (MCTS) algorithm~\cite{BrownePWLCRTPSC12} in the context of index configuration search.
It involves four basic steps: (1) selection, (2) expansion, (3) simulation, and (4) update.
Due to space limitation, we refer the readers to~\cite{WuWSWNCB22} for the full details of this procedure.
After all what-if calls are issued, we run \emph{vanilla greedy} again without making extra what-if calls to find the best configuration.
Our particular version of \emph{MCTS} here employs an $\epsilon$-greedy policy~\cite{sutton2018reinforcement} when selecting the next index to explore.
%\xy{state-of-the-art}
\iffalse
To bootstrap this $\epsilon$-greedy policy, there is a special \emph{warm-up} stage before running MCTS that aims for assigning a \emph{prior} to the reward distribution of singleton configurations, which consumes half of the budget on what-if calls and only targets singleton configurations.
Again, readers are referred to~\cite{WuWSWNCB22} for complete details. 
\fi

\section{What-If Call Interception}
\label{sec:framework}

We develop ``\sysname'' that can skip spurious what-if calls where their what-if costs and derived costs are close.
%improve budget allocation for existing budget-aware configuration search algorithms.
%, which operates at the level of individual what-if calls.
%Our basic idea is to develop lower and upper bounds for the what-if cost.
One key idea is to develop a \emph{lower bound} for the what-if cost: if the gap between the lower bound and the derived cost is small, then it is safe to skip the what-if call.
%without hurting the quality of the final index configuration found.
In this section, we present the \emph{generic} form of the lower bound, as well as a \emph{confidence}-based framework used by \sysname on top of the lower bound to skip spurious what-if calls.
We defer the discussion on further optimizations of the lower bound to Section~\ref{sec:refinements}.

\iffalse
In the following, we start with a problem formulation. 
We then present the details of the lower and upper bounds, followed by an optimization for greedy search given its importance in existing budget-aware configuration enumeration algorithms.
We end this section with practical considerations regarding (1) the computation overhead of the lower and upper bounds and (2) how to set the threshold on the gap between the lower and upper bounds.
\fi

%In the following, we start by presenting our techniques to establish lower/upper bounds for a single what-if call.
%We then extend the techniques to query-level. 

%and workload-level.

%\subsection{Call-level Lower/Upper Bounds}
%\label{sec:bounds:call-level}

\iffalse
\subsection{Problem Formulation}
\todo{Move this section to the confidence-based skipping mechanism.}

%We start by giving a problem formulation and then present our lower/upper bounds for a single what-if call.

Let $q$ be a query and $C$ be an index configuration, and let $c(q, C)$ be the what-if cost for the query-configuration pair $(q, C)$.
We develop a lower bound $L(q, C)$ and an upper bound $U(q, C)$ for $c(q, C)$, namely, 
$L(q, C)\leq c(q, C)\leq U(q, C).$
Given a threshold $\epsilon$, if $U(q, C)-L(q, C)\leq\epsilon$, then the what-if call to the pair $(q, C)$ can be skipped,
%, i.e., without going to the query optimizer to obtain $c(q, C)$, 
and \sysname returns $U(q, C)$ instead.
\fi

\subsection{Lower Bound of What-if Cost}
\label{sec:framework:lower-bound}

%\todo{Move this to the beginning of section 3.}

We use $L(q, C)$ to denote the lower bound of the what-if cost $c(q, C)$.
%To derive $L(q, C)$, 
In the following, we first introduce the notion of \emph{marginal cost improvement} (MCI) of an index, which indicates the additional benefit of adding this index to a configuration for a query. %with respect to a query-configuration pair. 
We then establish $L(q, C)$ by leveraging the \emph{upper bounds} of MCI.

%\subsubsection{Marginal Cost Improvement (MCI)}

\begin{definition}[Marginal Cost Improvement]\label{def:mci}
We define the \emph{marginal cost improvement} (MCI) of an index $z$ with respect to a query $q$  and a configuration $X$ as
\begin{equation}\label{eq:mci}
  \delta(q, z, X)=c(q, X)-c(q, X\cup\{z\}).  
\end{equation}
\end{definition}

\begin{definition}[Cost Improvement]
We define the \emph{cost improvement} (CI) of a query $q$ given a configuration $X$ as
\begin{equation}\label{eq:cost-improvement}
  \Delta(q, X)=c(q, \emptyset)-c(q, X). 
\end{equation}
\end{definition}

We can express CI in terms of MCI. Specifically, consider a query $q$ and a configuration $C=\{z_1, ..., z_m\}$.
% We can write $\Delta(q, C)$ as
The cost improvement $\Delta(q, C)$ can be seen as the sum of MCI's by adding the indexes from $C$ \emph{one by one}, namely,
$$\Delta(q, C)=\Big(c(q,\emptyset)-c(q, \{z_1\})\Big)+\Big(c(q,\{z_1\})-c(q,\{z_1,z_2\})\Big)$$
$$+\cdots+\Big(c(q,\{z_1, ..., z_{m-1}\})-c(q,C)\Big).$$
Let $C_0=\emptyset$ and $C_j=C_{j-1}\cup\{z_j\}$ for $1\leq j\leq m$.
It follows that $C_m=C$ and therefore,
$\Delta(q, C)=\sum\nolimits_{j=1}^m\delta(q, z_j, C_{j-1}).$

If we can have a \emph{configuration-independent} upper bound $u(q, z_j)$ for $\delta(q, z_j, C_{j-1})$, namely, $u(q, z_j)\geq \delta(q, z_j, X)$ for any $X$, then
$$\Delta(q, C)\leq \sum\nolimits_{j=1}^m u(q, z_j).$$
%Note that $u(q, z_j)$ is independent of $C_{j-1}$. 
%In theory, one could have a configuration-dependent upper bound $u(q, z_j, C_{j-1})$, but this would \emph{exponentially} increase the number of upper bounds that need to be maintained.
As a result,
$$c(q,\emptyset)-c(q,C)\leq \sum\nolimits_{j=1}^m u(q, z_j),$$
and it follows that
$$c(q,C)\geq c(q,\emptyset)-\sum\nolimits_{j=1}^m u(q, z_j).$$
We therefore can set the lower bound $L(q, C)$ as
\begin{equation}\label{eq:lower-bound:call-level:v1}
  L(q, C)=c(q,\emptyset)-\sum\nolimits_{j=1}^m u(q, z_j).  
\end{equation}

\vspace{-0.5em}
\paragraph*{Generalization}
This idea can be further generalized if we know the what-if costs of configurations that are subsets of $C$. 
Specifically, let $S\subset C$ be a subset of $C$ with known what-if cost $c(q, S)$.
Without loss of generality, let $C-S=\{z_1, ..., z_k\}$. We have
$$c(q, S)-c(q, C)=\sum\nolimits_{i=1}^k\Big(c(q, C_{i-1})-c(q, C_i)\Big)\leq \sum\nolimits_{i=1}^k u(q, z_i),$$
%where $C_0=S$, $C_i=C_{i-1}\cup \{z_i\}$, and $C_m=C$.
%By the submodularity assumption, we have
%$$\Delta(q, C_{i-1}, z_i)=c(q, C_{i-1})-c(q, C_i)\leq \Delta(q, \{z_i\})\leq u^{\max}(q, z_i).$$
where $C_0$ is now set to $S$. Therefore, 
%$c(q, S)-c(q, C)\leq \sum_{i=1}^m u^{\max}(q, z_i)$. As a result,
$$c(q, C)\geq c(q, S)-\sum\nolimits_{i=1}^k u(q, z_i).$$
Since $S$ is arbitrary, we conclude
$$c(q, C)\geq\max_{S\subset C}\Big(c(q, S)-\sum\nolimits_{z\in C-S} u(q, z)\Big).$$
As a result, it is safe to set
\begin{equation}\label{eq:lower-bound:call-level:v2}
L(q, C)=\max_{S\subset C}\Big(c(q, S)-\sum\nolimits_{z\in C-S} u(q, z)\Big).
\end{equation}
Since $\emptyset\subset C$, Equation~\ref{eq:lower-bound:call-level:v2} is a generalization of Equation~\ref{eq:lower-bound:call-level:v1}.

%\todo{Make the result into a formal statement, e.g., a theorem, and move the other text into the proof.}

%For ease of exposition and avoid clutter of notation, in the rest of the paper, we will refer to Equation ~\ref{eq:lower-bound:call-level:v1} as the lower bound of $c(q, C)$ while keeping in mind that there is an optimization/generation based on Equation~\ref{eq:lower-bound:call-level:v2}.

\subsection{Upper Bound of MCI}
\label{sec:bounds:call-level:mci}

The main question is then to maintain an upper bound $u(q, z)$ for the MCI of each query $q$ and each individual index $z$ so that $u(q, z)\geq \delta(q, z, X)$ for \emph{any} configuration $X$.
Below we discuss several such upper bounds. Our basic idea is to leverage the CIs of explored configurations that contain $z$, along with some well-known properties, such as \emph{monotonicity} and \emph{submodularity}, of the cost function used by the query optimizer.

\subsubsection{Naive Upper Bound}
Let $\Omega$ be the set of \emph{all} candidate indexes.
%In practical index tuning applications, we often have $c(q, \Omega)$ available, where 
\begin{definition}[Naive Upper Bound]
Under Assumption~\ref{assumption:monotone},
\begin{equation}\label{eq:mci:trivial}
    u(q, z)=c(q, \emptyset)-c(q, \Omega)=\Delta(q, \Omega)
\end{equation}
is a valid upper bound of $\delta(q,z,X)$ for any $X$.
\end{definition}
Intuitively, by the monotonicity property, the MCI of any single index $z$ cannot be larger than the CI of all candidate indexes in $\Omega$ combined. In practical index tuning applications, we often have $c(q, \Omega)$ available. However, if $c(q, \Omega)$ is unavailable, then we set $u(q, z)=c(q, \emptyset)$ as it always holds that $c(q, \Omega)\geq 0$.

%Under Assumption~\ref{assumption:monotone}, another idea is to set 
%$$u(q, z)=c(q, \emptyset)-c(q, \Omega)=\Delta(q, \Omega),$$
%where $\Omega$ is the set of \emph{all} candidate indexes. %assuming that the cost function is non-increasing and the query optimizer always picks the best set of indexes.

%In the rest of this paper, we assume the latter as the default value for $u(q, z)$ and call it the \emph{trivial upper bound}.

%\todo{State these assumptions more formally.}

\subsubsection{Upper Bound by Submodularity}
\label{sec:bounds:call-level:mci:submodularity}

%An interesting question is how to improve $u(q, z)$ over the trivial upper bound.
%, as it may be too loose in many situations.
We can improve over the naive upper bound by assuming that the cost function is \emph{submodular}, which has been studied by previous work~\cite{ChoenniBC93}.

\begin{assumption}[Submodularity]\label{assumption:submodular}
Given two configurations $X\subseteq Y$ and an index $z\not\in Y$, we have
\begin{equation}\label{eq:submodular}
c(q, Y)-c(q, Y\cup\{z\})\leq c(q, X)-c(q, X\cup\{z\}).
\end{equation}
Or equivalently, $\delta(q, z, Y)\leq \delta(q, z, X)$.
\end{assumption}
That is, the MCI of an index $z$ diminishes when $z$ is included into \emph{larger} configuration with more indexes. 
Submodularity does not hold often due to \emph{index interaction}~\cite{SchnaitterPG09}.
We also validated the submodularity assumption using Microsoft SQL Server and the same workloads that we used to validate the monotonicity assumption. 
% Our validation results show that submodularity holds with probability between 0.75 and 0.89 on the workloads tested~\cite{full-version}.
Our validation results show that submodularity holds with probability between 0.75 and 0.89 on the workloads tested~\cite{full-version}.%(Section~\ref{sec:evaluation:submodularity}).}

\begin{lemma}\label{lemma:bound:submodular}
Under Assumption~\ref{assumption:submodular}, we have
$$\delta(q, z, X)\leq \Delta(q, \{z\})$$
for any configuration $X$.
\end{lemma}
Due to space constraint, all proofs are postponed to the full version of this paper~\cite{full-version}. 
Intuitively, Lemma~\ref{lemma:bound:submodular} indicates that the CI of a singleton configuration $\{z\}$ can be used as an upper bound of the MCI of the index $z$. 
As a result, we can set
\begin{equation}\label{eq:bound:submodular}
  u(q, z)=\Delta(q, \{z\})=c(q, \emptyset)-c(q, \{z\}).  
\end{equation}
%\todo{Add a discussion on using super-set costs when $c(q, \{z\})$ is unknown.}
There are cases where $c(q, \{z\})$ is unknown but we know the cost of some configuration $X$ that contains $z$, e.g., in \emph{MCTS} where configurations are explored in random order.
By Assumption~\ref{assumption:monotone}, %we have
%$$c(q, \{z\})\geq \min_{z\in X}c(q, X).$$
$$c(q, \{z\})\geq \max_{z\in X}c(q, X).$$
Therefore, we can generalize Equation~\ref{eq:bound:submodular} to have
\begin{definition}[Submodular Upper Bound]
    %$$u(q, z)=c(q, \emptyset)-\min_{z\in X}c(q, X)=\min_{z\in X}\Big(c(q, \emptyset)-c(q, X)\Big)=\min_{z\in X}\Delta(q, X).$$
\begin{eqnarray*}
u(q, z)&=&c(q, \emptyset)-\max_{z\in X}c(q, X)\\
&=&\min_{z\in X}\Big(c(q, \emptyset)-c(q, X)\Big)\\
&=&\min_{z\in X}\Delta(q, X).
\end{eqnarray*}
\end{definition}

That is, the MCI of an index should be no larger than the minimum CI of all the configurations that contain it.

\subsubsection{Summary}

To summarize, assuming monotonicity and submodularity of the cost function $c$, we can set $u(q, z)$ as follows:
\begin{equation}\label{eq:upper-bound:mci:call-level}
    u(q,z)=\min\{c(q,\emptyset), \Delta(q,\Omega), \Delta(q, \{z\}), \min_{z\in X}\Delta(q, X)\}.
\end{equation}

%\xy{Maybe we can use
%\begin{eqnarray*} \delta(q, C, z) \leq \min_{z \notin S,S \subseteq C}\delta(q, S, z)\\
%= \min_{z \notin S,S \subseteq C}(c(q,S) - c(q,S \cup \{z\}))\\
%\leq \min_{z \notin S,z \in T,S \subseteq C,S \subseteq T}(c(q,S) - c(q,T)) = u(q,z)\end{eqnarray*} here since it is more general. The conservative case using empty set as basis ($S=\emptyset$) is a special case, and we can move it to the implementation details in section 2.5?}

%\xy{Or if we want to keep the most conservative case here (only use empty set as basis), maybe we can simplify the current to $u(q,z)=\min_{z\in S}\Delta(q,S)$ since both $\Omega$ and $\{z\}$ can be $S$ and $\Delta(q,\Omega) \leq c(q,\emptyset)$?}

%In the following, we discuss further refinements of $u(q, z)$ based on the specific search algorithms.
%In particular, we focus our discussion on greedy-style configuration search algorithms that represent the current state of the art.

%\subsubsection{Vanilla Greedy}

%\subsubsection{Two-phase Greedy}

%\subsubsection{MCTS Greedy}

\subsection{Confidence-based What-if Call Skipping}
\label{sec:integration:confidence}
%\todo{Should we put it here or in the section of integration?}

Intuitively, the \emph{confidence} of skipping the what-if call for a QCP $(q, C)$ depends on the \emph{closeness} between the lower bound $L(q, C)$ and the upper bound $U(q, C)$, i.e., the derived cost $d(q, C)$.
We define the \emph{gap} between $U(q, C)$ and $L(q, C)$ as
$$G(q, C)=U(q, C) - L(q, C).$$
Clearly, the larger the gap is, the lower the confidence is. Therefore, it is natural to define the confidence as
\begin{equation}\label{eq:confidence}
\alpha(q, C)=1-\frac{G(q, C)}{U(q, C)}=\frac{L(q, C)}{U(q, C)}.
\end{equation}
Following this definition, we have $0\leq \alpha(q, C)\leq 1$.
We further note two special cases: (1) $\alpha(q, C)=0$, which implies $L(q, C)=0$; and (2) $\alpha(q, C)=1$, which implies $L(q, C)=U(q, C)$.

Let $\alpha\in[0,1]$ be a threshold for the confidence, i.e., it is the minimum confidence for skipping a what-if call and we require $\alpha(q, C)\geq \alpha$.
Intuitively, the higher $\alpha$ is, the higher confidence that a what-if call can be skipped with.
%\todo{This intuition somehow cannot be backed up by the experiment results. Need more verification.}
In our experimental evaluation, we further varied $\alpha$ to test the effectiveness of this confidence-based interception mechanism (see Section~\ref{sec:evaluation}).
%\todo{Also mention what matters more is the one-sided error? When summing up the errors they are likely to cancel each other and only the value of the sum matters. Perhaps in the earlier section where confidence was defined?}
%(e.g., $\alpha\in\{0.8, 0.9, 0.95, 0.99\}$).

%\section{Coverage-based Refinements}
\section{Optimization}
\label{sec:refinements}

We present two optimization techniques for the \emph{generic} lower bound detailed in Section~\ref{sec:framework:lower-bound}, which is \emph{agnostic} to budget-aware configuration enumeration algorithms---it only relies on general assumptions (i.e., monotonicity and submodularity) of the cost function $c$.
% One optimization is dedicated to budget-aware greedy search~\cite{WuWSWNCB22},
One optimization is dedicated to budget-aware greedy search (i.e., \emph{vanilla/two-phase greedy}),
which is of practical importance due to its adoption in commercial index tuning software~\cite{dta} (Section~\ref{sec:bounds:mci:greedy}).
% The other optimization is more general and can also be used for other configuration enumeration algorithms such as MCTS~\cite{WuWSWNCB22} (Section~\ref{sec:optimization:coverage}).
The other optimization is more general and can also be used for other configuration enumeration algorithms mentioned in Section~\ref{sec:framework:existing} such as \emph{MCTS} (Section~\ref{sec:optimization:coverage}).

%In this section, we propose techniques that further refine the lower-/upper-bounds studied in the previous section.

%\subsection{Refinement by Coverage}
%\label{sec:refinement:coverage}

\subsection{MCI Upper Bounds for Greedy Search}
\label{sec:bounds:mci:greedy}

\iffalse
The MCI upper-bounds $u(q, z)$ studied so far are \emph{agnostic} to budget-aware configuration enumeration algorithms---they only rely on general assumptions of the cost function $c$.
Most existing algorithms, however, employ a \emph{greedy search} strategy (see Section~\ref{sec:preliminaries}), which offers opportunities for further optimization of $u(q, z)$ by utilizing special properties of greedy search.
\fi

%\todo{Add motivation for the importance of this special case  (i.e., greedy search). For example, we can say that our decision on skipping what-if calls seem ``local''? Can we use more global information to improve? This greedy-based refinement can be such an example of leveraging global information.}

%The basic idea remains similar to that used in the call-level lower bound.
%We maintain an upper bound $u(q, z)$ for the MCI of each individual index $z$ and use that to compute the lower bound $L(q, C_B^*)$.
%Given that we use greedy search to tune the query $q$, we can further refine $u(q, z)$ instead of always using the most conservative value (i.e., Equation~\ref{eq:upper-bound:mci:call-level}).

We propose the following optimization procedure for maintaining the MCI upper-bound $u(q, z)$, which is the basic building block of the lower bound presented in Section~\ref{sec:framework:lower-bound}, in \emph{vanilla greedy} and \emph{two-phase greedy} (see Section~\ref{sec:preliminaries}):
\begin{procedure}\label{proc:maintain-mci}
For each index $z$ that \underline{has not been selected} by greedy search, we can update $u(q, z)$ w.r.t. the current configuration selected by greedy search as follows:
\begin{enumerate}
    \item Initialize $u(q, z)=\min\{c(q,\emptyset), \Delta(q, \Omega)\}$ for each index $z$.
    \item During each greedy step $1\leq k\leq K$, update 
    $$u(q, z)=c(q, C_{k-1})-c(q, C_{k-1}\cup\{z\})=\delta(q, z, C_{k-1})$$
    if \underline{both} $c(q, C_{k-1})$ and $c(q, C_{k-1}\cup\{z\})$ are available.
\end{enumerate}
\end{procedure}
In step (2), $C_k$ is the configuration selected by greedy search in step $k$ and we set $C_0=\emptyset$.
A special case is when $k=1$, if we know $c(q, \{z\})$ then we can update $u(q, z)=c(q, \emptyset)-c(q, \{z\})=\Delta(q,\{z\})$, which reduces to the \emph{general} upper bound (see Lemma~\ref{lemma:bound:submodular}).

\begin{theorem}
\label{theorem:greedy:mci-update}
Under Assumptions~\ref{assumption:monotone} and~\ref{assumption:submodular}, Procedure~\ref{proc:maintain-mci} is correct, i.e., the $u(q, z)$ after each update remains an MCI upper bound w.r.t. any future configuration $X$ explored by greedy search.
\end{theorem}

\subsection{Coverage-based Refinement}
\label{sec:optimization:coverage}

The tightness of the MCI upper bounds in Section~\ref{sec:bounds:call-level:mci} largely depends on the knowledge about $c(q, \{z\})$, namely, what-if costs of \emph{singleton} configurations with one single index.
Unfortunately, such information is often unavailable, and the MCI upper bound in Equation~\ref{eq:upper-bound:mci:call-level} is reduced to its naive version (Equation~\ref{eq:mci:trivial}).
For \emph{vanilla greedy} and \emph{two-phase greedy}, this implies that none of the QCP's with singleton configurations can be skipped under a reasonable confidence threshold (e.g., 0.8), which can take a large fraction of the budget, although the bounds are effective at skipping what-if calls for multi-index configurations;
for \emph{MCTS} where configurations are explored in a random order, this further implies that skipping can be less effective for multi-index configurations as they are more likely to contain indexes with unknown what-if costs, in contrast to greedy search where multi-index configurations are always explored after singleton configurations.
%e.g. in MCTS-based configuration enumeration where configurations are explored in a random order.
%(e.g., in the first phase of \emph{two-phase greedy}).
 %in these situations.
%even $u^{\max}(q,z)$ often can only take the trivial values as in Section~\ref{sec:bounds:call-level:mci}, 
%\todo{Need to show how tight the bound is before motivating coverage.}
To overcome this limitation, we propose refinement techniques based on \emph{estimating} the what-if cost $c(q, \{z\})$ if it is unknown, by introducing the notion of ``coverage.''
%We start by defining the coverage measure and then discuss techniques for estimating it. 
\iffalse
\xy{for all algorithms, MCI is trivial until cost of singleton configuration on a query is knonw. For vanilla and two-phase greedy, it means it cannot skip in step 1 and Wii is useless when budget is small. For MCTS, it means 1. budget used to collect "prior reward" are wasted due to spurious singletons and 2. cannot have a non-trivial bound during exploration since most singletons are unseen.}
\fi
%One shall keep in mind that using coverage-based singleton configuration cost estimates makes the bounds no longer strict.
%Nonetheless, in our experiments, we have observed that the bounds almost always remain bounds (and become much tighter) after using the estimated costs.
%\todo{We actually don't have such an experiment? Need to add if so.}

\subsubsection{Definition of Coverage}

We assume that $c(q,\Omega)$ is known for each query $q$.
Moreover, we assume that we know the subset $\Omega_q\subset \Omega$ of indexes that appear in the optimal plan of $q$ by using indexes in $\Omega$.
Clearly, $c(q,\Omega)=c(q,\Omega_q)$.

For an index $z$, we define its \emph{coverage} on the query $q$ as
\begin{equation}\label{eq:coverage}
    \rho(q, z)=\frac{c(q, \emptyset)-c(q, \{z\})}{c(q, \emptyset)-c(q, \Omega_q)}=\frac{\Delta(q, \{z\})}{\Delta(q, \Omega_q)}.
\end{equation}
In other words, coverage measures the \emph{relative cost improvement} of $z$ w.r.t. the maximum possible cost improvement over $q$ delivered by $\Omega_q$.
If we know $\rho(q, z)$, the cost $c(q, \{z\})$ can be recovered as
\begin{eqnarray*}
c(q, \{z\})&=& c(q, \emptyset)-\rho(q, z)\cdot\Big(c(q, \emptyset)-c(q, \Omega_q)\Big)\\
&=&\Big(1-\rho(q, z)\Big)\cdot c(q, \emptyset) + \rho(q, z)\cdot c(q, \Omega_q).
\end{eqnarray*}
%Since we do not know $c(q, \{z\})$, we cannot directly compute the coverage $\rho(q, z)$. 
In the following, we present techniques to estimate $\rho(q, z)$ based on the similarities between index configurations, in particular $\{z\}$ and $\Omega_q$.

%\todo{Add a figure here (like the one in the slides) to illustrate the notion of coverage.}

%One is based on similarity between configurations, and the other is based on machine learning.

\subsubsection{Estimation of Coverage}

We estimate coverage based on the assumption that it depends on the \emph{similarity} between $\{z\}$ and $\Omega_q$.
Specifically, let $\Sim(\{z\}, \Omega_q)$ be some \emph{similarity measure} that is between 0 and 1, and we define 
$$\rho(q,z)=\Sim(\{z\}, \Omega_q).$$
The problem is then reduced to developing an appropriate similarity measure.
Our current solution is the following, while further improvement is possible and left for future work.
%\todo{Add a figure for illustration purpose.}

\vspace{-0.5em}
\paragraph*{Configuration Representation}

We use a representation similar to the one described in \emph{DBA bandits}~\cite{PereraORB21} that converts an index $z$ into a feature vector $\vec{\mathbf{z}}$.
Specifically, we use \emph{one-hot encoding} based on all indexable columns identified in the given workload $W$.
Let $\mathcal{D}=\{c_1, ..., c_L\}$ be the entire domain of these $L$ indexable columns.
For a given index $z$, $\vec{\mathbf{z}}$ is an $L$-dimensional vector.
If some column $c_l\in\mathcal{D}$ ($1\leq l\leq L$) appears in $z$, then $\vec{\mathbf{z}}[l]$ receives some nonzero weight $w_l$ based on the \emph{weighing policy} described below:
\begin{itemize}
    \item If $c_l$ is the $j$-th \emph{key column} of $z$, $w_l=\frac{1}{2^{j-1}}$;
    \item If $c_l$ is an \emph{included column} of $z$, $w_l=\frac{1}{2^J}$ where $J$ is the number of key columns contained by $z$.
\end{itemize}
Otherwise, we set $\vec{\mathbf{z}}[l]=0$. 
Note that the above weighing policy considers the columns contained by an index as well as their order. Intuitively, leading columns in index keys play a more important role than other columns (e.g., for a ``range predicate'', an access path chosen by the query optimizer needs to match the ``sort order'' specified in the index key columns).

%contains nonzero entries for indexable columns appearing in $z$ and zero entries for all other dimensions.
%The indexable columns in $z$ receive different weights in $\bar{z}$, based on the 

%\todo{Add details on the representation.}
We further combine feature vectors of individual indexes to generate a feature vector for the entire configuration. Specifically, consider a configuration $C=\{z_1, ..., z_m\}$ and let $\vec{\mathbf{z}}_i$ be the feature representation of the index $z_i$ ($1\leq i\leq m$).
The feature representation $\vec{\mathbf{C}}$ of $C$ is again an $L$-dimensional vector where
$$\vec{\mathbf{C}}[l]=\max\{\vec{\mathbf{z}}_1[l], ..., \vec{\mathbf{z}}_m[l]\}, \text{ for } 1\leq l\leq L.$$
That is, the weight $\vec{\mathbf{C}}[l]$ is the largest weight of the $l$-th dimension among the indexes contained by $C$. In particular, we generate the feature vector $\vec{\mathbf{\Omega}}_q$ for $\Omega_q$ in this way.

\vspace{-0.5em}
\paragraph*{Query Representation}
We further use a representation similar to the one described in \emph{ISUM}~\cite{SiddiquiJ00NC22} to represent a query $q$ as a feature vector $\vec{\mathbf{q}}$.
Specifically, we again use one-hot encoding for the query $q$ with the same domain $\mathcal{D}=\{c_1, ..., c_L\}$ of all indexable columns.
If some column $c_l\in\mathcal{D}$ appears in the query $q$, we assign a nonzero weight to $\vec{\mathbf{q}}[l]$; otherwise, $\vec{\mathbf{q}}[l]=0$. 
Here, we use the same weighing mechanism as used by \emph{ISUM}. That is, the weight of a column is computed based on its corresponding table size and the number of candidate indexes that contain it.
The intuition is that a column from a larger table and contained by more candidate indexes is more important and thus is assigned a higher weight.

% Here we use the same weighing mechanism as used by \emph{ISUM} (see~\cite{SiddiquiJ00NC22} for the details).

%\paragraph*{Configuration Representation Conditioned on Query}

\vspace{-0.5em}
\paragraph*{Similarity Measure}
% We project $\vec{\mathbf{z}}$ and $\vec{\mathbf{\Omega}}_q$ onto $\vec{\mathbf{q}}$ to get their \emph{images} under the projection.
Before measuring the similarity, we first project $\vec{\mathbf{z}}$ and $\vec{\mathbf{\Omega}}_q$ onto $\vec{\mathbf{q}}$ to get their \emph{images} under the context of the query $q$. 
The projection is done by taking the \emph{element-wise dot product}, i.e., $\Tilde{\mathbf{z}}=\vec{\mathbf{z}}\cdot \vec{\mathbf{q}}$ and $\Tilde{\mathbf{\Omega}}_q=\vec{\mathbf{\Omega}}_q\cdot\vec{\mathbf{q}}$.
Note that $\Tilde{\mathbf{z}}$ and $\Tilde{\mathbf{\Omega}}_q$ remain vectors.
We now define the similarity measure as
\begin{equation}\label{eq:coverage:estimated}
\Sim(\{z\}, \Omega_q)=\frac{\langle \Tilde{\mathbf{z}}, \Tilde{\mathbf{\Omega}}_q \rangle}{|\Tilde{\mathbf{\Omega}}_q|^2}=\frac{|\Tilde{\mathbf{z}}|\cdot|\Tilde{\mathbf{\Omega}}_q|\cdot\cos\theta}{|\Tilde{\mathbf{\Omega}}_q|^2}=\frac{|\Tilde{\mathbf{z}}|\cdot\cos\theta}{|\Tilde{\mathbf{\Omega}}_q|},  
\end{equation}
where $\theta$ represents the \emph{angle} between the two vectors $\Tilde{\mathbf{z}}$ and $\Tilde{\mathbf{\Omega}}_q$.

\begin{figure}
\centering
    \includegraphics[clip, trim=4cm 7.5cm 10cm 5cm, width=.7\columnwidth]{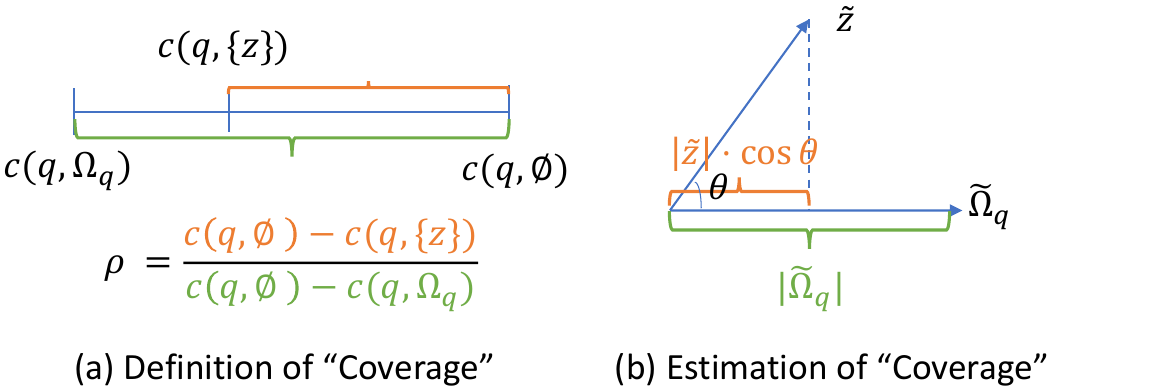}
\vspace{-1em}
\caption{The definition and estimation of ``coverage.''}
\label{fig:coverage}
%\vspace{0.5em}
\end{figure}

Figure~\ref{fig:coverage} illustrates and contrasts the definition and estimation of coverage.
Figure~\ref{fig:coverage}(a) highlights the observation that $c(q, \{z\})$ must lie between $c(q, \Omega_q)$ and $c(q,\emptyset)$, and coverage measures the cost improvement $\Delta(q, \Omega_q)$ of $\Omega_q$ (i.e., the green segment) that is \emph{covered} by the cost improvement $\Delta(q, \{z\})$ of $\{z\}$ (i.e., the orange segment).
On the other hand, Figure~\ref{fig:coverage}(b) depicts the geometric view involved in the estimation of coverage using the similarity metric $\Sim(\{z\}, \Omega_q)$.
%\todo{Add more description of the figure.}
Intuitively, the similarity measures how much ``length'' of the configuration $\Omega_q$ is \emph{covered} by the ``length'' of the index $z$ when projected to the (same) ``direction'' of $\Omega_q$ in the feature vector space.
Note that it is not important whether the lengths are close to the corresponding cost improvements---only their \emph{ratio} matters.
Based on our evaluation, the estimated coverage using Equation~\ref{eq:coverage:estimated} is close to the ground-truth coverage in Equation~\ref{eq:coverage} (see the full version of this paper~\cite{full-version} for details).%Section~\ref{sec:evaluation:coverage:accuracy}).
%\todo{Consider adding a concrete example for computing coverage?}

%\todo{Add some statement about how accurate the estimation of the ratio is in practice?}
%\todo{Add a figure to illustrate.}
%This matches our motivation of definition of \emph{coverage}. 
%\xy{Should we mention the assumptions made here like assuming the columns in each index are independent to the cost improvement?}

%\ww{We observe on TPC-H that the estimated cost of a singleton configuration by similarity-based coverage is typically lower than the actual cost, which indicates that it is a more conservative estimation that  more likely leads to a lower bound on the actual cost.}

%\todo{Add some experiments to validate the above claim.}

%\ww{Explanation of the above observation?}

%\ww{(1) We have a vector representation of each index based on indexable columns. (2) We merge index vectors into a configuration vector with a weighing mechanism. (3) We have a vector representation of each query based on ISUM. (4) We project the index/configuration vectors onto the query vectors. (5) We use this projection ratio as the similarity.}

\section{Integration}
\label{sec:integration}

In this section, we present design considerations and implementation details when integrating \sysname with existing budget-aware configuration search algorithms.
We start by presenting the API functions provided by \sysname.
We then illustrate 
%how the budget-aware greedy search procedure can use these \sysname API functions without any modification to the procedure itself.We end this section by describing 
how existing budget-aware configuration enumeration algorithms can leverage 
%the budget-aware greedy search procedure based on 
the \sysname API's without modification to the algorithms.
%We close this section by discussing how to set the threshold $\epsilon$ and other issues

%\subsection{Confidence-based Interception}
%\label{sec:integration:confidence}

\subsection{Wii API Functions}

As illustrated in Figure~\ref{fig:what-if-architecture}, \sysname sits between the index tuner and the query optimizer. It offers two API functions that can be invoked by a budget-aware configuration enumeration algorithm: (1) \texttt{InitMCIBounds} that initializes the MCI upper-bounds $u(q, z)$; and (2) \texttt{EvalCost} that obtains the cost of a QCP $(q, C)$ in a budget-aware manner by utilizing the lower bound $L(q, C)$ and the upper bound $U(q, C)$, i.e., the derived cost $d(q, C)$.
%We present the details of these two API functions in Algorithms~\ref{alg:init-mci-bounds} and~\ref{alg:get-cost-bound}.

\begin{algorithm}[t]
  \caption{\texttt{InitMCIBounds}($W$, $I$)}
  \label{alg:init-mci-bounds}
\small
  \SetAlgoLined
  \KwIn{$W$, the workload; $I$, the candidate indexes.}
  \KwOut{$u$, the initialized MCI upper bounds.}
  \SetAlgoLined
    \ForEach{$q \in W$}{
        $I_q\leftarrow$\texttt{GetCandidateIndexes}($q$, $I$)\;
        \ForEach{$z \in I_q$}{
            %\If{$u(q, z)$ has not been initialized}{
                \If{$c(q, \Omega_q)$ is available}{
                    $u(q,z) \leftarrow c(q,\emptyset) - c(q, \Omega_q)$\;
                }\Else{
                    $u(q,z) \leftarrow c(q,\emptyset)$\;
                }
            %}
        }
    }
\end{algorithm}

\subsubsection{The \texttt{InitMCIBounds} Function}

Algorithm~\ref{alg:init-mci-bounds} presents the details.
It initializes the MCI upper bound $u(q, z)$ for each query $q\in W$ and each of its candidate indexes $z\in I_q$.
If $c(q, \Omega_q)$ is available, it uses the naive upper bound (Equation~\ref{eq:mci:trivial}); otherwise, it uses $c(q, \emptyset)$.

\subsubsection{The \texttt{EvalCost} Function}

%Algorithm~\ref{alg:get-cost-bound} computes the cost of $(q, C)$ with respect to the current configuration $C^*$ selected by greedy search.
Algorithm~\ref{alg:get-cost-bound} presents the details.
If the what-if cost $c(q, C)$ is known, it simply uses that and updates the MCI upper-bounds %by calling the function \texttt{UpdateMCIBounds} 
(lines 1 to 3). Otherwise, it checks whether the budget $B$ on the number of what-if calls has been reached and returns the derived cost $d(q, C)$ if so (lines 4 to 5).
On the other hand, if there is remaining budget, i.e., $B>0$, it then tries to use the upper-bound $U(q, C)$ and the lower-bound $L(q, C)$ to see whether the what-if call for $(q, C)$ can be skipped; if so, the derived cost $d(q, C)$ is returned (lines 6 to 11)---the budget $B$ remains the same in this case.
Finally, if the confidence of skipping is low, we make one what-if call to obtain $c(q, C)$ (lines 12 to 13) and update the MCI upper-bounds %for the relevant indexes 
(line 14).
As a result, we deduct one from the current budget $B$ (line 15).
%\todo{Refine the above, and explain more.}

\begin{algorithm}[b]
\caption{\texttt{EvalCost}($q$, $C$, $B$, $\alpha$, $S\leftarrow\emptyset$)}
\label{alg:get-cost-bound}
\small
  \SetAlgoLined
  \KwIn{$q$, the query; $C$, the configuration; $B$, the budget on the number of what-if calls; $\alpha$, the threshold on the confidence $\alpha(q, C)$; $S$, an (optional) subset of $C$ with known what-if cost $c(q, S)$, which defaults to the existing configuration $\emptyset$.}
  \KwOut{$\cost(q, C)$, the cost of $q$ w.r.t. $C$; $B'$, the remaining budget.}
  \SetAlgoLined
  %$\cost(q, C)\leftarrow d(q, C)$\;
  \uIf{$c(q, C)$ is known} {
    \texttt{UpdateMCIBounds}($C$, $S$)\;
    \Return{\Big($\cost(q, C)\leftarrow c(q, C)$, $B'\leftarrow B$\Big)\;}
  }
  \uIf{$B$ is zero}{
    \Return{\Big($\cost(q, C)\leftarrow d(q, C)$, $B'\leftarrow 0$\Big)\;}
  }
  \#\# $c(q, C)$ is unknown and we still have budget.\\
  $U(q, C)\leftarrow d(q, C)$\;
  $L(q, C)\leftarrow\max\{0, c(q, \Omega_q), c(q, S)-\sum_{x\in C-S}u(q, x)\}$\;
  \If{$\alpha(q, C)=\frac{L(q, C)}{U(q, C)}\geq\alpha$}{
    \#\# The confidence is high enough.\\
    \Return{\Big($\cost(q, C)\leftarrow d(q, C)$, $B'\leftarrow B$\Big)\;}
  }
  \#\# Need to go to the query optimizer to get $c(q, C)$.\\
  $c(q, C)\leftarrow$\texttt{WhatIfCall}($q$, $C$)\;
  \texttt{UpdateMCIBounds}($C$, $S$)\;
  \Return{\Big($\cost(q, C)\leftarrow c(q, C)$, $B'\leftarrow B-1$\Big);}
  
\quad\\
\underline{\textbf{UpdateMCIBounds}}($C$, $S$)\\
    \#\# Update the MCI bounds based on $c(q, C)$ and $c(q, S)$.\\
  \ForEach{$x\in C-S$}{
    $u(q,x) \leftarrow\min\{u(q,x), c(q, S)-c(q, C)\}$\;
  }
\end{algorithm}

One may have noticed the optional input parameter $S$ in Algorithm~\ref{alg:get-cost-bound}, which represents some subset configuration of $C$ and is set to be the existing configuration $\emptyset$ by default.
We will discuss how to specify this parameter when using \sysname in existing budget-aware configuration enumeration algorithms (e.g., greedy search and MCTS) shortly.

\subsection{Budget-aware Greedy Search}

To demonstrate how to use the \sysname API's without modifying the existing budget-aware configuration search algorithms, Algorithm~\ref{alg:greedy-bound} showcases how these API's can be used by budget-aware greedy search, a basic building block of the existing algorithms.
% Notice that the \texttt{InitMCIBounds} API is invoked at line 1, whereas the \texttt{EvalCost} API is invoked at line 9, which are the only two differences compared to regular budget-aware greedy search (ref. Algorithm 1 in~\cite{WuWSWNCB22}).
Notice that the \texttt{InitMCIBounds} API is invoked at line 1, whereas the \texttt{EvalCost} API is invoked at line 9, which are the only two differences compared to regular budget-aware greedy search.
Therefore, there is no \emph{intrusive} change to the greedy search procedure itself.
%In other words, there is no \emph{intrusive} change to the greedy search procedure itself---only the cost evaluation call is delegated to \sysname, and the call to initialize the MCI bounds is completely external to greedy search.

%Algorithm~\ref{alg:greedy-bound} 
%Compared to regular budget-aware greedy search, it only introduces one additional input parameter $\epsilon$, which represents the threshold on the gap between the lower-bound $L(q, C)$ and the upper-bound $U(q, C)$ of $c(q, C)$.
%We will discuss how to set $\epsilon$ in Section~\ref{sec:skipping:threshold}.
%In addition, it needs to 
%Apart from that, there is no change to the greedy search algorithm except for the implementation of %the \texttt{GetCost} function detailed in Algorithm~\ref{alg:get-cost-bound}.

\vspace{-0.5em}
\paragraph*{Remarks}
We have two remarks here. First, when calling \sysname to evaluate cost at line 9, we pass $C^*$ to the optional parameter $S$ in Algorithm~\ref{alg:get-cost-bound}.
Note that this is just a special case of Equation~\ref{eq:lower-bound:call-level:v2} for greedy search, as stated by the following theorem:
%, since $C^*$ is the subset of $C$ with the minimum cost, assuming submodularity of the cost function $c$ (Assumption~\ref{assumption:submodular}).
%\todo{Maybe we can add a formal proof here that Equation~\ref{eq:lower-bound:call-level:v2} reduces to just using $C^*$ in the case of greedy search?}
\begin{theorem}\label{theorem:lower-bound:greedy-search}
In the context of greedy search, Equation~\ref{eq:lower-bound:call-level:v2} reduces to
$$L(q, C_z)=c(q, C^*)-\sum\nolimits_{x\in C_z-C^*}u(q, x)=c(q, C^*)-u(q, z),$$
where $C_z=C^*\cup\{z\}$ and $C^*$ is the latest configuration selected by budget-aware greedy search (as shown in Algorithm~\ref{alg:greedy-bound}).
\end{theorem}

Second, in the context of greedy search, the update step at line 20 of Algorithm~\ref{alg:get-cost-bound} becomes
$$u(q, x)\leftarrow\min\{u(q,x), c(q, C^*)-c(q,C)\}.$$
The correctness of this update has been given by Theorem~\ref{theorem:greedy:mci-update}.
%\todo{Algorithm~\ref{alg:get-cost-bound} should be refined to make it usable for both greedy search and MCTS.}

\begin{algorithm}[h]
  \caption{\texttt{GreedySearch}($W$, $I$, $K$, $B$, $\alpha$)}
  \label{alg:greedy-bound}
\small
  \SetAlgoLined
  \KwIn{$W$, the workload; $I$, the candidate indexes; $K$, the cardinality constraint; $B$, the budget on the number of what-if calls; $\alpha$, the confidence threshold.}
  \KwOut{$C^*$, the best configuration; $B'$, the remaining budget.}
  \SetAlgoLined
  \texttt{InitMCIBounds}($W$, $I$)\;
  $C^*\leftarrow\emptyset$, $\cost^*\leftarrow\cost(W,\emptyset)$,
  $B'\leftarrow B$\;
  \While{$I\neq\emptyset$ and $|C^*|<K$} {
    $C\leftarrow C^*$, $\cost\leftarrow\cost^*$\;
    \ForEach{index $z\in I$} {
        $C_z\leftarrow C^*\cup\{z\}$\;
        $\cost(W, C_z)\leftarrow 0$\;
        \ForEach{$q\in W$}{
            $\Big(\cost(q, C_z), B'\Big)\leftarrow$\texttt{EvalCost}$(q, C_z, B', \alpha, C^*)$\;
            $\cost(W, C_z)\leftarrow\cost(W, C_z)+\cost(q, C_z)$\;
        %$\Big(\cost(W, C_z), B'\Big)\leftarrow\sum_{q\in W}$\texttt{EvalCost}$(q, C_z, B', \epsilon, C^*)$\;
        }
        \uIf{$\cost(W, C_z)<\cost$}{
            $C\leftarrow C_z$, $\cost\leftarrow\cost(W, C_z)$\;
        }
    }
    \uIf{$\cost\geq\cost^*$} {
        \Return{\Big($C^*$, $B'$\Big)\;}
    }\uElse{
        $C^*\leftarrow C$, $\cost^*\leftarrow\cost$, $I\leftarrow I-C^*$\;
    }
  }
  \Return{\Big($C^*$, $B'$\Big)\;}
%\vspace{-0.5em}
\end{algorithm}

\vspace{-1em}
\subsection{Budget-aware Configuration Enumeration}

We now outline the skeleton of existing budget-aware configuration enumeration algorithms after integrating \sysname. We use the integrated budget-aware greedy search procedure in Algorithm~\ref{alg:greedy-bound} as a building block in our illustration.

\subsubsection{Vanilla Greedy} The \emph{vanilla greedy} algorithm after integrating \sysname is exactly the same as the \texttt{GreedySearch} procedure presented by Algorithm~\ref{alg:greedy-bound}.

\subsubsection{Two-phase Greedy}

Algorithm~\ref{alg:two-phase-greedy} presents the details of the \emph{two-phase greedy} algorithm after integrating \sysname. 
There is no change to \emph{two-phase greedy} 
% (compared to Algorithm 2 in~\cite{WuWSWNCB22}) 
except for using the version of \texttt{GreedySearch} in Algorithm~\ref{alg:greedy-bound}.
The function \texttt{GetCandidateIndexes} selects a subset of candidate indexes $I_q$ from $I$, considering only the indexable columns contained by the query $q$~\cite{ChaudhuriN97}.

\begin{algorithm}[ht]
  \caption{\texttt{TwoPhaseGreedy}($W$, $I$, $K$, $B$, $\alpha$)}
  \label{alg:two-phase-greedy}
\small
  \SetAlgoLined
  \KwIn{$W$, the workload; $I$, the candidate indexes; $K$, the cardinality constraint; $B$, the budget on the number of what-if calls; $\alpha$, the confidence threshold.}
  \KwOut{$C^*$, the best configuration; $B'$, the remaining budget.}
  \SetAlgoLined
  $I_W\leftarrow\emptyset$,
  $B'\leftarrow B$\;
  \ForEach{$q\in W $} {
    $I_q\leftarrow$\texttt{GetCandidateIndexes}($q$, $I$)\;
    $\Big(C_q, B'\Big)\leftarrow$\texttt{GreedySearch}($\{q\}$, $I_q$, $K$, $B'$, $\alpha$)\;
    $I_W\leftarrow I_W\cup C_q$\;
  }
  $\Big(C^*, B'\Big)\leftarrow$\texttt{GreedySearch}($W$, $I_W$, $K$, $B'$, $\alpha$)\;
  \Return{$\Big(C^*, B'\Big)$}\;
\end{algorithm}

\subsubsection{MCTS}

%can be used in either \emph{vanilla greedy} or the second phase of \emph{two-phase greedy}.
%The \texttt{GetCost} procedure in Algorithm~\ref{alg:greedy-bound} can be further used in the first phase of \emph{two-phase greedy} as well as \emph{MCTS greedy}, by setting the parameter $C^*=\emptyset$.

Algorithm~\ref{alg:mcts-greedy} presents the skeleton of \emph{MCTS} after \sysname is integrated. The details of the three functions \texttt{InitMCTS}, \texttt{SelectQueryConfigByMCTS}, and \texttt{UpdateRewardForMCTS} can be found in~\cite{WuWSWNCB22}.
% The details of the three functions \texttt{InitMCTS}, \texttt{SelectQueryConfigByMCTS}, and \texttt{UpdateRewardForMCTS} can be found in~\cite{WuWSWNCB22}.
Again, there is no change to the \emph{MCTS} algorithm except for that cost evaluation at line 5 is delegated to the \texttt{EvalCost} API of \sysname (Algorithm~\ref{alg:get-cost-bound}).
%Moreover, the final greedy search step at line 7 is now the version presented in Algorithm~\ref{alg:greedy-bound}. However, since $B'$ is zero here, 

Note that here we pass the existing configuration $\emptyset$ to the optional parameter $S$ in Algorithm~\ref{alg:get-cost-bound},
which makes line 8 of Algorithm~\ref{alg:get-cost-bound} on computing $L(q, C)$ become
$$L(q, C)\leftarrow\max\{0, c(q, \Omega_q), c(q, \emptyset)-\sum\nolimits_{x\in C}u(q, x)\}.$$
Essentially, this means that we use Equation~\ref{eq:lower-bound:call-level:v1} for $L(q, C)$, instead of its generalized version shown in Equation~\ref{eq:lower-bound:call-level:v2}.
Although we could have used Equation~\ref{eq:lower-bound:call-level:v2}, it was our design decision to stay with Equation~\ref{eq:lower-bound:call-level:v1}, not only for simplicity but also because of the inefficacy of Equation~\ref{eq:lower-bound:call-level:v2} in the context of \emph{MCTS}.
This is due to the fact that in \emph{MCTS} configurations and queries are explored in random order. Therefore, the subsets $S$ w.r.t. a given pair of $q$ and $C$ with known what-if costs $c(q, S)$ are \emph{sparse}.
As a result, Equation~\ref{eq:lower-bound:call-level:v2} often reduces to Equation~\ref{eq:lower-bound:call-level:v1} when running \sysname underlying \emph{MCTS}.

\begin{algorithm}[ht] %[H]
  \caption{\texttt{MCTS}($W$, $I$, $K$, $B$, $\tau$)}
  \label{alg:mcts-greedy}
\small
  \SetAlgoLined
  \KwIn{$W$, the workload; $I$, the candidate indexes; $K$, the cardinality constraint; $B$, the budget on the number of what-if calls; $\alpha$, the confidence threshold.}
  \KwOut{$C^*$, the best configuration; $B'$, the remaining budget.}
  \SetAlgoLined
  $B'\leftarrow B$\;
  \texttt{InitMCTS}($W$, $I$)\;
  \While{$B'>0$}{
    $(q,C)\leftarrow$\texttt{SelectQueryConfigByMCTS}($W$, $I$, $K$)\;
    $\Big(\cost(q, C), B'\Big)\leftarrow$\texttt{EvalCost}$(q, C, B', \alpha, \emptyset)$\;
    \texttt{UpdateRewardForMCTS}($q$, $C$, $\cost(q, C)$)\;
  }
  $\Big(C^*, B'\Big)\leftarrow$\texttt{GreedySearch}($W$, $I$, $K$, $B'$, $\alpha$)\;
  \Return{$\Big(C^*,B'\Big)$}\;
\end{algorithm}

%\subsection{Discussion}

%\todo{Add discussion on replacing the cost derivation API, as discussed with Vivek?}

%Although we chose to use the derived cost $d(q, C)$ as the upper-bound $U(q, C)$ in IQ-booster, this is just an implementation
\section{Experimental Evaluation}
\label{sec:evaluation}

%\ww{Plan for the experimental evaluation: Let's finish skipping experiments first and then finish the rest of the early-exit experiments.}

%\ww{In addition to the optimal threshold setting problem (for both call-level and query-level skipping), another question is that they are competitors? Maybe we shall omit query-level skipping in this paper (and move it to the early-exit paper)?}

We now report experimental results on evaluating \sysname when integrated with existing budget-aware configuration search algorithms.
We perform all experiments using Microsoft SQL Server 2017 under Windows Server 2022, running on a workstation equipped with 2.6 GHz multi-core AMD CPUs and 256 GB main memory.

\subsection{Experiment Settings}

\paragraph*{Datasets} 

We used standard benchmarks and real workloads in our study.
Table~\ref{tab:databases} summarizes the information of the workloads. %\footnote{The filters here are those that cannot be pushed down into table scans.}
For benchmark workloads, we use 
both the \textbf{TPC-H} and \textbf{TPC-DS} benchmarks with scaling factor 10.
We also use two real workloads, denoted by \textbf{Real-D} and \textbf{Real-M} in Table~\ref{tab:databases}, which are significantly more complicated compared to the benchmark workloads, in terms of schema complexity (e.g., the number of tables), query complexity (e.g., the average number of joins and table scans contained by a query), and database/workload size.
Moreover, we report the number of candidate indexes of each workload, which serves as an indicator of the size of the corresponding search space faced by an index configuration search algorithm.

\vspace{-0.5em}
\paragraph*{Algorithms Evaluated}
We focus on two state-of-the-art budget-aware configuration search algorithms described in Section~\ref{sec:preliminaries}: (1) \emph{two-phase greedy}, which has been adopted by commercial index tuning software~\cite{dta}; 
% and (2) \emph{MCTS}, which shows better performance than \emph{two-phase greedy}~\cite{WuWSWNCB22}.
and (2) \emph{MCTS}, which shows better performance than \emph{two-phase greedy}.
We omit \emph{vanilla greedy} as it is significantly inferior to \emph{two-phase greedy}~\cite{WuWSWNCB22}. 
Both \emph{two-phase greedy} and \emph{MCTS} use derived cost as an estimate for the what-if cost when the budget on what-if calls is exhausted.
We evaluate \sysname when integrated with the above configuration search algorithms.

\vspace{-0.5em}
\paragraph*{Other Experimental Settings}
In our experiments, we set the cardinality constraint $K\in\{$10, 20$\}$.
Since the \textbf{TPC-H} workload is relatively small compared to the other workloads, we varied the budget $B$ on the number of what-if calls in $\{500, 1000\}$; for the other workloads, we varied the budget $B$ in $\{500, 1000, 2000, 5000\}$.
%Moreover, we varied the confidence threshold $\alpha$, which measures the gap between the lower and upper bounds of what-if cost, as \revision{$\alpha\in\{0.5, 0.8, 0.9, 0.95\}$}.

\begin{table}[h]%[b]%[!htb]
\small
\centering
%\begin{tabularx}{\columnwidth}{|l|l|X|X|X|X|X|}
\begin{tabular}{|l|r|r|r|r|r|r|}
\hline
\textbf{Name} & \textbf{DB Size} & \textbf{\#} \textbf{Queries} & \textbf{\#} \textbf{Tables} & \textbf{Avg. \#} \textbf{Joins} 
%& \textbf{Avg. \#} \newline \textbf{Filters} 
& \textbf{Avg. \#} \textbf{Scans}
& \textbf{\# Candidate} \newline \textbf{Indexes}\\
\hline
\hline
%\textbf{JOB} & 9.2GB & 33 & 21 & 7.9 & 2.5 & 8.9\\
\textbf{TPC-H} & \emph{sf}=10 & 22 & 8 & 2.8 %& 0.3 
& 3.7 & 168\\
\textbf{TPC-DS} & \emph{sf}=10 & 99 & 24 & 7.7 %& 0.5 
& 8.8 & 848\\
\hline
\hline
\textbf{Real-D} & 587GB & 32 & 7,912 & 15.6 %& 0.2 
& 17 & 417\\
\textbf{Real-M} & 26GB & 317 & 474 & 20.2 %& 1.5 
& 21.7 & 4,490\\
%Real-3 & 100GB & 20\\
\hline
%\end{tabularx}
\end{tabular}
\caption{Summary of database and workload statistics.}
\vspace{-3.5em}
\label{tab:databases}
\end{table}

%\vspace{-0.5em}
\iffalse
\paragraph*{Remark}
Due to space constraints, we defer the experimental results on \textbf{TPC-H}, to~\cite{full-version}.
The observations on \textbf{TPC-H} are similar.
\fi

\subsection{End-to-End Improvement}
\label{sec:evaluation:e2e}

The evaluation metric used in our experiments is the \emph{percentage improvement} of the workload based on the final index configuration found by a search algorithm, defined as
%$\eta(W, C)=\Big(1-\frac{c(W,C)}{c(W,\emptyset)}\Big)\times 100\%,$
\begin{equation}\label{eq:percentage-improvement:workload}
    \eta(W, C)=\Big(1-\frac{c(W,C)}{c(W,\emptyset)}\Big)\times 100\%,
\end{equation}
where $c(W,C)=\sum_{q\in W}c(q, C)$.
Note that here we use the query optimizer's what-if cost estimate $c(q, C)$ as the gold standard of query execution cost, instead of using the actual query execution time, to be in line with previous work on evaluating index configuration enumeration algorithms~\cite{ChaudhuriN97,KossmannHJS20}.

%Due to space limitation, we only include the results on \textbf{TPC-DS}, \textbf{Real-D}, and \textbf{Real-M} in the paper. We refer the readers to the full version~\cite{full-version} for the results on \textbf{TPC-H}.

%\todo{Add TPC-H results back.}

\subsubsection{Two-phase Greedy}

%%%%%%%%%%%% Two-phase Greedy (call-level) %%%%%%%%%%

\begin{figure*}
\centering
\subfigure[\textbf{TPC-H}, $K=10$]{ \label{fig:two-phase:call-level:tpch:K10}
    \includegraphics[width=0.23\textwidth]{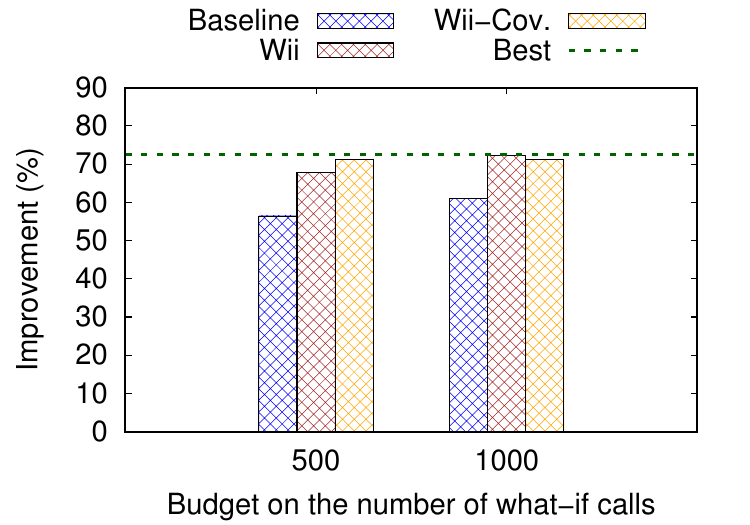}}
\subfigure[\textbf{TPC-DS}, $K=10$]{ \label{fig:two-phase:call-level:tpcds:K10}
    \includegraphics[width=0.23\textwidth]{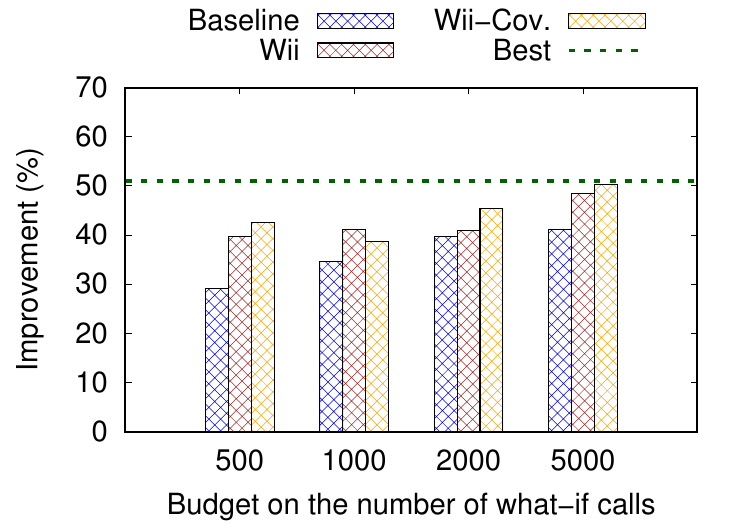}}
\subfigure[\textbf{Real-D}, $K=10$]{ 
\label{fig:two-phase:call-level:real-d:K10}
    \includegraphics[width=0.23\textwidth]{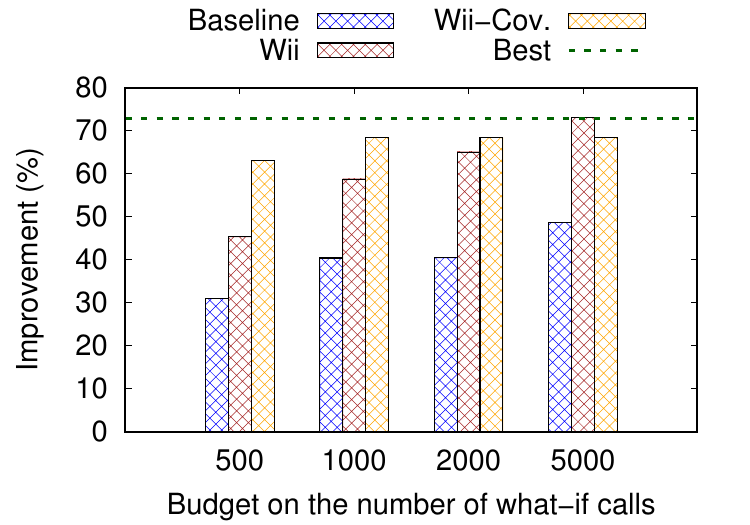}}
\subfigure[\textbf{Real-M}, $K=10$]{ \label{fig:two-phase:call-level:real-m:K10}
    \includegraphics[width=0.23\textwidth]{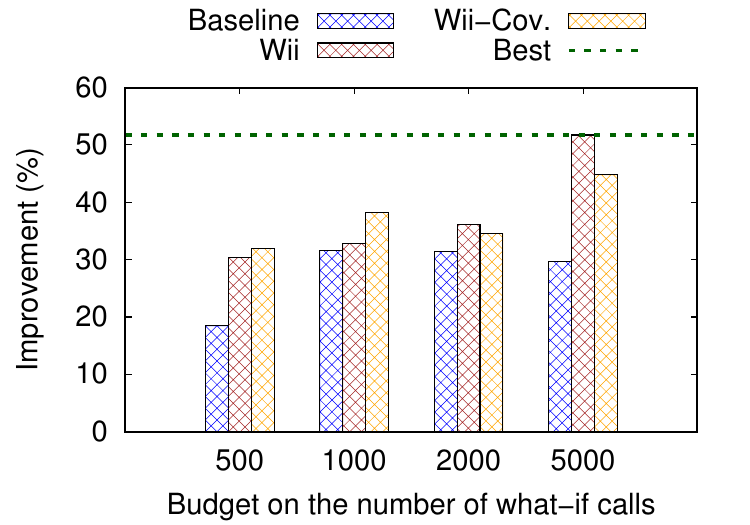}}
\subfigure[\textbf{TPC-H}, $K=20$]{ \label{fig:two-phase:call-level:tpch:K20}
    \includegraphics[width=0.23\textwidth]{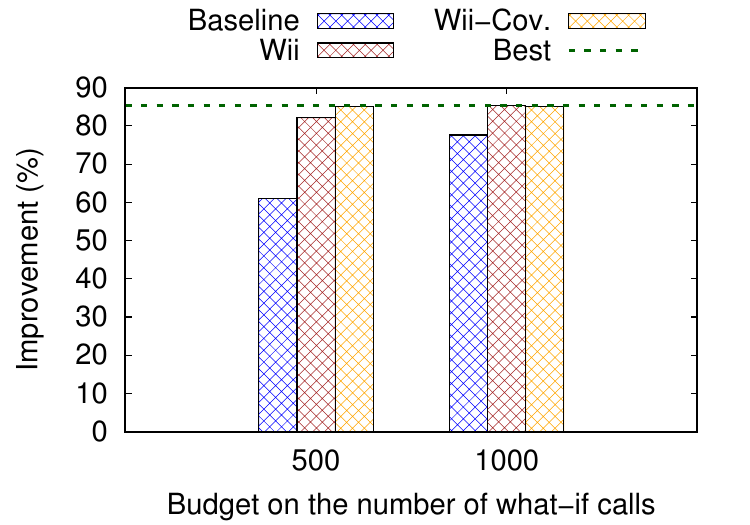}}
\subfigure[\textbf{TPC-DS}, $K=20$]{ \label{fig:two-phase:call-level:tpcds:K20}
    \includegraphics[width=0.23\textwidth]{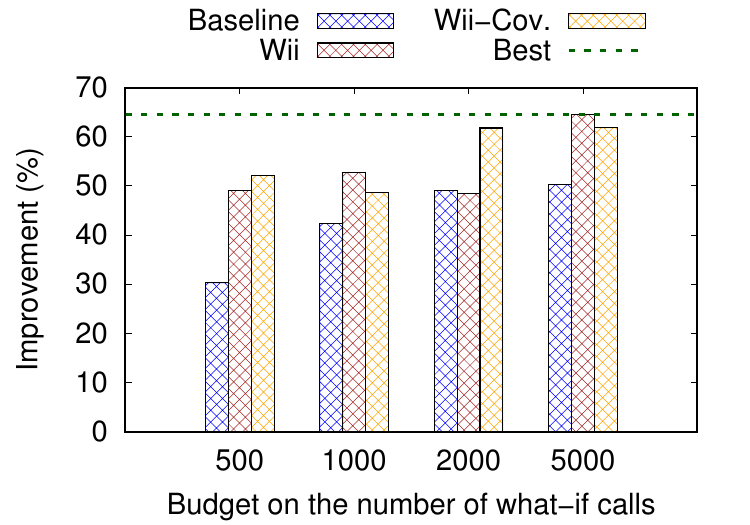}}
\subfigure[\textbf{Real-D}, $K=20$]{ \label{fig:two-phase:call-level:real-d:K20}
    \includegraphics[width=0.23\textwidth]{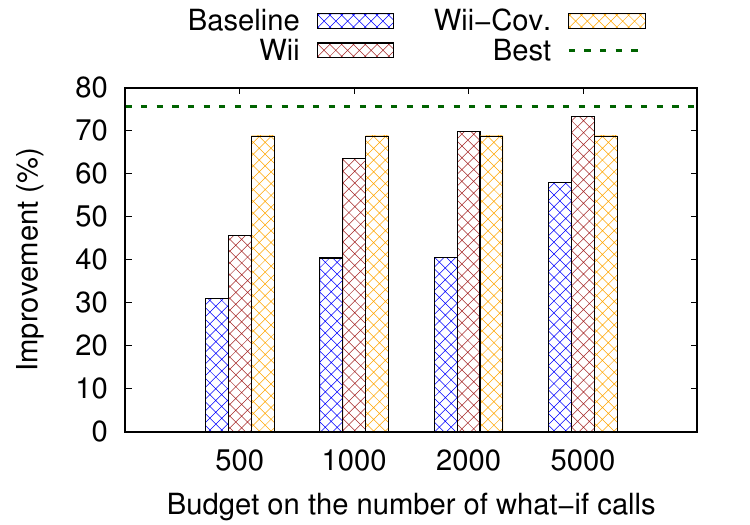}}
\subfigure[\textbf{Real-M}, $K=20$]{ \label{fig:two-phase:call-level:real-m:K20}
    \includegraphics[width=0.23\textwidth]{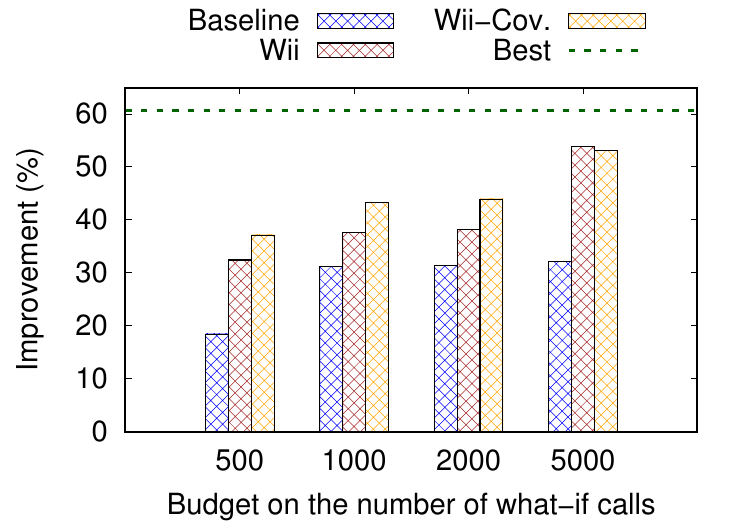}}
\vspace{-1.5em}
\caption{Results for \emph{two-phase greedy} with confidence threshold $\alpha=0.9$ (``Cov.'' is shorthand for ``Coverage'').}
%\xy{split two figures (one per budget)? Coverage -> Wii-cov?}}
\label{fig:two-phase:call-level}
\vspace{-1.5em}
\end{figure*}

%%%%%%%%%%%% Two-phase Greedy No MCI Optimization %%%%%%%%%%

\begin{figure*}
\centering
\subfigure[\textbf{TPC-H}, $B=1,000$]{ \label{fig:two-phase:no-opt-mci:tpch}
    \includegraphics[width=0.23\textwidth]{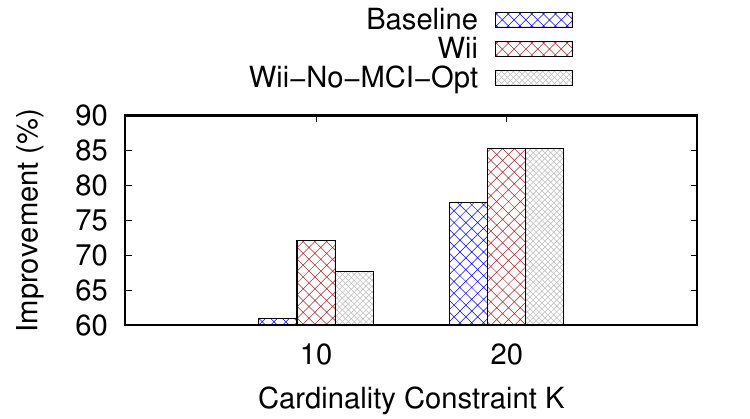}}
\subfigure[\textbf{TPC-DS}, $B=5,000$]{ \label{fig:two-phase:no-opt-mci:tpcds}
    \includegraphics[width=0.23\textwidth]{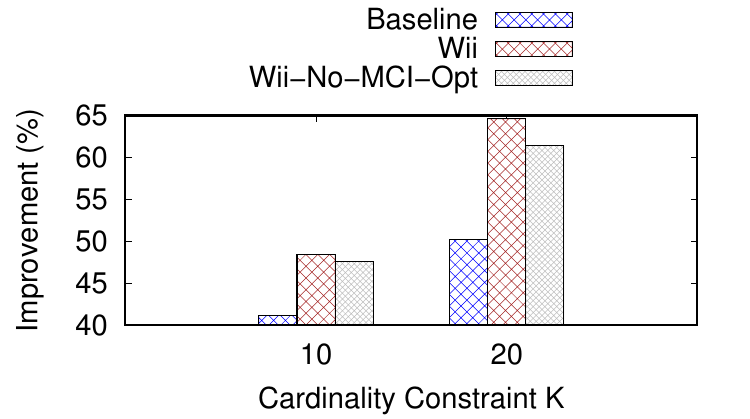}}
\subfigure[\textbf{Real-D}, $B=5,000$]{ 
\label{fig:two-phase:no-opt-mci:real-d}
    \includegraphics[width=0.23\textwidth]{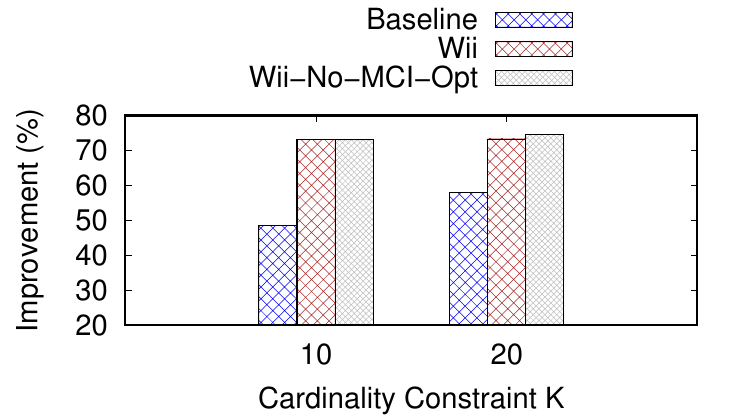}}
\subfigure[\textbf{Real-M}, $B=5,000$]{ \label{fig:two-phase:no-opt-mci:real-m}
    \includegraphics[width=0.23\textwidth]{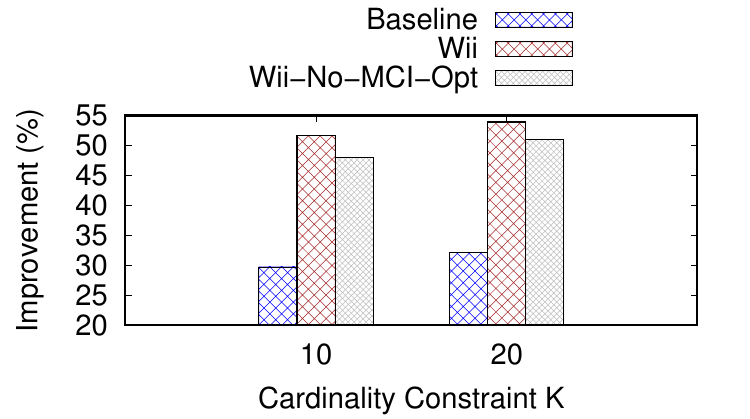}}
\vspace{-1.5em}
\caption{Impact on the performance of \sysname with or without the optimization for the MCI upper bounds ($\alpha=0.9$).}
\label{fig:two-phase:no-opt-mci}
\vspace{-1.5em}
\end{figure*}

%%%%%%%%%%%% MCTS Greedy %%%%%%%%%%

\begin{figure*}
\centering
\subfigure[\textbf{TPC-H}, $K=10$]{ \label{fig:mcts:tpch:K10}
    \includegraphics[width=0.23\textwidth]{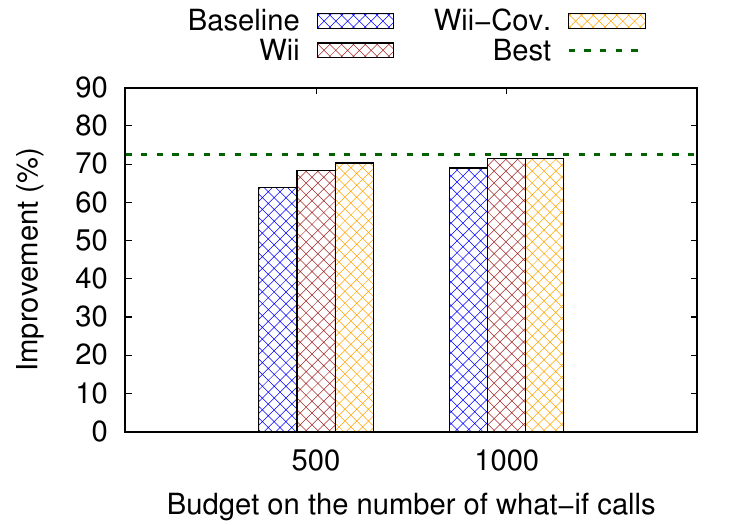}}
\subfigure[\textbf{TPC-DS}, $K=10$]{ \label{fig:mcts:tpcds:K10}
    \includegraphics[width=0.23\textwidth]{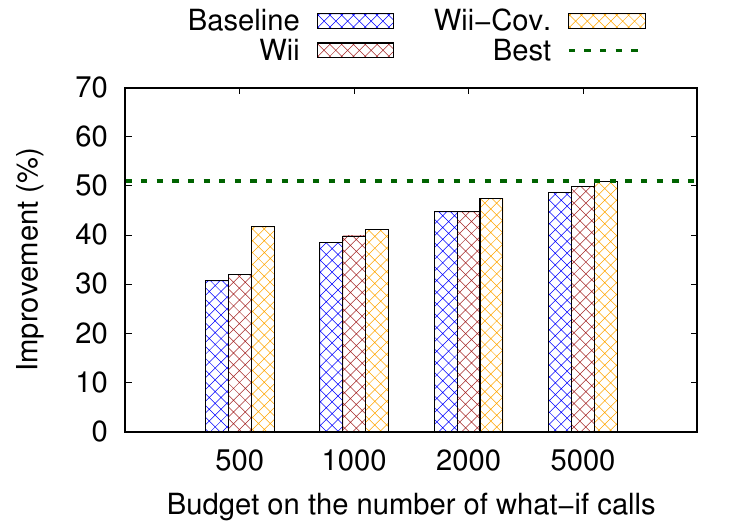}}
\subfigure[\textbf{Real-D}, $K=10$]{ \label{fig:mcts:real-d:K10}
    \includegraphics[width=0.23\textwidth]{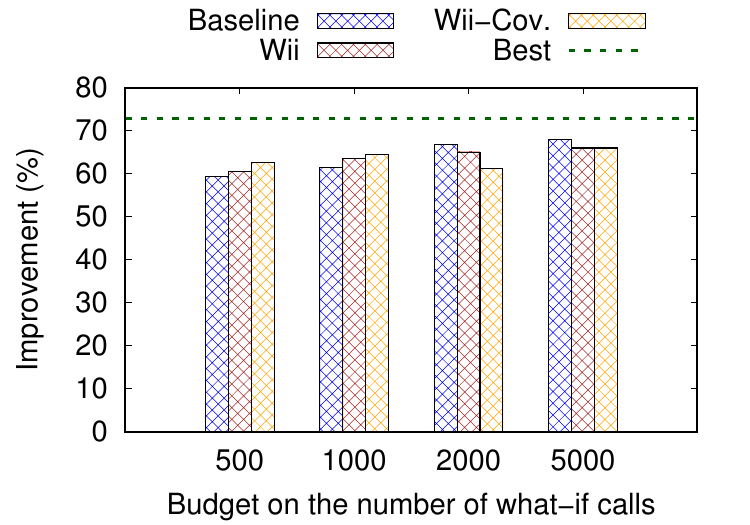}}
\subfigure[\textbf{Real-M}, $K=10$]{ \label{fig:mcts:real-m:K10}
    \includegraphics[width=0.23\textwidth]{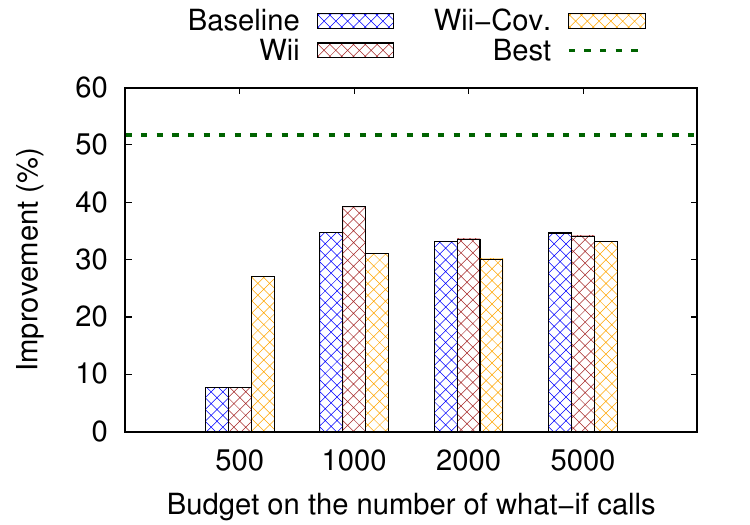}}
\subfigure[\textbf{TPC-H}, $K=20$]{ \label{fig:mcts:tpch:K20}
    \includegraphics[width=0.23\textwidth]{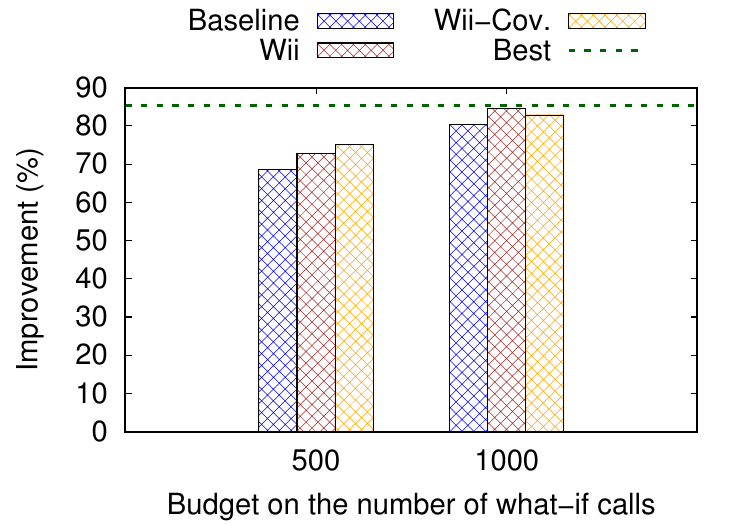}}
\subfigure[\textbf{TPC-DS}, $K=20$]{ \label{fig:mcts:tpcds:K20}
    \includegraphics[width=0.23\textwidth]{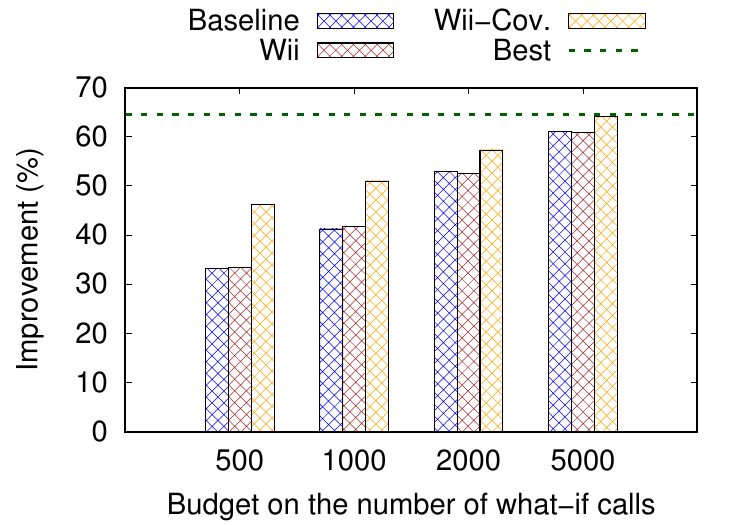}}
\subfigure[\textbf{Real-D}, $K=20$]{ \label{fig:mcts:real-d:K20}
    \includegraphics[width=0.23\textwidth]{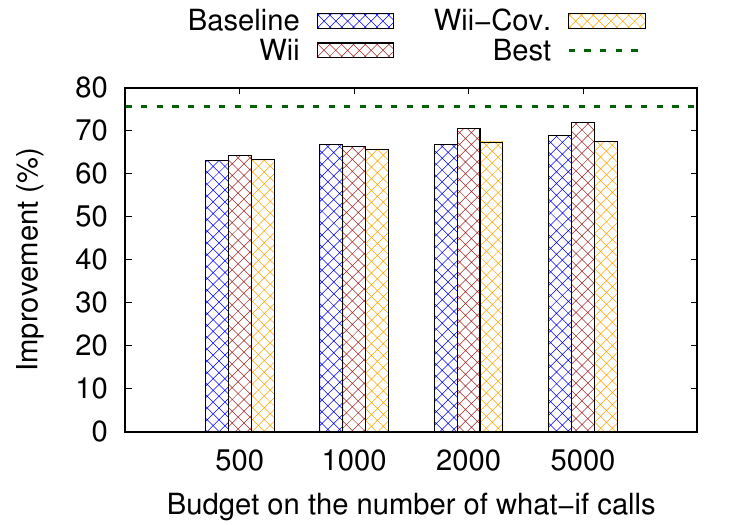}}
\subfigure[\textbf{Real-M}, $K=20$]{ \label{fig:mcts:real-m:K20}
    \includegraphics[width=0.23\textwidth]{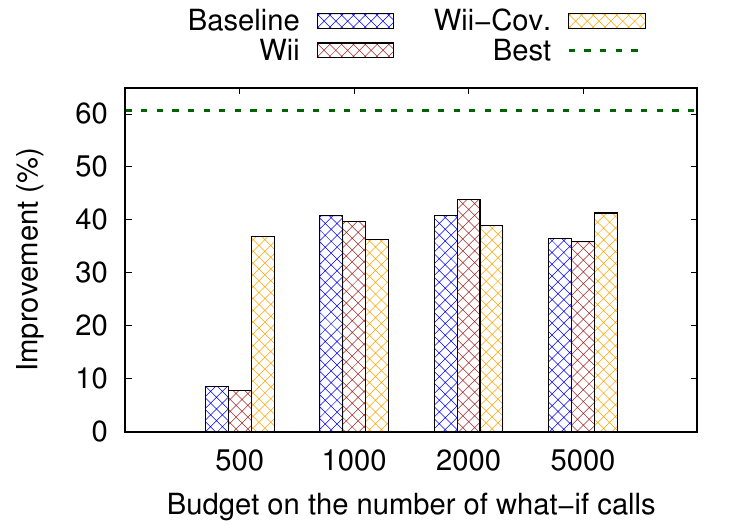}}
\vspace{-1.5em}
\caption{Results for \emph{MCTS} with confidence threshold $\alpha=0.9$ (``Cov.'' is shorthand for ``Coverage'').}%\xy{threshold issue? combine figures of two-phase and MCTS?}}
\label{fig:mcts}
%\vspace{-1em}
\end{figure*}

%\todo{Report results for both \emph{bound}-based and coverage-based techniques.}

%(for both query-level and call-level skipping). We haven't tested \emph{coverage-singleton} and \emph{coverage-all}.

%For \emph{bound}, in general, improvements are higher than query-level skipping.

%\todo{Revise after having all results.}
Figure~\ref{fig:two-phase:call-level} presents the evaluation results of \sysname for \emph{two-phase greedy} when setting the confidence threshold $\alpha=0.9$ (see Section~\ref{sec:eval:e2e:best-possible} for details of the `Best' lines).
%and the cardinality constraint $K\in\{10, 20\}$.
%$K=20$. The results when setting $K=10$ are similar~\cite{full-version}.
We observe that %both the bound-based and the coverage-refined 
\sysname significantly outperforms the baseline (i.e., \emph{two-phase greedy} without what-if call interception).
%on all the workloads tested.
For example, when setting $K=20$ and $B=5,000$, \sysname improves over the baseline by increasing the percentage improvement from 50\% to 65\% on \textbf{TPC-DS} (Figure~\ref{fig:two-phase:call-level:tpcds:K20}), from 58\% to 74\% on \textbf{Real-D} (Figure~\ref{fig:two-phase:call-level:real-d:K20}), and from 32\% to 54\% on \textbf{Real-M} (Figure~\ref{fig:two-phase:call-level:real-m:K20});
even for the smallest workload \textbf{TPC-H}, when setting $K=20$ and $B=1,000$, \sysname improves over the baseline from 78\% to 86\% (Figure~\ref{fig:two-phase:call-level:tpch:K20}).
Note that here \sysname has used the optimization for greedy search (Section~\ref{sec:bounds:mci:greedy}).
%Even if we just use $\tau=0$, i.e., the most conservative setting of $\tau$, the improvements remain significant: from 50\% to 61\% on \textbf{TPC-DS}, from 58\% to 76\% on \textbf{Real-D}, and from 32\% to 44\% on \textbf{Real-M}.
\iffalse
Integrating \sysname with \emph{two-phase greedy} is of special interest and importance due to its adoption in commercial database tuning software such as Microsoft's DTA~\cite{dta,dta-utility}.
\fi

We also observe that incorporating the coverage-based refinement described in Section~\ref{sec:optimization:coverage} can further improve \sysname in certain cases.
For instance, on \textbf{TPC-DS} when setting $K=20$ and $B=2,000$, it improves \sysname by 13\%, i.e., from 49\% to 62\%, whereas \sysname and the baseline perform similarly (Figure~\ref{fig:two-phase:call-level:tpcds:K20}); on \textbf{Real-D} when setting $K=10$ and $B=500$ (Figure~\ref{fig:two-phase:call-level:real-d:K10}), it improves \sysname by an additional percentage improvement of 17.8\% (i.e., from 45.3\% to 63.1\%), which translates to 32.2\% improvement over the baseline (i.e., from 30.9\% to 63.1\%).

%\vspace{-0.5em}
\paragraph*{Impact of Optimization for MCI Upper Bounds}

We further study the impact of the optimization proposed in Section~\ref{sec:bounds:mci:greedy} for \emph{two-phase greedy}.
In our experiment, we set $\alpha=0.9$, $B=1,000$ for \textbf{TPC-H} and $B=5,000$ for the other workloads.
Figure~\ref{fig:two-phase:no-opt-mci} presents the results.
We observe that the optimization for MCI upper bounds offers a differentiable benefit in \emph{two-phase greedy} on \textbf{TPC-H}, \textbf{TPC-DS}, and \textbf{Real-M}.
Given its negligible computation overhead, this optimization is warranted to be enabled by default in \sysname.

%\todo{Add details.}

\subsubsection{MCTS}

%\todo{Report results for both \emph{bound}-based and coverage-based techniques.}

%\todo{Revise after having all results.}
Figure~\ref{fig:mcts} presents the results of \sysname for \emph{MCTS}, again by setting the confidence threshold $\alpha=0.9$.
%and $K=20$. The results when setting $K=10$ are similar~\cite{full-version}.
Unlike the case of \emph{two-phase greedy}, for \emph{MCTS} \sysname often performs similarly to the baseline (i.e., \emph{MCTS} without what-if call interception).
This is not surprising, given that \emph{MCTS} already significantly outperforms \emph{two-phase greedy} in many (but not all) cases, which can be verified by comparing the corresponding charts in Figure~\ref{fig:two-phase:call-level} and Figure~\ref{fig:mcts}---further improvement on top of that is more challenging.
However, there are noticeable cases where we do observe significant improvement as we incorporate the coverage-based refinement into \sysname.
For instance, on \textbf{Real-M}, when setting $K=10$ and $B=500$ (Figure~\ref{fig:mcts:real-m:K10}), it improves over the baseline by increasing the percentage improvement of the final index configuration found by \emph{MCTS} from 7.8\% to 27.1\%; similar observation holds when we increasing $K$ to 20 (Figure~\ref{fig:mcts:real-m:K20}), where
we observe an even higher boost on the percentage improvement
(i.e., from 8.5\% to 36.9\%).
%when varying $K$ to 20 while keeping the same $\alpha$ and $B$, it improves over the baseline from 37\% to 47\%. 
In general, we observe that \sysname is more effective on the two larger workloads (\textbf{TPC-DS} and \textbf{Real-M}), which have more complex queries and thus much larger search spaces (ref. Table~\ref{tab:databases}). In such situations, the number of configurations that \emph{MCTS} can explore within the budget constraint is too small compared to the entire search space.
\sysname increases the opportunity for \emph{MCTS} to find a better 
configuration by skipping spurious what-if calls.
%and leveraging domain knowledge with the coverage-based refinement.}
%\todo{Add some numbers about the number of candidate indexes so that we can use $|W|\times|I|$ to measure the size of the search space?}
Nevertheless, compared to \emph{two-phase greedy}, \emph{MCTS} has its own limitations (e.g., its inherent usage of randomization) that require more research to pave its way of being adopted by commercial index tuners~\cite{ml-index-tuning-overview}. Moreover, \emph{MCTS} is not suitable for the ``unlimited budget'' case (Section~\ref{sec:eval:no-budget-limit}) as it requires a budget constraint as input.

\iffalse
on \textbf{Real-M}, the coverage-refined \sysname significantly outperforms the others, especially when the budget on the number of what-if calls is small. 
For example, when setting $K=20$, $\tau=10$, and $B=5,000$, the coverage-refined \sysname increases the percentage improvement from 37\% to 55\% (Figure~\ref{fig:mcts:real-m:K20}).
For \emph{MCTS}, the bound-based \sysname is not effective at skipping what-if calls of multi-index configurations, because such configurations usually contain indexes (i.e., singleton configurations) with unknown what-if costs that force the bound-based \sysname to use the most conservative MCI upper-bounds (Equation~\ref{eq:upper-bound:mci:call-level}).
This is different from either \emph{vanilla greedy} or \emph{two-phase greedy} where we can further apply the greedy-specific refinement procedure (Procedure~\ref{proc:maintain-mci}) to improve the MCI upper-bounds.
\fi

\subsubsection{Discussion}
%\todo{comparable performance of two-phase and MCTS after using \sysname.}

Comparing Figures~\ref{fig:two-phase:call-level} and~\ref{fig:mcts}, while the baseline version of \emph{two-phase greedy} clearly underperforms that of \emph{MCTS}, the \sysname-enhanced version of \emph{two-phase greedy} performs similarly or even better than that of \emph{MCTS}. %due to the further optimization designed for greedy search (Section~\ref{sec:bounds:mci:greedy}).
Existing budget allocation policies are largely \emph{macro-level} optimization mechanisms, meaning that they deem what-if calls as \emph{atomic} black-box operations that are out of their optimization scopes.
However, our results here reveal that \emph{micro-level} optimization mechanisms like \sysname that operate at the granularity of individual what-if calls can interact with and have profound impact on the performance of those \emph{macro-level} optimization mechanisms.
An in-depth study and understanding of such macro-/micro-level interactions may lead to invention of better budget allocation policies.
%We leave this as a very interesting direction for future exploration.

Moreover, based on our evaluation results, the coverage-based refinement does not always improve \sysname's performance.
A natural question is then how users would choose whether or not to use it. Are there some simple tests that can indicate whether or not it will be beneficial? 
Since the motivation of the coverage-based refinement is to make \sysname work more effectively in the presence of unknown singleton-configuration what-if costs, one idea could be to measure the fraction of such singleton-configurations and enable the coverage-based refinement only when this fraction is high. However, this measurement can only be monitored ``during'' index tuning and there are further questions if index tuning is budget-constrained (e.g., how much budget should be allocated for monitoring this measurement). Thus, there seems to be no simple answer and we leave its investigation for future work.

\iffalse
Furthermore, from Figures~\ref{fig:two-phase:call-level} and~\ref{fig:mcts}, we can see that the coverage optimization can boost \sysname on small budgets in general, and becomes less effective or may even have negative impact when budget increases. It is because coverage is only used to skip singleton configurations, which are explored mainly in the early stages of both algorithms. Hence, this optimization could largely improve the result by advancing the tuning progress in the beginning. However, it may also mistakenly skip some important indexes due to inaccurate estimations, and the impact of which becomes more obvious with more budgets are allocated on multi-index configurations. Therefore, one should enable coverage optimization when the fraction of singleton configurations in the budget allocation is large. However, this fraction can only be monitored \emph{during} index tuning and there are further questions such as how much budget should be allocated for monitoring this fraction. Thus, there seems to be no simple way to choose between \sysname and \sysname-Cov beforehand and we intend to leave this investigation for future work.
\fi

%%%%%%%%%%%% Two-phase Greedy Low Confidence %%%%%%%%%%

\begin{figure*}
\centering
\subfigure[\textbf{TPC-H}, $B=1,000$]{ \label{fig:two-phase:tpch:low-conf}
    \includegraphics[width=0.23\textwidth]{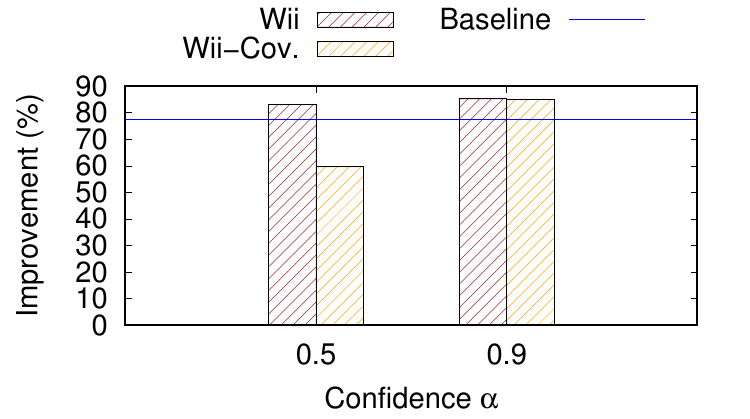}}
\subfigure[\textbf{TPC-DS}, $B=5,000$]{ \label{fig:two-phase:tpcds:low-conf}
    \includegraphics[width=0.23\textwidth]{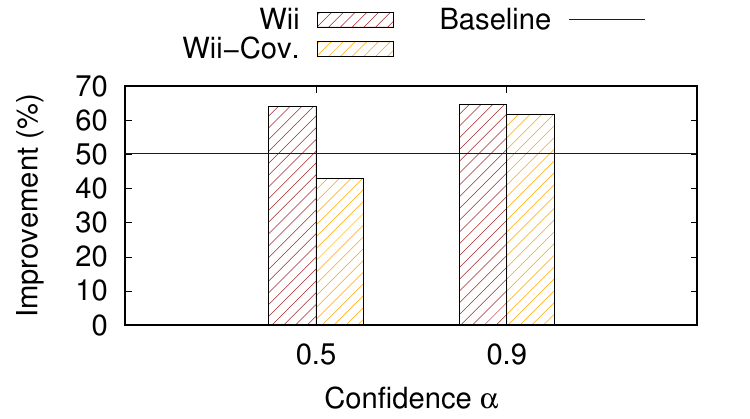}}
\subfigure[\textbf{Real-D}, $B=5,000$]{ \label{fig:two-phase:real-d:low-conf}
    \includegraphics[width=0.23\textwidth]{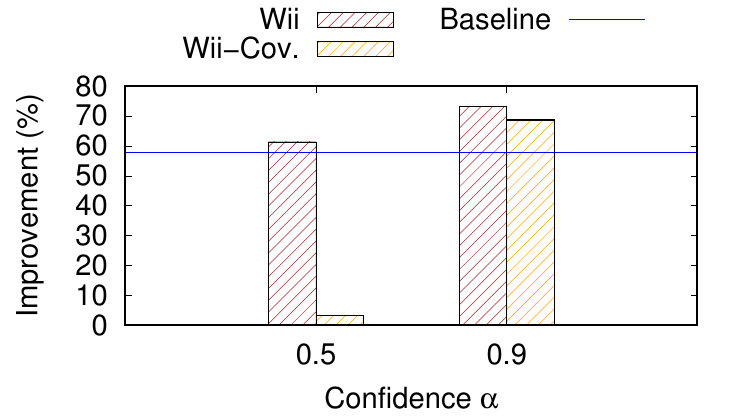}}
\subfigure[\textbf{Real-M}, $B=5,000$]{ \label{fig:two-phase:real-m:low-conf}
\includegraphics[width=0.23\textwidth]{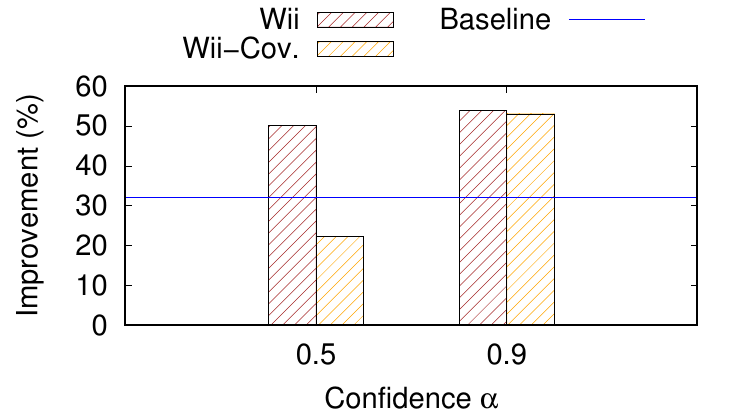}}
\vspace{-1.5em}
\caption{Performance impact when lowering the confidence threshold $\alpha$ of \sysname for \emph{two-phase greedy} ($K=20$).}
\label{fig:two-phase:low-conf}
%\vspace{-1em}
\end{figure*}

%%%%%%%%%%%% Two-phase Greedy Confidence Threshold %%%%%%%%%%

\begin{figure*}
\centering
\subfigure[\textbf{TPC-H}, $B=1,000$]{ \label{fig:two-phase:tpch:conf}
    \includegraphics[width=0.23\textwidth]{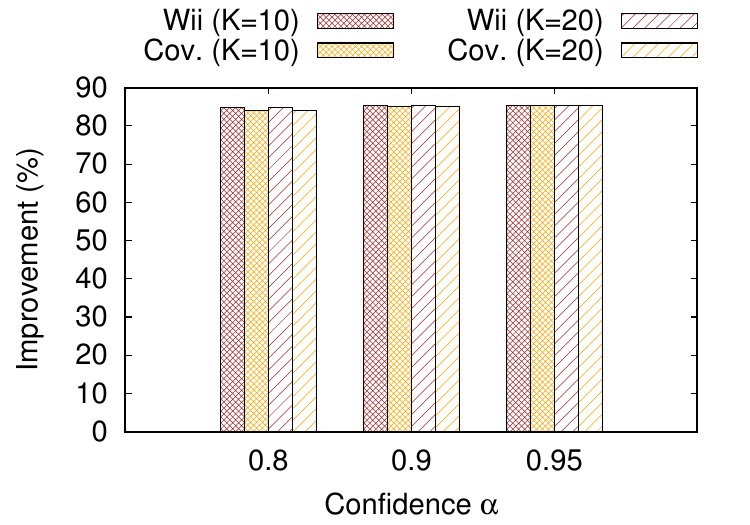}}
\subfigure[\textbf{TPC-DS}, $B=5,000$]{ \label{fig:two-phase:tpcds:conf}
    \includegraphics[width=0.23\textwidth]{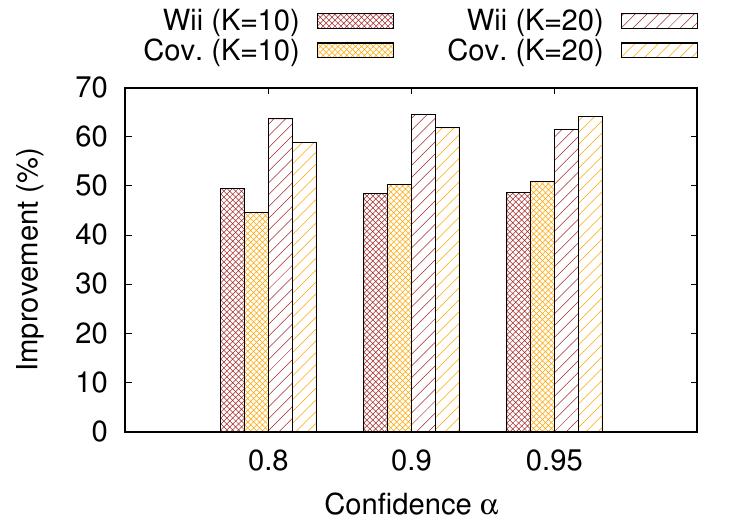}}
\subfigure[\textbf{Real-D}, $B=5,000$]{ \label{fig:two-phase:real-d:conf}
    \includegraphics[width=0.23\textwidth]{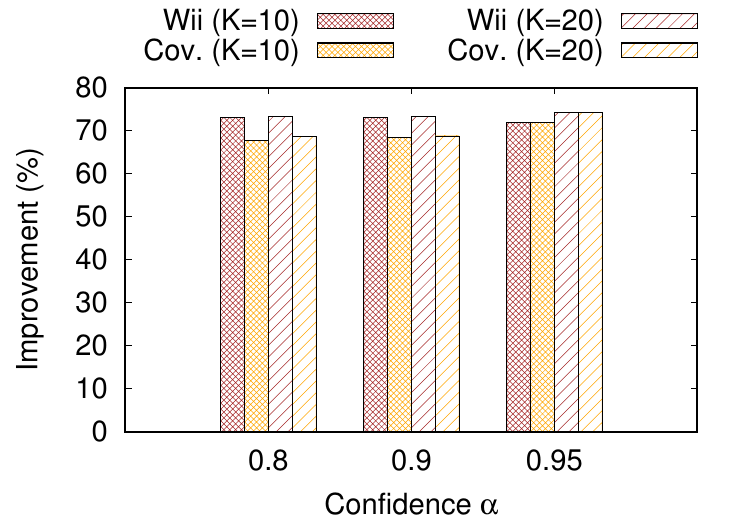}}
\subfigure[\textbf{Real-M}, $B=5,000$]{ \label{fig:two-phase:real-m:conf}
\includegraphics[width=0.23\textwidth]{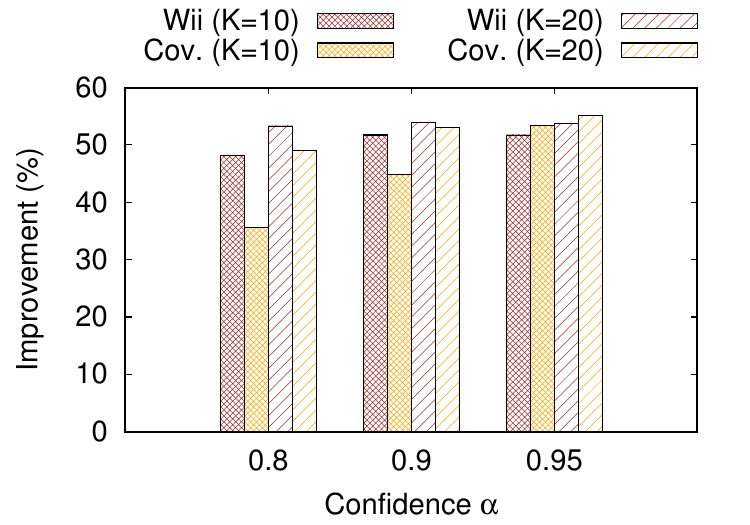}}
\vspace{-1.5em}
\caption{Impact of the confidence threshold for \emph{two-phase greedy} (``Cov.'' is shorthand for ``\sysname-Coverage'').} %\xy{split k = 10 and 20?}}
\label{fig:two-phase:conf}
\vspace{-1em}
\end{figure*}

%%%%%%%%%%%% MCTS Confidence Threshold %%%%%%%%%%

\begin{figure*}
\centering
\subfigure[\textbf{TPC-H}, $B=1,000$]{ \label{fig:mcts:tpch:conf}
    \includegraphics[width=0.23\textwidth]{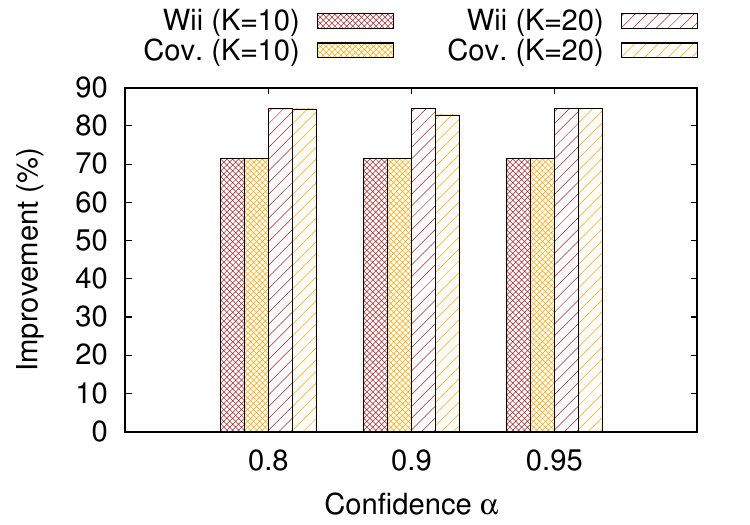}}
\subfigure[\textbf{TPC-DS}, $B=5,000$]{ \label{fig:mcts:tpcds:conf}
    \includegraphics[width=0.23\textwidth]{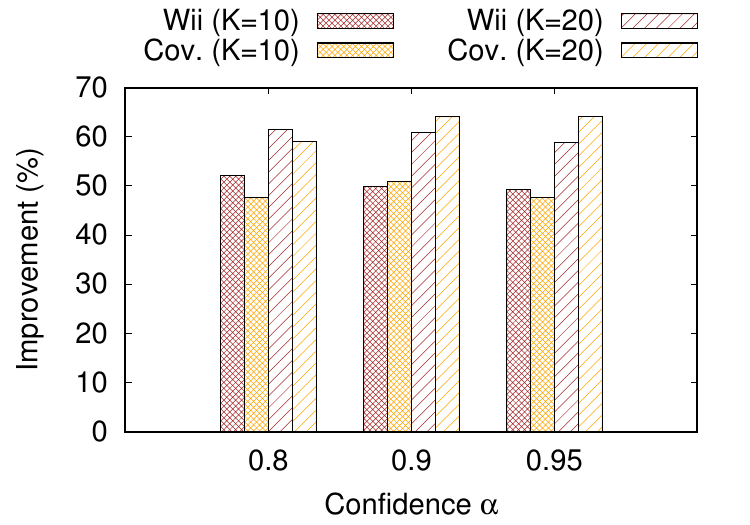}}
\subfigure[\textbf{Real-D}, $B=5,000$]{ \label{fig:mcts:real-d:conf}
    \includegraphics[width=0.23\textwidth]{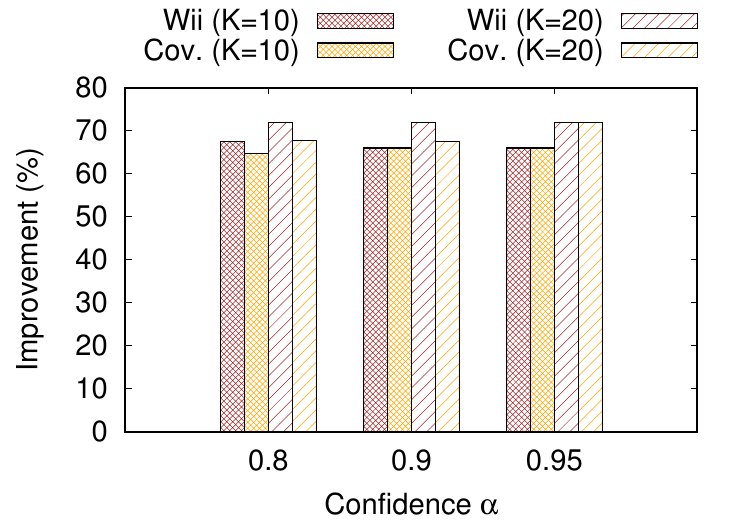}}
\subfigure[\textbf{Real-M}, $B=5,000$]{ \label{fig:mcts:real-m:conf}
\includegraphics[width=0.23\textwidth]{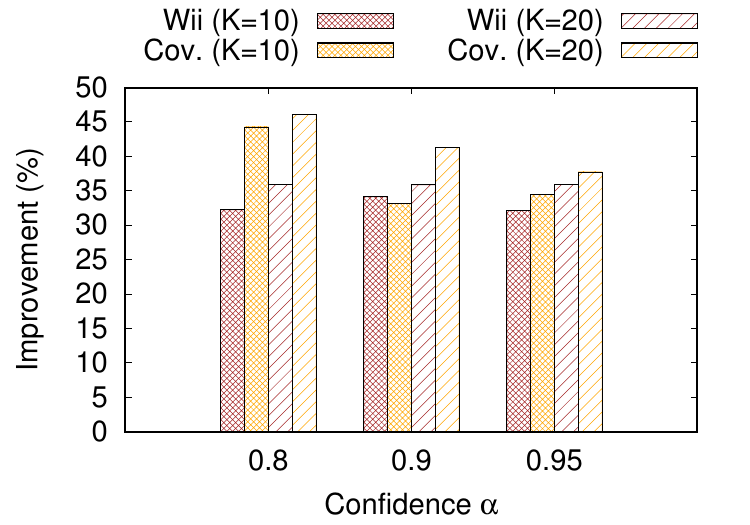}}
\vspace{-1.5em}
\caption{Impact of the confidence threshold for \emph{MCTS} (``Cov.'' is shorthand for ``\sysname-Coverage'').}
\label{fig:mcts:conf}
\vspace{-1.5em}
\end{figure*}

%%%%%%%%%%%% MCTS Low Confidence %%%%%%%%%%

\begin{figure*}
\centering
\subfigure[\textbf{TPC-H}, $B=1,000$]{ \label{fig:mcts:tpch:low-conf}
    \includegraphics[width=0.23\textwidth]{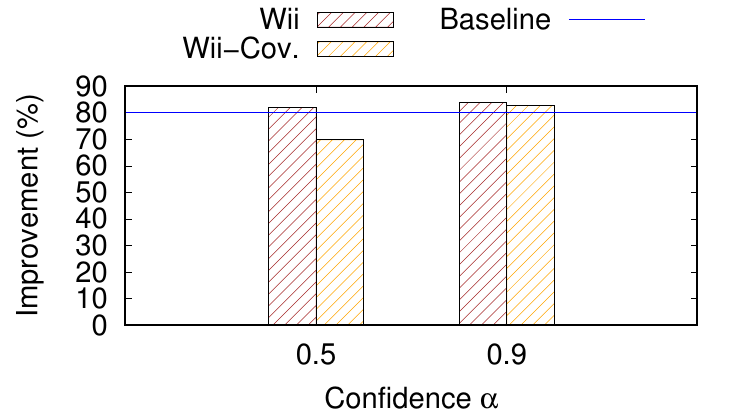}}
\subfigure[\textbf{TPC-DS}, $B=5,000$]{ \label{fig:mcts:tpcds:low-conf}
    \includegraphics[width=0.23\textwidth]{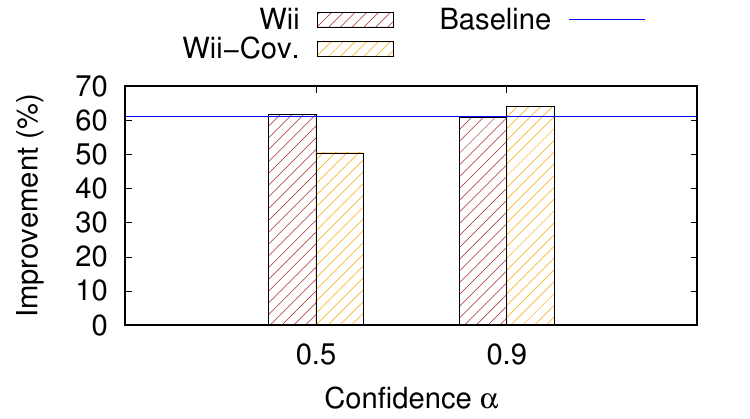}}
\subfigure[\textbf{Real-D}, $B=5,000$]{ \label{fig:mcts:real-d:low-conf}
    \includegraphics[width=0.23\textwidth]{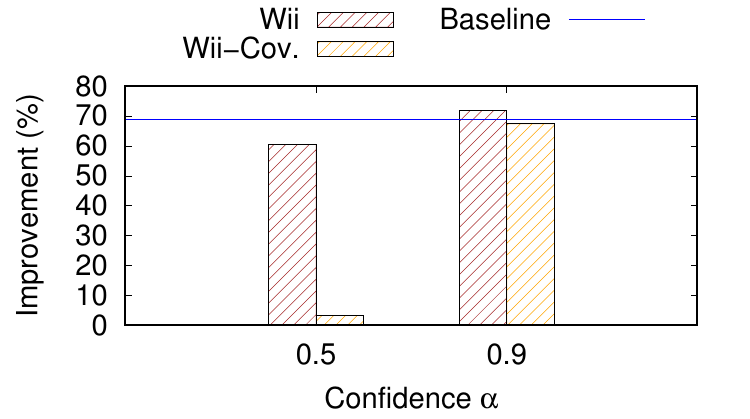}}
\subfigure[\textbf{Real-M}, $B=5,000$]{ \label{fig:mcts:real-m:low-conf}
\includegraphics[width=0.23\textwidth]{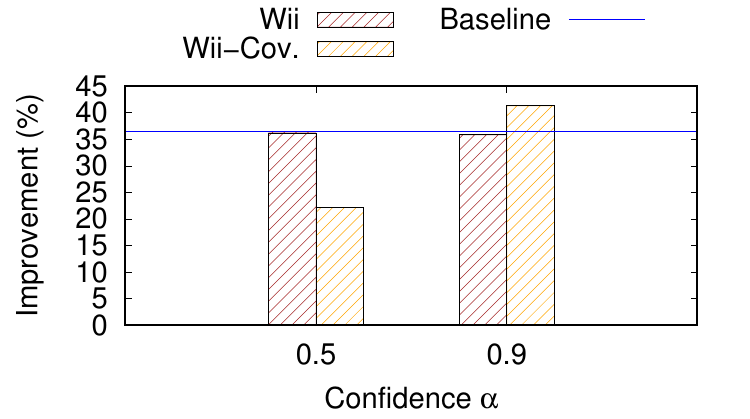}}
\vspace{-1.5em}
\caption{Performance impact when lowering the confidence threshold $\alpha$ used by \sysname for \emph{MCTS} ($K=20$).} 
\label{fig:mcts:low-conf}
\vspace{-1.5em}
\end{figure*}

\subsubsection{Evaluation of Confidence-based What-if Call Skipping}
We start by investigating the impact of the confidence threshold $\alpha$ on \sysname.
For this set of experiments, we use the budget $B=1,000$ for \textbf{TPC-H} and use $B=5,000$ for the other workloads, and we vary $\alpha\in\{0.8, 0.9, 0.95\}$.
%Figures~\ref{fig:two-phase:conf} and~\ref{fig:mcts:conf} present the evaluation results for \emph{two-phase greedy} and \emph{MCTS}, respectively, by varying $\alpha\in\{0.8, 0.9, 0.95\}$.
%\todo{Add observations.}
%Due to space limitation, the details are deferred to~\cite{full-version}.
Figures~\ref{fig:two-phase:conf} and~\ref{fig:mcts:conf} present the evaluation results.
%the detailed evaluation results on the impact of the confidence threshold $\alpha$. 
%A more formal analysis can be found in~\cite{full-version}.
We observe that \sysname is not sensitive to the threshold $\alpha$ within the range that we tested, for both \emph{two-phase greedy} and \emph{MCTS}.
On the other hand, coverage-based refinement is more sensitive to $\alpha$.
For instance, for \emph{two-phase greeedy} on \textbf{Real-M} with cardinality constraint $K=10$ (ref. Figure~\ref{fig:two-phase:real-m:conf}), the end-to-end percentage improvement of the final configuration found increases from 35.6\% to 53.3\% when raising $\alpha$ from 0.8 to 0.95.
This suggests both opportunities and risks of using the coverage-based refinement for \sysname, as one needs to choose the confidence threshold $\alpha$ more carefully.
A more formal analysis can be found in~\cite{full-version}.
%A quantitative analysis on the impact of coverage over the confidence can be found in~\cite{full-version}.

%\vspace{-0.5em}
\noindent
\paragraph*{Low Confidence Threshold}

An interesting question is the performance impact of using a relatively lower confidence threshold compared to the ones used in the previous evaluations.
To investigate this question, we further conduct experiments by setting the confidence threshold $\alpha=0.5$.
Figures~\ref{fig:two-phase:low-conf} and~\ref{fig:mcts:low-conf} present results for \emph{two-phase greedy} and \emph{MCTS} with the cardinality constraint $K=20$.
%while setting $B=1,000$ for \textbf{TPC-H} and $B=5,000$ for the other workloads.}
%\xy{to save space, maybe figure \ref{fig:two-phase:low-conf} and \ref{fig:mcts:low-conf} can be compressed into one line?}
We have the following observations. First, the performance of \sysname often becomes much worse compared to using a high confidence threshold like the $\alpha=0.9$ in the charts---it is sometimes even worse than the baseline, e.g., in the case of \emph{MCTS} on \textbf{Real-D}, as shown in Figure~\ref{fig:mcts:real-d:low-conf}.
Second, coverage-based refinement seems more sensitive to the use of a low confidence threshold, due to its inherent uncertainty of estimating singleton-configuration what-if costs.

%\xy{explain a bit? the result is not surprising because ...}

%%%%%%%%%%%% Random Skipping %%%%%%%%%%

\begin{figure*}
\centering
\subfigure[\textbf{TPC-H}, $B=1,000$]{ \label{fig:tpch:random-skip}
    \includegraphics[width=0.23\textwidth]{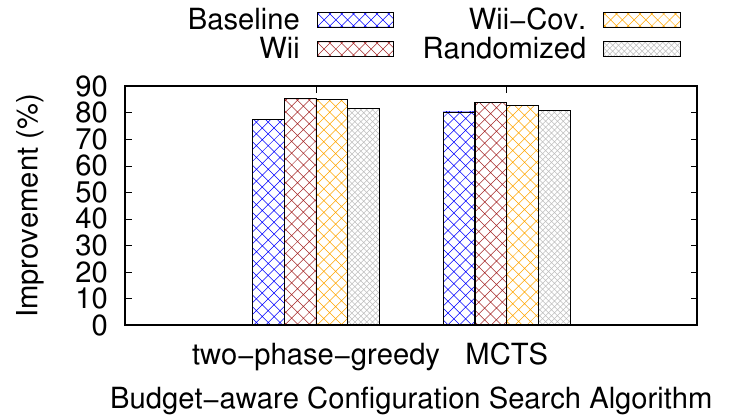}}
\subfigure[\textbf{TPC-DS}, $B=5,000$]{ \label{fig:tpcds:random-skip}
    \includegraphics[width=0.23\textwidth]{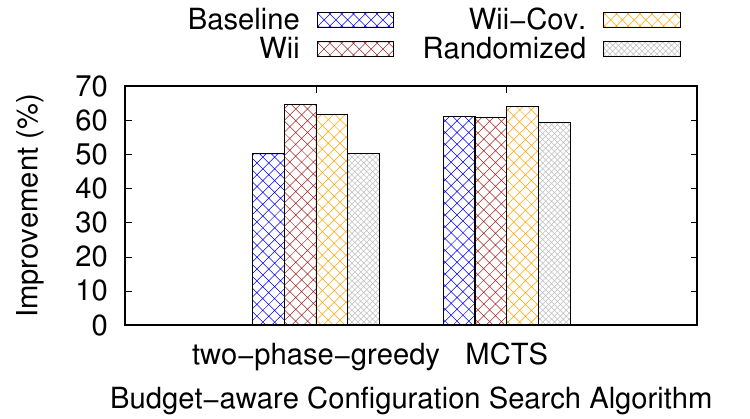}}
\subfigure[\textbf{Real-D}, $B=5,000$]{ \label{fig:real-d:random-skip}
    \includegraphics[width=0.23\textwidth]{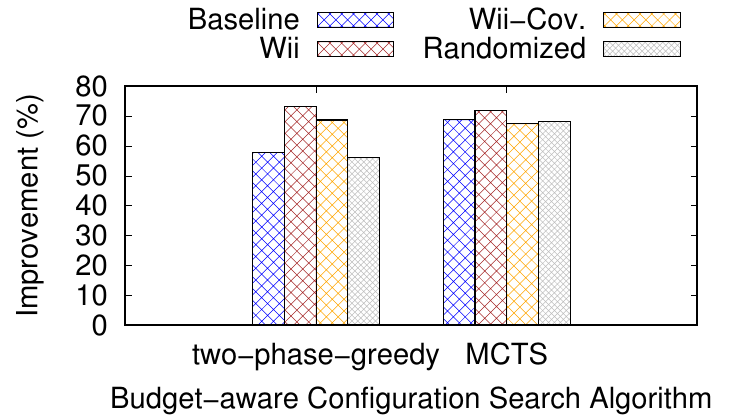}}
\subfigure[\textbf{Real-M}, $B=5,000$]{ \label{fig:real-m:random-skip}
\includegraphics[width=0.23\textwidth]{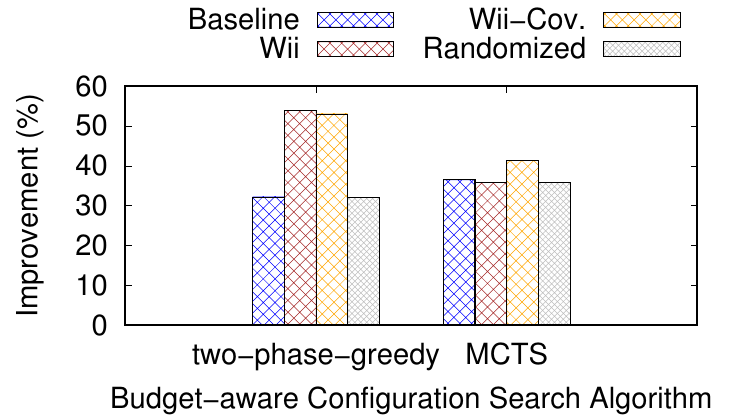}}
\vspace{-1.5em}
\caption{Confidence-based skipping vs. randomized skipping of what-if calls ($p=\alpha=0.9$, $K=20$).}
\label{fig:random-skip}
%\vspace{0.5em}
\end{figure*}

%%%%%%%%%%%% Two-phase Greedy What-if Call Skipping %%%%%%%%%%

\begin{figure*}
\centering
\subfigure[\textbf{TPC-H}, $B=1,000$]{ \label{fig:two-phase:call-level:tpch:skip}
    \includegraphics[width=0.23\textwidth]{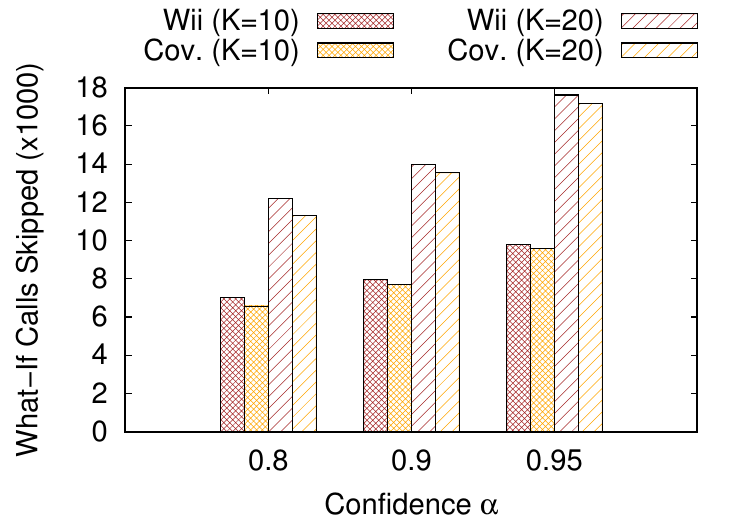}}
\subfigure[\textbf{TPC-DS}, $B=5,000$]{ \label{fig:two-phase:call-level:tpcds:skip}
    \includegraphics[width=0.23\textwidth]{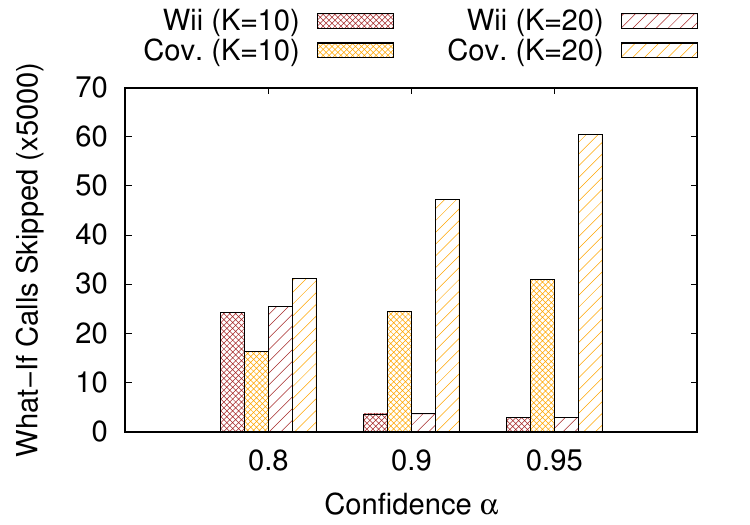}}
\subfigure[\textbf{Real-D}, $B=5,000$]{ 
\label{fig:two-phase:call-level:real-d:skip}
    \includegraphics[width=0.23\textwidth]{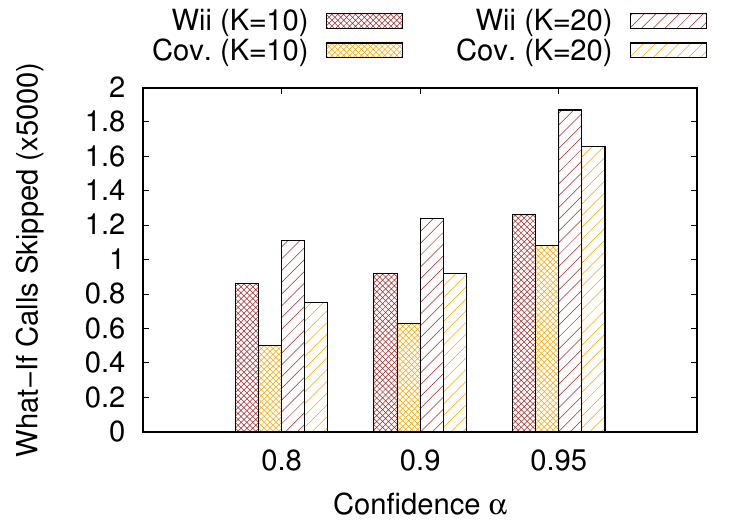}}
\subfigure[\textbf{Real-M}, $B=5,000$]{ 
\label{fig:two-phase:call-level:real-m:skip}
\includegraphics[width=0.23\textwidth]{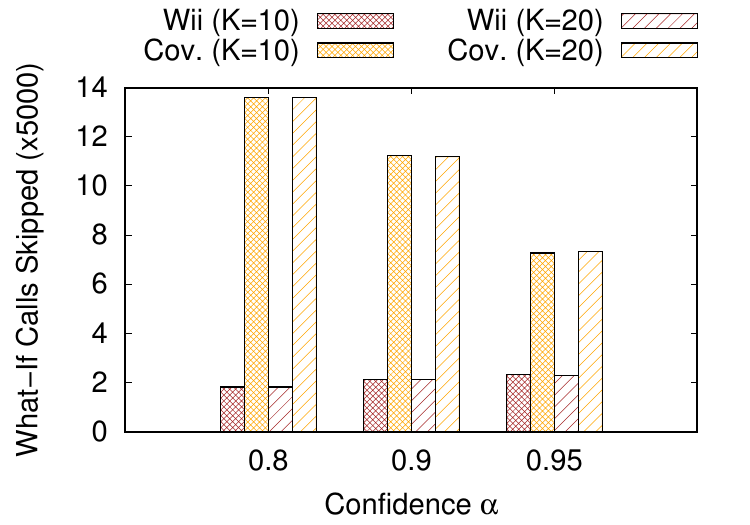}}
\vspace{-1.5em}
\caption{Amount of what-if calls skipped by \sysname for \emph{two-phase greedy} (``Cov.'' is shorthand for ``\sysname-Coverage'').} %\xy{split k = 10 and 20?}}
\label{fig:two-phase:call-level:skip}
\vspace{-1.5em}
\end{figure*}

%%%%%%%%%%%% MCTS What-if Call Skipping %%%%%%%%%%

\begin{figure*}
\centering
\subfigure[\textbf{TPC-H}, $B=1,000$]{ \label{fig:mcts:tpch:skip}
    \includegraphics[width=0.23\textwidth]{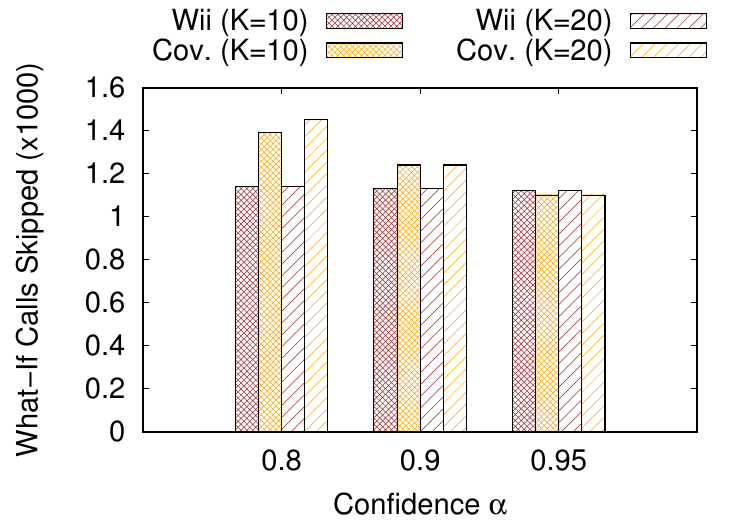}}
\subfigure[\textbf{TPC-DS}, $B=5,000$]{ \label{fig:mcts:tpcds:skip}
    \includegraphics[width=0.23\textwidth]{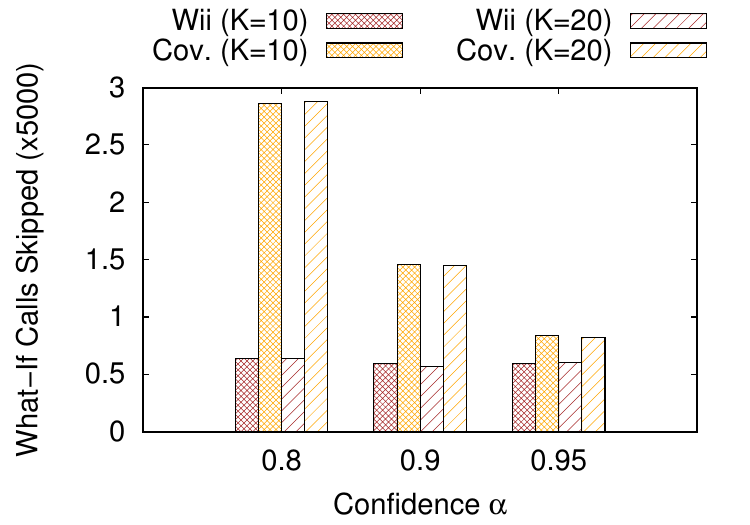}}
\subfigure[\textbf{Real-D}, $B=5,000$]{ 
\label{fig:mcts:real-d:skip}
    \includegraphics[width=0.23\textwidth]{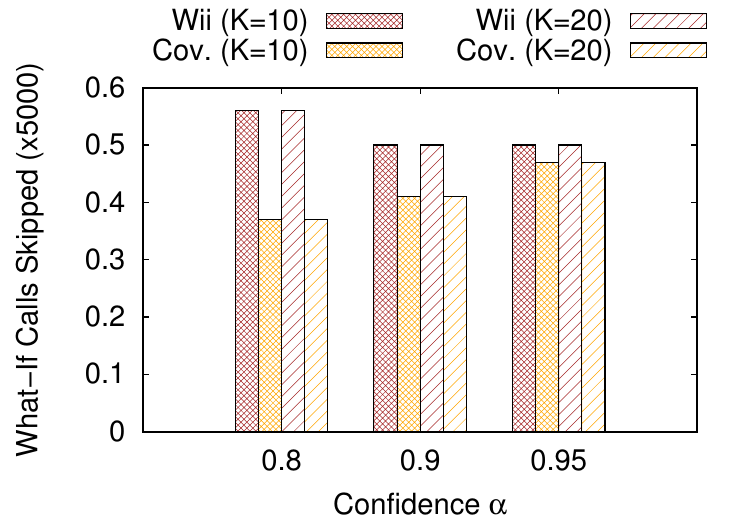}}
\subfigure[\textbf{Real-M}, $B=5,000$]{ 
\label{fig:mcts:real-m:skip}
\includegraphics[width=0.23\textwidth]{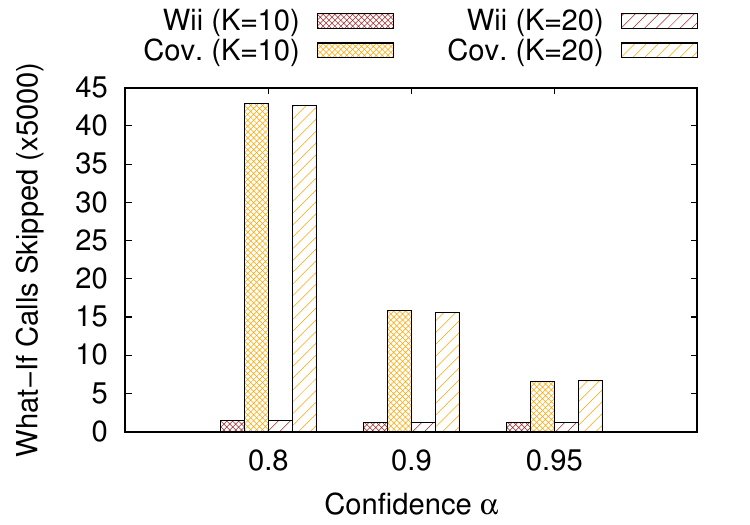}}
\vspace{-1.5em}
\caption{Amount of what-if calls skipped by \sysname for \emph{MCTS} (``Cov.'' is shorthand for ``\sysname-Coverage'').}
\label{fig:mcts:skip}
\vspace{-1.5em}
\end{figure*}

\begin{figure*}
\centering
\subfigure[\textbf{TPC-H}, $B=1,000$]{ \label{fig:bound-computation:tpch}
    \includegraphics[width=0.23\textwidth]{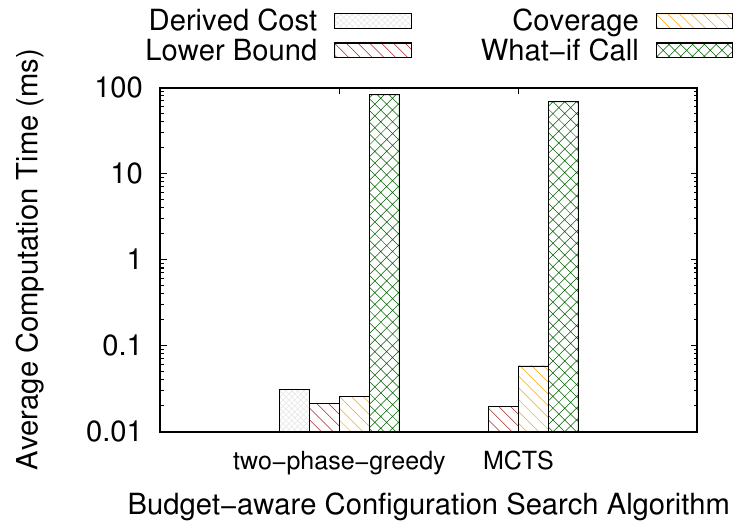}}
\subfigure[\textbf{TPC-DS}, $B=5,000$]{ \label{fig:bound-computation:tpcds}
    \includegraphics[width=0.23\textwidth]{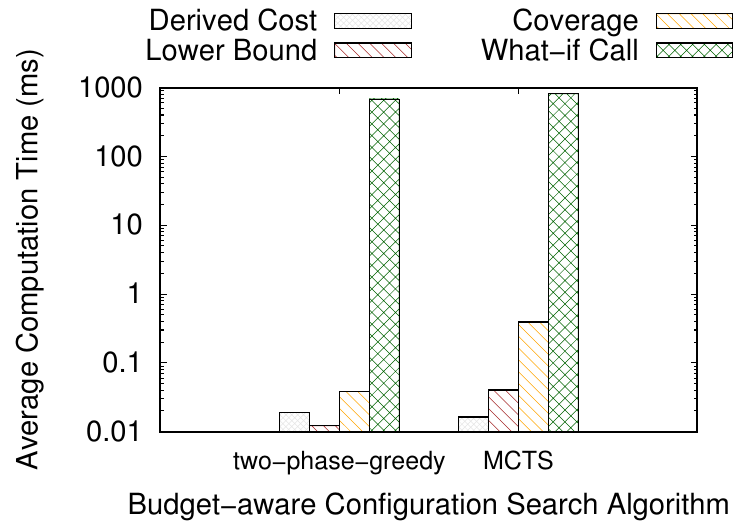}}
%\hspace{0.1\columnwidth}
\subfigure[\textbf{Real-D}, $B=5,000$]{ \label{fig:bound-computation:real-d}
    \includegraphics[width=0.23\textwidth]{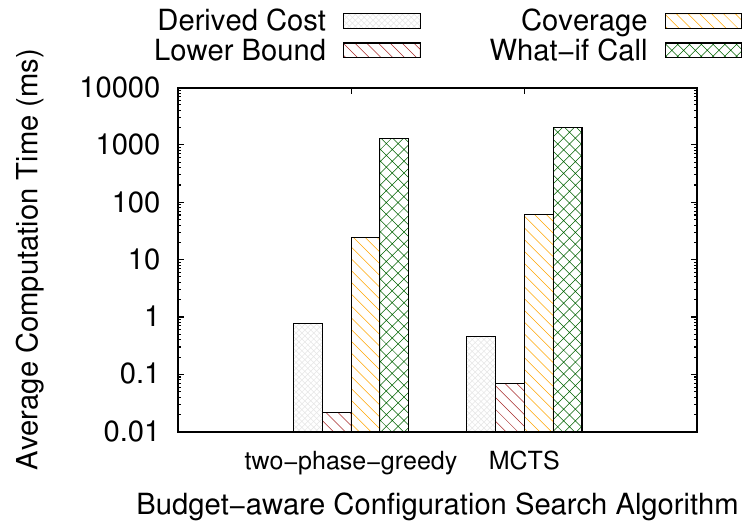}}
%\hspace{0.1\columnwidth}
\subfigure[\textbf{Real-M}, $B=5,000$]{ \label{fig:bound-computation:real-m}
    \includegraphics[width=0.23\textwidth]{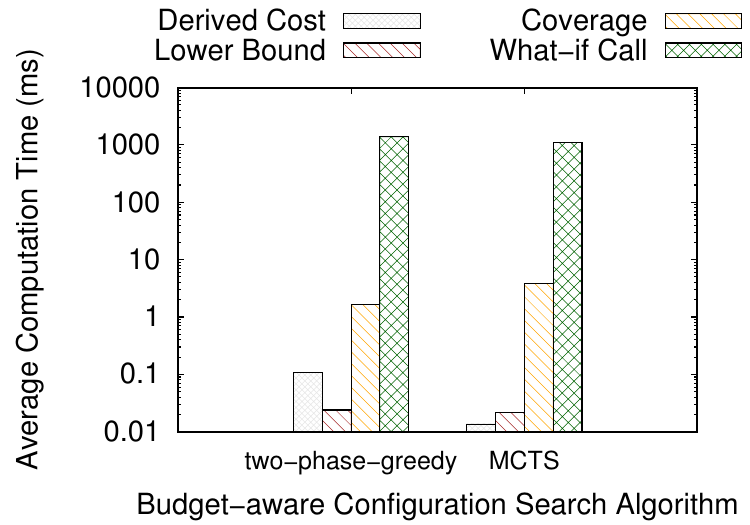}}
\vspace{-1.5em}
\caption{Average computation time (in milliseconds) of the lower bound on the what-if cost ($K=20$, $\alpha=0.9$).}
\label{fig:bound-computation}
%\vspace{-1.5em}
\end{figure*}

%\vspace{-0.5em}
\noindent
\paragraph*{Necessity of Confidence-based Mechanism}

Since the confidence-based skipping mechanism comes with additional overhead of computing the lower and upper bounds of what-if cost (Section~\ref{sec:evaluation:computation:overhead}), it is natural to ask whether such complexity is necessary.
To justify this, we compare the confidence-based mechanism with a simple \emph{randomized} mechanism that skips what-if calls randomly w.r.t. a given skipping probability threshold $p$.
Figure~\ref{fig:random-skip} presents the results when setting $p=\alpha=0.9$---we use the same confidence threshold for a fair comparison.
%Here we set $K=20$, and we use $B=1,000$ for \textbf{TPC-H} and $B=5,000$ for the other workloads.
We observe that the randomized mechanism performs similarly to the baseline but often much worse than \sysname.

%\xy{to save space, maybe we can merge figure \ref{fig:two-phase:no-opt-mci} K=20 into \ref{fig:random-skip}'s two-phase group, and put K=10 to appendix?}

%\vspace{-0.5em}
\subsubsection{Best Possible Improvement}
\label{sec:eval:e2e:best-possible}
It is difficult to know the best possible improvement without making a what-if call for \emph{every} QCP enumerated during configuration search, which is infeasible in practice. 
We provide an approximate assessment by using a much larger budget $B$ in \emph{two-phase greedy}. Specifically, we use $B=5,000$ for \textbf{TPC-H} and $B=20,000$ for the other workloads.
For each workload, we run both \emph{two-phase greedy} without and with \sysname, and we take the best improvement observed in these two runs.
The `Best' line in Figures~\ref{fig:two-phase:call-level} and~\ref{fig:mcts} presents this result.

%(this is different from vanilla greedy or two-phase greedy). As a result, the bounds are usually loose.

%\todo{Add details.}

%\todo{More results are in the slides. Need to be dumped here.}

%\todo{Change all figures to 3-dimensional ones.}

%\vspace{-0.5em}
\subsection{Efficacy of What-If Call Interception}
\label{sec:evaluation:skipping}

%\subsubsection{Amount of What-If Calls Skipped}

We measure the \emph{relative} amount of what-if calls skipped by \sysname, namely, the \emph{ratio} between the number of what-if calls skipped and the budget allowed.
Figures~\ref{fig:two-phase:call-level:skip} and~\ref{fig:mcts:skip} present the results for \emph{two-phase greedy} and \emph{MCTS} when varying $\alpha\in\{0.8, 0.9, 0.95\}$.
%choosing budget constraint $B=1,000$ for \textbf{TPC-H} and $B=5,000$ for the other workloads.

We have several observations.
First, in general, \sysname is more effective at skipping spurious what-if calls for \emph{two-phase greedy} than \emph{MCTS}. For example, when setting $K=20$ and $\alpha=0.9$, \sysname is able to skip $3.6B$ (i.e., $3.6 \times 5,000 = 18,000$) what-if calls for \emph{two-phase greedy} whereas only $0.57B$ (i.e., 2,850) what-if calls for \emph{MCTS}. 
%\xy{(since MCTS lacks enough information about singleton cost)} 
This is correlated with the observation that \sysname exhibits more significant end-to-end improvement in terms of the final index configuration found for \emph{two-phase greedy} than \emph{MCTS}, as we highlighted in Section~\ref{sec:evaluation:e2e}.
Second, the coverage-based refinement often enables \sysname to skip more what-if calls.
For instance, for \emph{MCTS} on \textbf{Real-M} when setting $K=20$ and $\alpha=0.8$, \sysname is able to skip only $1.48B$ (i.e., 7,400) what-if calls, which leads to no observable end-to-end improvement over the baseline; with the coverage-based refinement enabled, however, the number of what-if calls that \sysname can skip rises to $42.7B$ (i.e., 213,500), which results in nearly 10\% boost on the end-to-end improvement (ref. Figure~\ref{fig:mcts:real-m:conf}).
Third, while one would expect that the amount of what-if calls skipped decreases when we increase the confidence threshold $\alpha$, this is sometimes not the case, especially for \emph{two-phase greedy}.
As shown in Figures~\ref{fig:two-phase:call-level:tpch:skip},~\ref{fig:two-phase:call-level:tpcds:skip}, and~\ref{fig:two-phase:call-level:real-d:skip}, the number of skipped calls can actually increase when raising $\alpha$.
The reason for this unexpected phenomenon is the special structure of the \emph{two-phase greedy} algorithm: lowering $\alpha$ allows for more what-if calls to be skipped in the \emph{first phase} where the goal is to find good candidate indexes for each individual query. Skipping more what-if calls in the first phase therefore can result in fewer candidate indexes being selected because, without what-if calls, the derived costs for the candidate indexes will have the same value (as the what-if cost with the existing index configuration, i.e., $c(q, \emptyset)$) and thus exit early in Algorithm~\ref{alg:greedy-bound} (line 14). As a result, it eventually leads to a smaller search space for the \emph{second phase} and therefore fewer opportunities for what-if call interception.

\vspace{-0.5em}
\subsection{Computation Overhead}
\label{sec:evaluation:computation:overhead}

%\todo{Add TPC-H results.}
We measure the average computation time of the lower bound of the what-if cost.
For comparison, we also report the average time of cost derivation as well as making a what-if call.
Figure~\ref{fig:bound-computation} summarizes the results when running \emph{two-phase greedy} and \emph{MCTS} with $K=20$ and $\alpha=0.9$. 
%We set $B=1,000$ for \textbf{TPC-H} and $B=5,000$ for the other workloads.

We have the following observations.
First, the computation time of the lower bound is similar to cost derivation, both of which are orders of magnitude less than the time of making a what-if call---the $y$-axis of Figure~\ref{fig:bound-computation} is in logarithmic scale.
Second, the coverage-based refinement increases the computation time of the lower-bound, but it remains negligible compared to a what-if call. 

Table~\ref{tab:additional-overhead} further presents the additional overhead of \sysname w.r.t. the baseline configuration search algorithm without \sysname, measured as a percentage of the baseline execution time. We observe that \sysname's additional overhead, with or without the coverage-based refinement, 
is around 3\% at maximum, while the typical additional overhead is less than 0.5\%.
%is typically less than 3\% of the baseline execution time.

\iffalse
\begin{table} %[b]%[!htb]
\small
\centering
\begin{tabularx}{0.9\columnwidth}{|l|X|X|}
\hline
\multicolumn{3}{|c|}{\textbf{TPC-H} ($K=20$, $\alpha=0.9$, and $B=1,000$)}\\
\hline
\textbf{Baseline} & \textbf{\sysname} (time \%) & \textbf{\sysname-Cov} (time \%) \\
\hline
\emph{Two-phase greedy} & 0.199\% & 0.273\% \\
\emph{MCTS} & 0.064\% & 0.106\% \\
\hline
\hline
\multicolumn{3}{|c|}{\textbf{TPC-DS} ($K=20$, $\alpha=0.9$, and $B=5,000$)}\\
\hline
\textbf{Baseline} & \textbf{\sysname} (time \%) & \textbf{\sysname-Cov} (time \%) \\
\hline
\emph{Two-phase greedy} & 0.016\% & 0.345\% \\
\emph{MCTS} & 0.015\% & 0.164\% \\
\hline
\hline
\multicolumn{3}{|c|}{\textbf{Real-D} ($K=20$, $\alpha=0.9$, and $B=5,000$)}\\
\hline
\textbf{Baseline} & \textbf{\sysname} (time \%) & \textbf{\sysname-Cov} (time \%) \\
\hline
\emph{Two-phase greedy} & 0.087\% & 2.354\% \\
\emph{MCTS} & 0.029\% & 3.165\% \\
\hline
\hline
\multicolumn{3}{|c|}{\textbf{Real-M} ($K=20$, $\alpha=0.9$, and $B=5,000$)}\\
\hline
\textbf{Baseline} & \textbf{\sysname} (time \%) & \textbf{\sysname-Cov} (time \%) \\
\hline
\emph{Two-phase greedy} & 0.055\% & 2.861\% \\
\emph{MCTS} & 0.003\% & 2.544\% \\
\hline
\end{tabularx}
\caption{Additional overhead of \sysname and \sysname-Coverage, measured as percentage of the baseline execution time.}
\vspace{-1.5em}
\label{tab:additional-overhead}
\end{table}
\fi

\begin{table}%[b]%[!htb]
\small
\centering
%\begin{tabularx}{.7\columnwidth}{|l|X|X|}
\begin{tabular}{|l|r|r|}
\hline
\textbf{\sysname (\sysname-Cov.)} & \emph{two-phase greedy} & \emph{MCTS}\\
\hline
\hline
\textbf{TPC-H} ($B=1,000$) & 0.199\% (0.273\%) & 0.064\% (0.106\%)\\
\textbf{TPC-DS} ($B=5,000$) & 0.016\% (0.345\%) & 0.015\% (0.164\%)\\
\hline
\hline
\textbf{Real-D} ($B=5,000$) & 0.087\% (2.354\%) & 0.029\% (3.165\%)\\
\textbf{Real-M} ($B=5,000$) & 0.055\% (2.861\%) & 0.003\% (2.544\%)\\
\hline
%\end{tabularx}
\end{tabular}
\caption{Additional overhead of \sysname and \sysname-Coverage, measured as percentage of the execution time of the baseline configuration search algorithm ($K=20$, $\alpha=0.9$).}
%\xy{maybe compress table \ref{tab:additional-overhead} like this?}}
\vspace{-1em}
\label{tab:additional-overhead}
\end{table}

%\xy{end-to-end runing time?}
%\todo{revise}
%\todo{Add more discussion.}

%%%%%%%%%%%% Two-phase Greedy (storage constraint) %%%%%%%%%%

%\vspace{-0.5em}
\subsection{Storage Constraints}
As mentioned earlier, one may have other constraints in practical index tuning in addition to the cardinality constraint. One common constraint is the \emph{storage constraint} (SC) that limits the maximum amount of storage taken by the recommended indexes~\cite{KossmannHJS20}. To demonstrate the robustness of \sysname w.r.t. other constraints, we evaluate its efficacy by varying the SC as well. In our evaluation, we fix $K=20$, $\alpha=0.9$, $B=1,000$ for \textbf{TPC-H} and $B=5,000$ for the other workloads, while varying the allowed storage size as 2$\times$ and 3$\times$ of the database (3$\times$ is the default setting of DTA~\cite{dta-utility}).

\begin{figure*}
\centering
\subfigure[\textbf{TPC-H}, $B=1,000$]{ \label{fig:two-phase:sc:tpch}
    \includegraphics[width=0.23\textwidth]{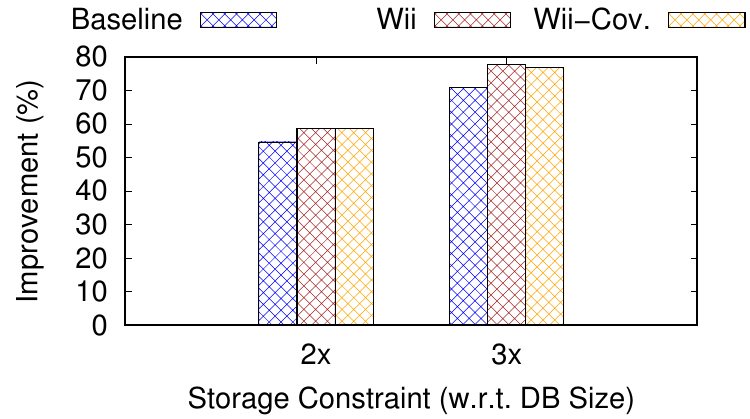}}
\subfigure[\textbf{TPC-DS}, $B=5,000$]{ \label{fig:two-phase:sc:tpcds}
    \includegraphics[width=0.23\textwidth]{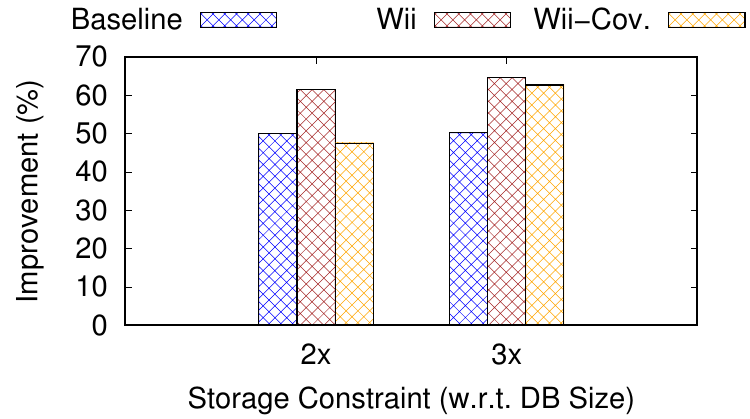}}
\subfigure[\textbf{Real-D}, $B=5,000$]{ 
\label{fig:two-phase:sc:real-d}
    \includegraphics[width=0.23\textwidth]{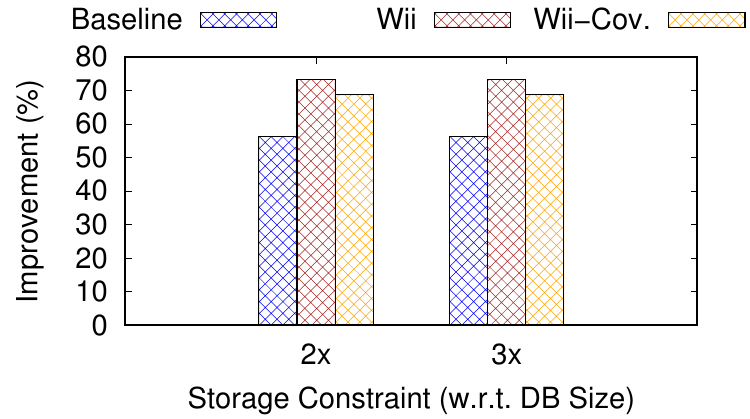}}
\subfigure[\textbf{Real-M}, $B=5,000$]{ \label{fig:two-phase:sc:real-m}
    \includegraphics[width=0.23\textwidth]{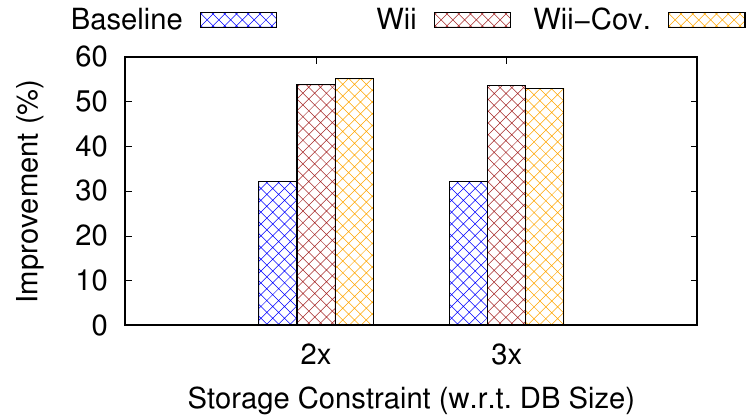}}
\vspace{-1.5em}
\caption{Evaluation results of \sysname for \emph{two-phase greedy} with varying storage constraints ($K=20$, $\alpha=0.9$).}
\label{fig:two-phase:sc}
\vspace{-1.5em}
\end{figure*}

\begin{figure*}
\centering
\subfigure[\textbf{TPC-H}, $B=1,000$]{ \label{fig:mcts:sc:tpch}
    \includegraphics[width=0.23\textwidth]{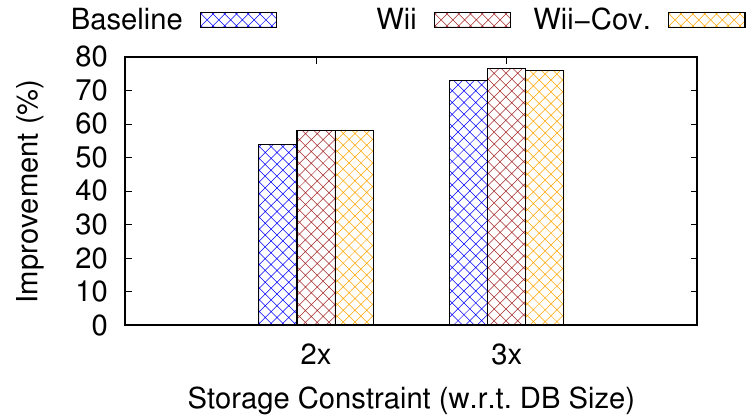}}
\subfigure[\textbf{TPC-DS}, $B=5,000$]{ \label{fig:mcts:sc:tpcds}
    \includegraphics[width=0.23\textwidth]{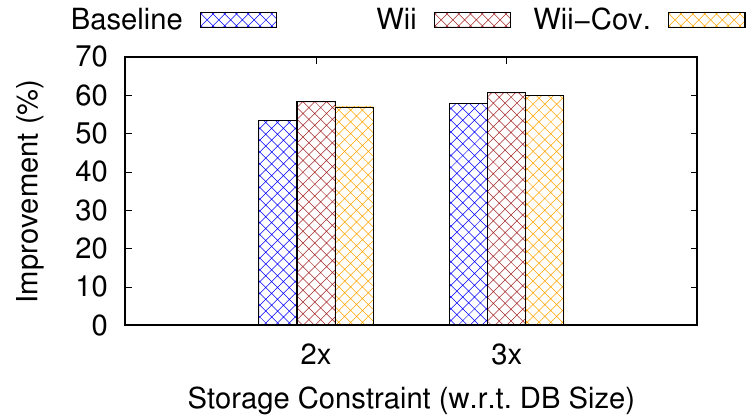}}
\subfigure[\textbf{Real-D}, $B=5,000$]{ 
\label{fig:mcts:sc:real-d}
    \includegraphics[width=0.23\textwidth]{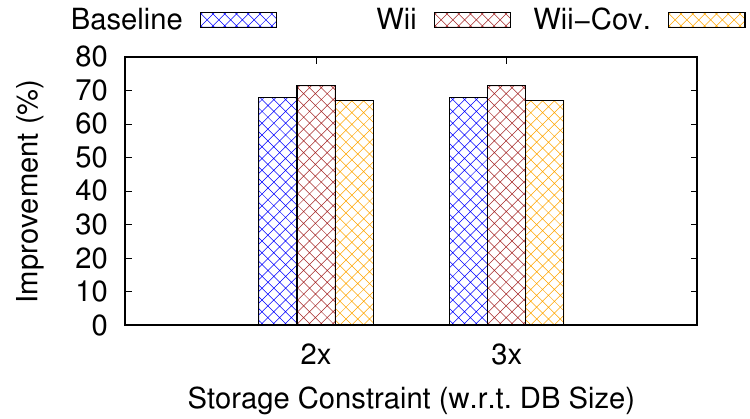}}
\subfigure[\textbf{Real-M}, $B=5,000$]{ \label{fig:mcts:sc:real-m}
    \includegraphics[width=0.23\textwidth]{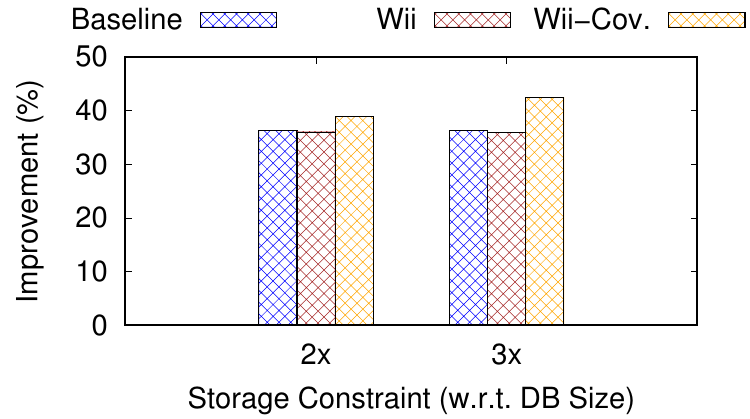}}
\vspace{-1.5em}
\caption{Evaluation results of \sysname for \emph{MCTS} with varying storage constraints ($K=20$, $\alpha=0.9$).}
\label{fig:mcts:sc}
\vspace{-1.5em}
\end{figure*}

Figures~\ref{fig:two-phase:sc} and~\ref{fig:mcts:sc} present the evaluation results for \emph{two-phase greedy} and \emph{MCTS}.
Overall, we observe similar patterns in the presence of SC. That is, \sysname, with or without the coverage-based refinement, often significantly outperforms the baseline approaches, especially for \emph{two-phase greedy}.

%\xy{two figures may be able to merge into one line.}

%%%%%%%%%%%% MCTS (storage constraint) %%%%%%%%%%

%\vspace{-0.5em}
%\vspace{-1em}

%%%%%%%%%%%% Two-phase Greedy Avg Bounds %%%%%%%%%%

\begin{figure*}
\centering
\subfigure[\textbf{TPC-H}, $B=1,000$]{ \label{fig:two-phase:tpch:avg-bounds}
    \includegraphics[width=0.23\textwidth]{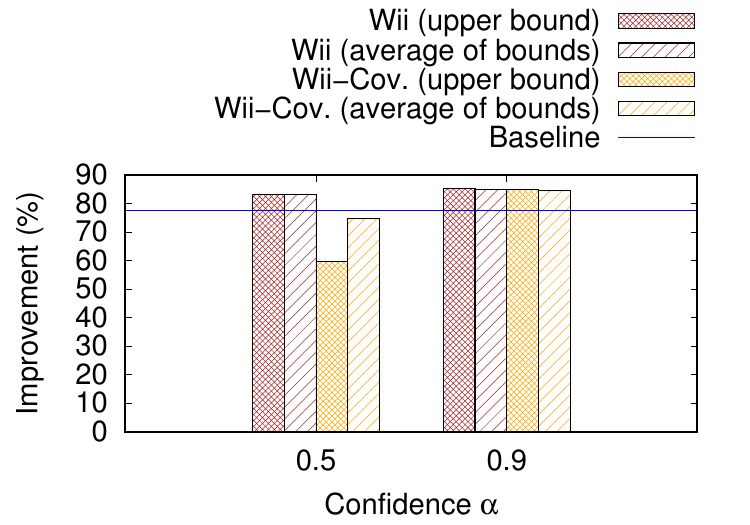}}
\subfigure[\textbf{TPC-DS}, $B=5,000$]{ \label{fig:two-phase:tpcds:avg-bounds}
    \includegraphics[width=0.23\textwidth]{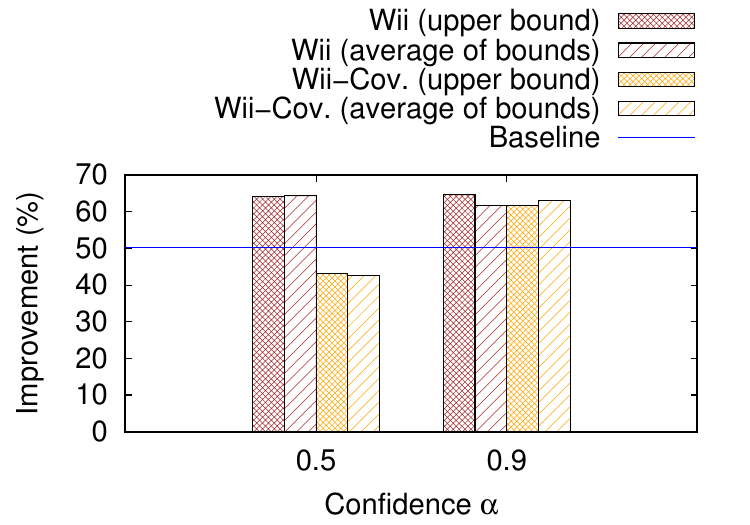}}
\subfigure[\textbf{Real-D}, $B=5,000$]{ \label{fig:two-phase:real-d:avg-bounds}
    \includegraphics[width=0.23\textwidth]{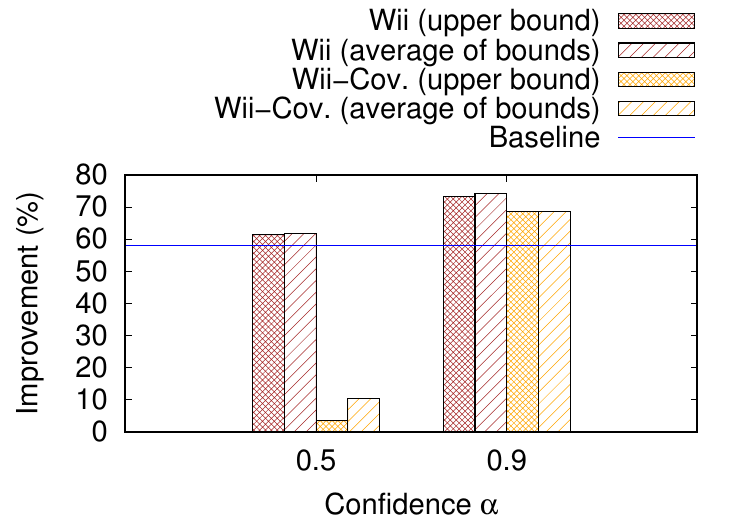}}
\subfigure[\textbf{Real-M}, $B=5,000$]{ \label{fig:two-phase:real-m:avg-bounds}
\includegraphics[width=0.23\textwidth]{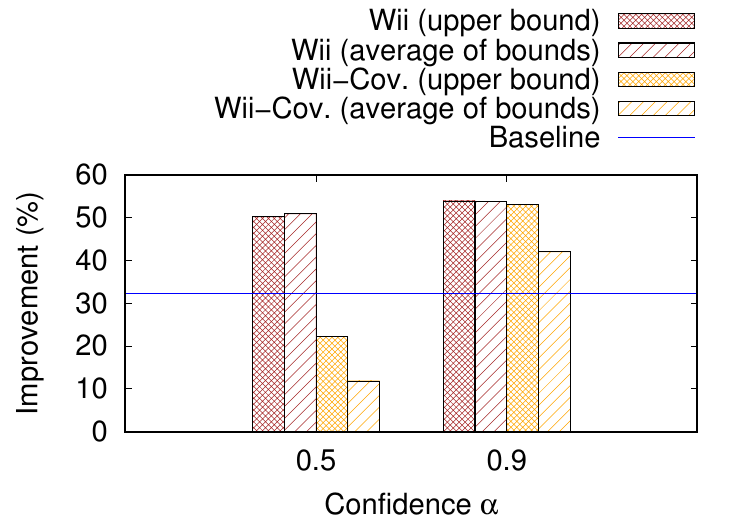}}
\vspace{-1.5em}
\caption{Using derived cost vs. the average of lower and upper bounds for \emph{two-phase greedy} ($K=20$).}
\label{fig:two-phase:avg-bounds}
\vspace{-1.5em}
\end{figure*}

%%%%%%%%%%%% MCTS Avg Bounds %%%%%%%%%%

\begin{figure*}
\centering
\subfigure[\textbf{TPC-H}, $B=1,000$]{ \label{fig:mcts:tpch:avg-bounds}
    \includegraphics[width=0.23\textwidth]{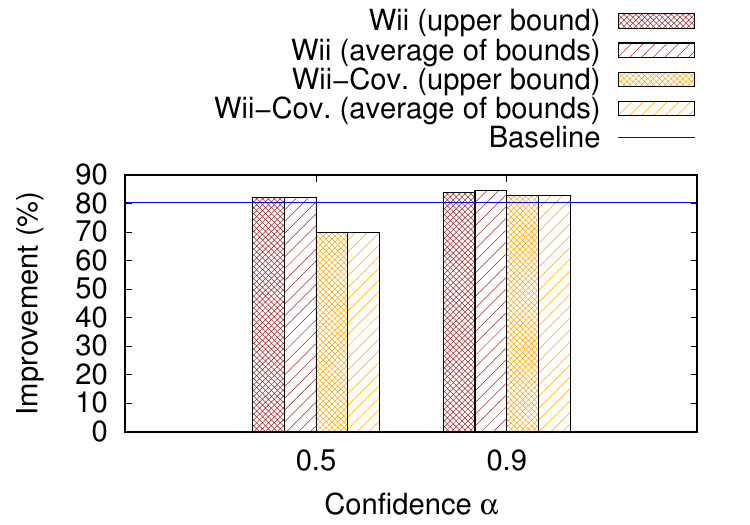}}
\subfigure[\textbf{TPC-DS}, $B=5,000$]{ \label{fig:mcts:tpcds:avg-bounds}
    \includegraphics[width=0.23\textwidth]{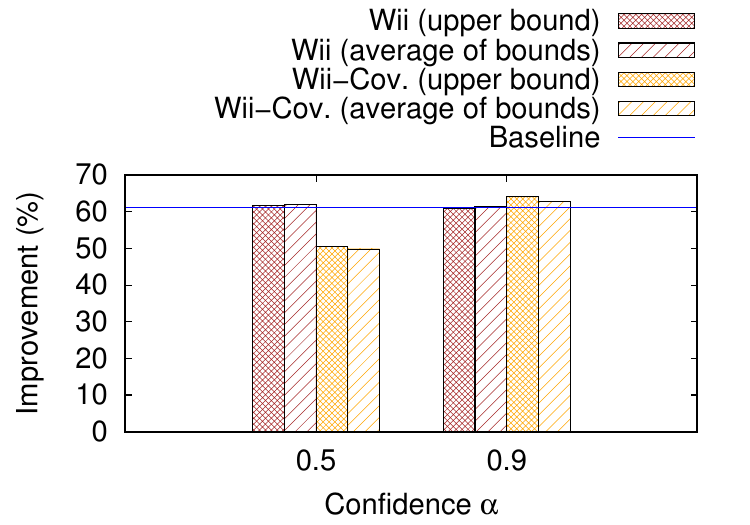}}
\subfigure[\textbf{Real-D}, $B=5,000$]{ \label{fig:mcts:real-d:avg-bounds}
    \includegraphics[width=0.23\textwidth]{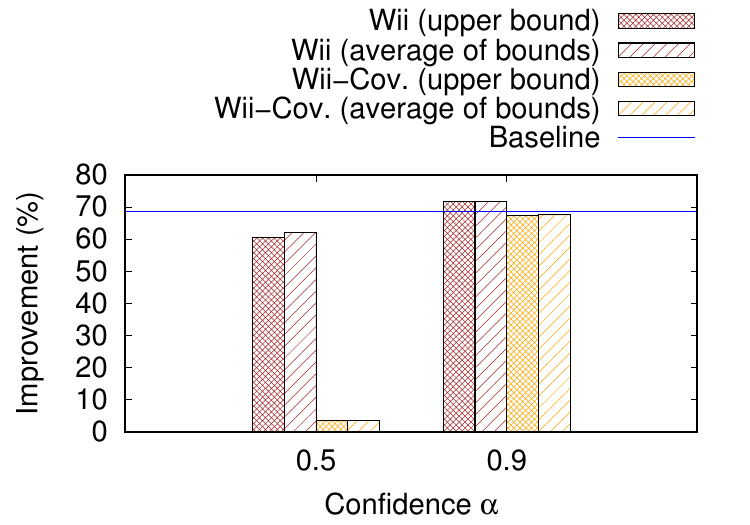}}
\subfigure[\textbf{Real-M}, $B=5,000$]{ \label{fig:mcts:real-m:avg-bounds}
\includegraphics[width=0.23\textwidth]{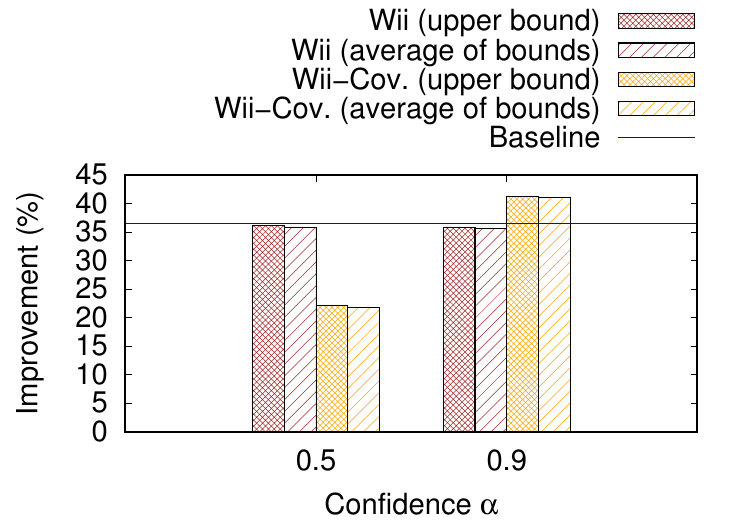}}
\vspace{-1.5em}
\caption{Using derived cost vs. the average of lower and upper bounds for \emph{MCTS} ($K=20$).} 
\label{fig:mcts:avg-bounds}
%\vspace{-1em}
\end{figure*}

\subsection{Beyond Derived Cost}

When \sysname decides to skip a what-if call, it returns the derived cost (i.e., the upper bound) as an approximation of the what-if cost. This is not mandatory, and there are other options.
For example, one can instead return the \emph{average} of the lower and upper bounds.
We further evaluate this idea below.
%In our experiments, we set $K=20$, $B=1,000$ for \textbf{TPC-H}, and $B=5,000$ for the other workloads. \xy{maybe to set this setup as the default setup in the sec 6.1, and refer to it directly later.}
Figures~\ref{fig:two-phase:avg-bounds} and~\ref{fig:mcts:avg-bounds} present the results.
While both options perform similarly most of the time, we observe that they perform quite differently in a few cases; moreover, one may outperform the other in these cases.
For example, with the coverage-based refinement enabled in \sysname, when setting $\alpha=0.5$, on \textbf{TPC-H} returning the average significantly outperforms returning the upper bound (74.7\% vs. 59.7\%); however, on \textbf{Real-M} returning the average loses 10.5\% in percentage improvement compared to returning the upper bound (11.8\% vs. 22.3\%).
As a result, the question of having a better cost approximation than the upper bound (i.e., the derived cost) remains open, and we leave it for future exploration.

%\xy{should we explain the result a bit here?}

%\vspace{1em}
\subsection{Impact of Submodularity Assumption}
\label{sec:evaluation:submodularity}

\begin{table}%[!htb]
\small
\centering
%\begin{tabularx}{.7\columnwidth}{|l|X|X|X|}
\begin{tabular}{|l|r|r|r|}
\hline
\textbf{Workload} & \textbf{Average} & \textbf{Median} & \textbf{$95^{\text{th}}$ Percentile} \\
\hline
\hline
\textbf{TPC-H} & 0.209 & 0.001 & 1.498 \\
\textbf{TPC-DS} & 2.203 & 0.001 & 10.532\\
\hline
\hline
\textbf{Real-D} & 7.658 & 0.010 & 38.197\\
\textbf{Real-M} & 4.125 & 0.001 & 31.358\\
\hline
%\end{tabularx}
\end{tabular}
\caption{Magnitude of violation (of submodularity).}
\vspace{-1.5em}
\label{tab:validation:submodularity:violation}
\vspace{0.5em}
\end{table}

\iffalse
\begin{table}%[!htb]
\small
\centering
\begin{tabularx}{\columnwidth}{|l|X|X|X|X|X|}
\hline
\textbf{Workload} & \textbf{\# Total} & \textbf{\# Yes} & \textbf{\# No} & \textbf{\% Yes} & \textbf{\% No} \\
\hline
\hline
\textbf{TPC-H} & 444 & 389 & 55 & 87.6\% & 12.4\% \\
\textbf{TPC-DS} & 3,120 & 2,349 & 771 & 75.3\% & 24.7\% \\
\hline
\hline
\textbf{Real-D} & 3,282 & 2,896 & 386 & 88.2\% & 11.8\% \\
\textbf{Real-M} & 48,166 & 40,976 & 7,190 & 85.1\% & 14.9\% \\
\hline
\end{tabularx}
\caption{Validation Results of Submodularity Assumption.}
\label{tab:validation:submodularity}
\end{table}
\fi

%\todo{Add a discussion on \sysname's performance when submodularity does not hold.}

Although our validation results show that submodularity holds with probability between 0.75 and 0.89 on the workloads tested~\cite{full-version}, it remains an interesting question to understand the impact on \sysname when submodularity does not hold.
As we mentioned in Section~\ref{sec:bounds:call-level:mci:submodularity}, submodularity does not hold often due to \emph{index interaction}~\cite{SchnaitterPG09}.
For example, the query optimizer may choose an \emph{index-intersection} plan with two indexes available \emph{at the same time} but utilizing neither if only one of them is present.
In this example, submodularity does not hold, because the MCI of either index will \emph{increase} after the other index is selected.
As a result, Equation~\ref{eq:bound:submodular} is no longer an MCI upper-bound---it will be \emph{smaller} than the actual MCI upper-bound.
Consequently, the $L(q, C)$ computed by Equation~\ref{eq:lower-bound:call-level:v1} will be \emph{larger} than the actual lower-bound of the what-if cost, which implies an \emph{overconfident} situation for \sysname where the confidence is computed by Equation~\ref{eq:confidence}.
The \emph{degree of overconfidence} depends on the \emph{magnitude of violation} of the submodularity assumption, which we further measured in our evaluation (see~\cite{full-version} for details).

Table~\ref{tab:validation:submodularity:violation} summarizes the key statistics of the magnitude of violation measured.
Among the four workloads, we observe that \textbf{Real-D} and \textbf{Real-M} have relatively higher magnitude of violation, which implies that \sysname tends to be more overconfident on these two workloads.
As a result, \sysname is more likely to skip what-if calls that should not have been skipped, especially when the confidence threshold $\alpha$ is relatively low.
Correspondingly, we observe more sensitive behavior of \sysname on \textbf{Real-D} and \textbf{Real-M} when increasing $\alpha$ from 0.5 to 0.9 (ref. Figures~\ref{fig:two-phase:low-conf} and~\ref{fig:mcts:low-conf}).

%\revision{Moreover, we analyzed the cases when submodularity was violated and we observe that the violation is typically very small (see~\cite{full-version} for details).}

%\vspace{-0.5em}
\subsection{The Case of Unlimited Budget}
\label{sec:eval:no-budget-limit}

%While we focus on budget-aware index tuning in this paper, 
As we noted in the introduction, \sysname can also be used in a special situation where one does not enforce a budget on the index tuner, namely, the tuner can make unlimited number of what-if calls.
This situation may make sense if one has a relatively small workload.
Although \sysname cannot improve the quality of the final configuration found, by skipping unnecessary what-if calls it can significantly reduce the overall index tuning time.

\iffalse
\sysname plays a different role here.
Since there is no budget constraint, \sysname cannot improve the quality of the final configuration found as the best quality can eventually be achieved by keeping on issuing what-if calls. 
Instead, by skipping unnecessary what-if calls, \sysname can improve the \emph{efficiency} of index tuning by significantly reducing the tuning time.
\fi

To demonstrate this, we tune the two relatively small workloads, namely \textbf{TPC-H} with 22 queries and \textbf{Real-D} with 32 queries, using \emph{two-phase greedy} without enforcing a budget constraint on the number of what-if calls.
% We do not use \emph{MCTS} as it explicitly leverages the budget constraint by design and cannot work without the budget information~\cite{WuWSWNCB22}.
We do not use \emph{MCTS} as it explicitly leverages the budget constraint by design and cannot work without the budget information.
We set $K=20$ for \textbf{TPC-H} and $K=5$ for \textbf{Real-D} in our experiments to put the total execution time under control.
We also vary the confidence threshold $\alpha\in\{0.8, 0.9\}$ for \sysname.
Table~\ref{tab:no-budget-limit} summarizes the evaluation results.

We observe significant reduction of index tuning time by using \sysname.
For instance, on \textbf{TPC-H} when setting the confidence threshold $\alpha=0.9$, the final configurations returned by \emph{two-phase greedy}, with or without \sysname, achieve (the same) 85.2\% improvement over the existing configuration.
However, the tuning time is reduced from 8.2 minutes to 1.9 minutes (i.e., 4.3$\times$ speedup) when \sysname is used.
%We also tune the \textbf{Real-D} workload that contains 32 queries (where the queries are much more complex than \textbf{TPC-H} queries) by setting $K=5$ and $\alpha=0.9$, again using \emph{two-phase greedy}.
As another example, on \textbf{Real-D} when setting $\alpha=0.9$, the final configurations returned, with or without \sysname, achieve similar improvements over the existing configuration (64\% vs. 62.3\%).
However, the tuning time is reduced from 380.6 minutes to 120 minutes (i.e., 3.2$\times$ speedup) by using \sysname.
The index tuning time on \textbf{Real-D} is considerably longer than that on \textbf{TPC-H}, since the \textbf{Real-D} queries are much more complex. %than the \textbf{TPC-H} queries.
%\todo{Revise. Add a table perhaps.}

\begin{table}[t] %[b]%[!htb]
\small
\centering
\begin{tabularx}{.7\columnwidth}{|l|X|X|X|X|}
\hline
\multicolumn{5}{|c|}{\textbf{TPC-H}, $K=20$}\\
\hline
\textbf{Method} & \textbf{Time} \newline ($\alpha=0.8$) & \textbf{Impr.} \newline ($\alpha=0.8$) & \textbf{Time} \newline ($\alpha=0.9$)& \textbf{Impr.} \newline ($\alpha=0.9$) \\
\hline
\emph{Baseline} & 8.22 min & 85.22\% & 8.22 min & 85.22\%\\
\emph{\sysname} & 1.62 min & 84.74\% & 1.95 min & 85.26\% \\
\emph{\sysname-Cov.} & 0.94 min & 83.95\% & 1.67 min & 85.02\% \\
\hline
\hline
\multicolumn{5}{|c|}{\textbf{Real-D}, $K=5$}\\
\hline
\textbf{Method} & \textbf{Time} \newline ($\alpha=0.8$) & \textbf{Impr.} \newline ($\alpha=0.8$) & \textbf{Time} \newline ($\alpha=0.9$)& \textbf{Impr.} \newline ($\alpha=0.9$) \\
\hline
\emph{Baseline} & 380.63 min & 62.32\% & 380.63 min & 62.32\% \\
\emph{\sysname} & 118.95 min & 64.10\% & 119.99 min & 64.10\% \\
\emph{\sysname-Cov.} & 31.42 min & 62.90\% & 53.38 min & 59.63\% \\
\hline
\end{tabularx}
\caption{Index tuning time with unlimited budget.}
\label{tab:no-budget-limit}
\vspace{-1em}
\end{table}
%\vspace{1em}
\section{Related Work}
\label{sec:related-work}

\iffalse
\begin{itemize}
    \item Related work on index tuning.
    \item Related work on saving what-if calls.
    \item Related work on ML-based approach for encoding index feature representations.
\end{itemize}
\fi

%\vspace{-0.5em}
\paragraph*{Index Tuning}

Index tuning has been studied extensively by previous work (e.g.,~\cite{Whang85,ChaudhuriN97,dta,ValentinZZLS00,BrunoC05,DashPA11,dexter-2,SchlosserK019,KossmannKS22,WuWSWNCB22,SiddiquiJ00NC22,SiddiquiWNC22,Wred}).
The recent work by Kossmann et al.~\cite{KossmannHJS20} conducted a survey as well as a benchmark study of existing index tuning technologies.
Their evaluation results show that \emph{DTA} with the \emph{two-phase greedy} search algorithm~\cite{ChaudhuriN97,dta} can yield the state-of-the-art performance, which has been the focus of our study in this paper as well.
%\todo{Add comments on using what-if cost as gold standard. Inaccuracy in optimizer cost estimation may lead to query performance regression, which is beyond the scope of this paper.}

%\vspace{-0.5em}
\paragraph*{Budget-aware Configuration Enumeration}

Configuration enumeration is one core problem of index tuning.
The problem is \emph{NP-hard} and \emph{hard to approximate}~\cite{Comer78,ChaudhuriDN04}.
%Among existing technologies, the benchmark study~\cite{KossmannHJS20} shows that \emph{DTA} with the greedy search algorithm~\cite{ChaudhuriN97,dta} can yield the state-of-the-art performance in terms of the quality of the final index configuration found.
%Nonetheless, 
Although \emph{two-phase greedy} is the current state-of-the-art~\cite{KossmannHJS20}, it remains inefficient on large and/or complex workloads, due to the large amount of what-if calls made to the query optimizer during configuration enumeration~\cite{PapadomanolakisDA07,KossmannHJS20,ShiCL22,SiddiquiWNC22}.
Motivated by this, \cite{WuWSWNCB22} studies a constrained configuration enumeration problem, called \emph{budget-aware configuration enumeration}, that limits the number of what-if calls allowed in configuration enumeration.
Budget-aware configuration enumeration introduces a new \emph{budget allocation} problem, regarding which query-configuration pairs (QCP's) deserve what-if calls.
%so that the what-if cost of the final configuration found can be minimized.
\iffalse
There is a natural \emph{exploration}/\emph{exploitation} trade-off in budget allocation, and \cite{WuWSWNCB22} further proposes an MCTS algorithm based on reinforcement learning (RL) to systematically address this trade-off~\cite{sutton2018reinforcement}.
\fi

%\vspace{-0.5em}
\paragraph*{Application of Data-driven ML Technologies}
There has been a flurry of recent work on applying data-driven machine learning (ML) technologies to various aspects of index tuning~\cite{ml-index-tuning-overview}, such as reducing the chance of performance regression on the recommended indexes~\cite{DingDM0CN19,ZhaoCSM22}, configuration search algorithms based on deep learning and reinforcement learning~\cite{PereraORB21,abs-1801-05643,LanBP20,PereraORB22}, using learned cost models to replace what-if calls~\cite{SiddiquiWNC22,ShiCL22}, and so on.
While we do not use ML technologies in this work,
%for the reasons stated in Section~\ref{sec:evaluation:discussion}, 
it remains interesting future work to consider using ML-based technologies, for example, to improve the accuracy of the estimated coverage.
\iffalse
Using ML technologies, however, may significantly increase the index tuning overhead if we want to collect data and train ML models \emph{during} configuration enumeration.
One potential direction is to use pre-trained ML models~\cite{BERT,GPT-1}, though it raises new questions such as their generalization capability across workloads.
\fi

%though it raises a set of new challenges and research questions that we have discussed in~\ref{sec:evaluation:discussion}.

%\vspace{-0.5em}
\paragraph*{Cost Approximation and Modeling}

From an API point of view, \sysname returns an approximation (i.e., derived cost) of the what-if cost whenever a what-if call is saved.
%In this sense, it plays the same role as cost derivation.
There have been various other technologies on cost approximation and modeling, focusing on replacing query optimizer's cost estimate by actual prediction of query execution time (e.g.,~\cite{Ganapathi-berkeley09,AkdereCRUZ12-brown-icde,LiKNC12,WuCZTHN13,WuCHN13,WuWHN14,WuNS16,MarcusP19,MarcusNMZAKPT19,SunL19,SiddiquiJQPL20,PaulCLS21,hilprecht2022zero}). This line of effort is orthogonal to our work, which uses optimizer's cost estimate as the gold standard of query execution cost, to be in line with previous work on evaluating index configuration enumeration algorithms~\cite{ChaudhuriN97,KossmannHJS20}.

%\todo{Add more on recent ML work for QO, especially for cost modeling and approximation.}

\iffalse
\paragraph*{Comparison to Cost Derivation}

From an API point of view, \sysname plays the same role as cost derivation by providing the index tuner with approximated what-if costs whenever possible.
Unlike cost derivation, \sysname puts more guards on the approximated costs---it only delegates to cost derivation when it believes that the derived costs will be close to the actual what-if costs.
This, however, comes with an overhead---\sysname is computationally more expensive than cost derivation as it requires computing an upper bound of the what-if cost as well.
In this regard, one can think of \sysname as a compromise between using cost derivation and using true what-if cost, with the trade-off between computation overhead and cost approximation accuracy in mind.
\todo{This discussion is a bit out of nowhere. We need to demonstrate that the overheads of our techniques are between cost derivation and true what-if cost.}
\fi

%\todo{A question here is if we need to compare against DISTILL, which claims to be able to use some what-if calls to build an ML model to predict the what-if costs.}

%\textbf{Notes on the DISTILL paper:} It actually trains a model per query template, which cannot be applied to our setting.

\section{Conclusion}
\label{sec:conclusion}

%\todo{Revise, copied from the SIGMOD paper.}

In this paper, we proposed \sysname that can be seamlessly integrated into existing configuration enumeration algorithms to improve budget allocation and ultimately quality of the final index configuration found.
\sysname develops and leverages lower and upper bounds of the what-if cost to skip unnecessary what-if calls during configuration enumeration.
%We developed lower-bound and upper-bound for 
%two versions of \sysname: one is based on strict lower and upper bounds of the what-if cost, whereas the other is based on relaxed/approximate bounds by the notion of coverage.
%We further improve the lower bound using coverage-based refinement when what-if costs of singleton configurations are unavailable.
Our evaluation results on both industrial benchmarks and real workloads demonstrate the effectiveness of \sysname.
%of both the bound-based and coverage-refined \sysname.

%\todo{Mention future work such as integrating with DTA.}

\vspace{0.5em}
\noindent
\textbf{Acknowledgments:} We thank the anonymous reviewers, Arnd Christian König, Anshuman Dutt, Bailu Ding, and Tarique Siddiqui for their valuable and constructive feedback.
This work was done when Xiaoying Wang was at Microsoft Research.

\clearpage
\bibliographystyle{ACM-Reference-Format}
\bibliography{paper}

\clearpage
\appendix
\section{Proofs}

\subsection{Proof of Lemma~\ref{lemma:bound:submodular}}

\begin{proof}
By Assumption~\ref{assumption:submodular}, we have 
$\delta(q, z, C)\leq \delta(q, z, \emptyset)$, since $\emptyset\subseteq C$. On the other hand, by the definition of $\delta(q, z, \emptyset)$ we have
$$\delta(q, z, \emptyset)=c(q, \emptyset)-c(q, \emptyset\cup\{z\})=c(q, \emptyset)-c(q, \{z\})=\Delta(q, \{z\}).$$
This completes the proof.
\end{proof}

\subsection{Proof of Theorem~\ref{theorem:greedy:mci-update}}

\begin{proof}
We use $u^{(k)}(q, z)$ to represent the $u(q, z)$ after the greedy step $k$.
We prove by induction on the greedy step $k$ ($1\leq k\leq K$):
\begin{itemize}[leftmargin=*]
    \item \textbf{(Base)} When $k=0$, by the update step (1) in Procedure~\ref{proc:maintain-mci},
    $$u^{(0)}(q, z)=\min\{c(q,\emptyset), \Delta(q, \Omega)\}$$
    is clearly an MCI upper-bound.
    \item \textbf{(Induction)} Suppose that $u^{(k)}(q, z)$ remains an MCI upper-bound. Consider $u^{(k+1)}(q, z)$. There are two cases. First, if either $c(q, C_k)$ or $c(q, C_k\cup\{z\})$ is unavailable, then there is no update to $u(q, z)$ and therefore $u^{(k+1)}(q, z)=u^{(k)}(q, z)$. Otherwise, by the update step (2) in Procedure~\ref{proc:maintain-mci},
    $$u^{(k+1)}(q, z)=c(q, C_k)-c(q, C_k\cup\{z\})=\delta(q, z, C_k).$$ 
    Due to the nature of the greedy search procedure, we can restrict the configuration $C$ in the MCI $\delta(q, z, C)$ to those configurations \emph{selected by each greedy step}. Here, it means that we only need to consider $\delta(q, z, C_j)$ where $j>k$. By definition of $\delta(q, z, C_j)$,
    $$\delta(q, z, C_j)=c(q, C_j)-c(q, C_j\cup\{z\}).$$
    By Assumption~\ref{assumption:submodular}, we have 
    $$\delta(q, z, C_j)\leq \delta(q, z, C_k)=u^{(k+1)}(q, z).$$
    As a result, $u^{(k+1)}(q, z)$ remains an MCI upper-bound.
\end{itemize}
This completes the proof.
\end{proof}

\subsection{Proof of Theorem~\ref{theorem:lower-bound:greedy-search}}

\begin{proof}
By Equation~\ref{eq:lower-bound:call-level:v2}, we have
$$L(q, C_z)=\max_{S\subset C_z}\Big(c(q, S)-\sum\nolimits_{x\in C_z-S} u(q, x)\Big).$$
Since $C_z=C^*\cup\{z\}$, there are two cases for $S\subset C_z$: (1) $S\subseteq C^*$ and (2) $S=S^*\cup\{z\}$ where $S^*\subset C^*$. 
In either case, we need to show
$$c(q, S)-\sum\nolimits_{x\in C_z-S} u(q, x)\leq c(q, C^*)-u(q,z),$$
or equivalently,
$$c(q, S)-c(q, C^*)\leq \sum\nolimits_{x\in C_z-S} u(q, x)-u(q,z).$$
Without loss of generality, let $C^*=C_k=\{x_1, ..., x_k\}$, where $x_i$ is the index selected in the $i$-th step of greedy search.
We now discuss each of these two cases below:
\begin{itemize}[leftmargin=*]
\item \textbf{(Case 1)} If $S\subseteq C^*$, then let $|S|=l$ for some $l<k$ and denote $C^*-S=\{x_{i_1}, ..., x_{i_{k-l}}\}$.
We have
    \begin{eqnarray*}
    c(q, S)-c(q, C^*)&=&c(q, C_l)-c(q, C_k)\\  
    &=&\sum\nolimits_{j=l}^{k-1}\Big(c(q, C_j)-c(q, C_{j+1})\Big)\\
    &=&\sum\nolimits_{j=l}^{k-1}\Big(c(q, C_j)-c(q, C_j\cup\{x_{i_{j-l+1}}\})\Big)\\
    &=&\sum\nolimits_{j=l}^{k-1}\delta(q, x_{i_{j-l+1}}, C_j)\\
    &\leq& \sum\nolimits_{j=l}^{k-1} u(q, x_{i_{j-l+1}})\\
    &\leq& \sum\nolimits_{x\in C^*-S} u(q, x)\\
    &=&\sum\nolimits_{x\in C_z-S} u(q, x)-u(q,z).
    \end{eqnarray*}
The last step holds because 
$$C_z-S=(C^*\cup\{z\})-S=(C^*-S)\cup\{z\}=\{x_{i_1}, ..., x_{i_{k-l}}, z\}.$$
\item \textbf{(Case 2)} If $S=S^*\cup\{z\}$ where $S^*\subset C^*$, then it follows that
    \begin{eqnarray*}
    c(q, S)-c(q, C^*)=c(q, S^*\cup\{z\})-c(q, C^*)
    =\Big(c(q, S^*\cup\{z\})-c(q, S^*)\Big)
    +\Big(c(q, S^*) - c(q, C^*)\Big).
    \end{eqnarray*}
On one hand, we have
\begin{eqnarray*}
c(q, S^*\cup\{z\})-c(q, S^*)=-\Big(c(q, S^*)-c(q, S^*\cup\{z\})\Big)
=-\delta(q, z, S^*).
\end{eqnarray*}
On the other hand, let $|S^*|=l$ for some $l<k$ and denote $C^*-S^*=\{x_{i_1}, ..., x_{i_{k-l}}\}$, following the proof of \textbf{Case 1} we have
    \begin{eqnarray*}
    c(q, S^*)-c(q, C^*)&=&c(q, C_l)-c(q, C_k)\\  
    &=&\sum\nolimits_{j=l}^{k-1}\delta(q, x_{i_{j-l+1}}, C_j)\\
    &\leq& \sum\nolimits_{j=l}^{k-1} u(q, x_{i_{j-l+1}})\\
    &=&\sum\nolimits_{x\in C^*-S^*} u(q, x)\\
    &=&\sum\nolimits_{x\in C_z-S} u(q, x).
    \end{eqnarray*}
The last step holds by noticing
$$C_z-S=(C^*\cup\{z\})-(S^*\cup\{z\})=C^*-S^*.$$
As a result, it follows that
$$c(q, S)-c(q, C^*)\leq \sum\nolimits_{x\in C_z-S} u(q, x)-\delta(q, z, S^*).$$
Moreover, notice that $\delta(q, z, S^*)\geq u(q, z)$, due to the update step (2) in Procedure~\ref{proc:maintain-mci}. Specifically, here $S^*$ cannot be just a subset of $C^*$; rather, it must be some ``prefix'' of $C^*$.
To see this, since $z$ has not been selected by greedy search yet, it must have been considered with \emph{any} prefix of $C^*$ but nothing else. That is, we only have what-if costs for configurations that contain $z$ and some prefix of $C^*$---we do not have what-if cost for any other configuration that contains $z$.
Note that this does not need to hold for the $S$ in \textbf{Case 1}, namely, $S$ is not necessarily a prefix of $C^*$ there. However, the $S$ in \textbf{Case 1} must also contain some prefix of $C^*$---in fact, $S$ must be either a prefix of $C^*$ or a prefix of $C^*$ plus one additional index from $C^*$, due to the structure of greedy search (ref. Figure~\ref{fig:greedy}). To summarize, we conclude
$$c(q, S)-c(q, C^*)\leq \sum\nolimits_{x\in C_z-S} u(q, x)-u(q, z).$$
\end{itemize}
This completes the proof of the theorem.
\end{proof}

\section{More Evaluation Results}

\subsection{Accuracy of Estimated Coverage}
\label{sec:evaluation:coverage:accuracy}

%\todo{Add TPC-H results.}
One critical factor for the efficacy of the coverage-based refinement is the accuracy of estimated coverage.
We test this by measuring the \emph{absolute error} of the estimated coverage in terms of the ground truth.
Specifically, let $\hat{\rho}$ be the estimated coverage using Equation~\ref{eq:coverage:estimated} and let $\rho$ be the ground-truth coverage defined by Equation~\ref{eq:coverage}.
The absolute error $\epsilon(\hat{\rho},\rho)$ is defined as
$\epsilon(\hat{\rho},\rho)=|\hat{\rho}-\rho|.$

Note that $0\leq \epsilon(\hat{\rho},\rho)\leq 1$ and a smaller $\epsilon(\hat{\rho},\rho)$ means that the estimated coverage is more accurate.
We collect data points for this investigation as follows.
For each query $q$ in a workload, we collect all of its candidate indexes and treat each of them as a singleton configuration $\{z\}$.
We then make a what-if call for each such query-index pair $(q, \{z\})$ to the query optimizer and obtain its what-if cost $c(q, \{z\})$.
We compute $\hat{\rho}$ and $\rho$ for each pair $(q, \{z\})$ based on Equations~\ref{eq:coverage:estimated} and~\ref{eq:coverage}.
%, respectively.

\begin{figure*}
\centering
\subfigure[\textbf{TPC-H}]{ \label{fig:coverage-accuracy:tpch}
    \includegraphics[width=0.23\textwidth]{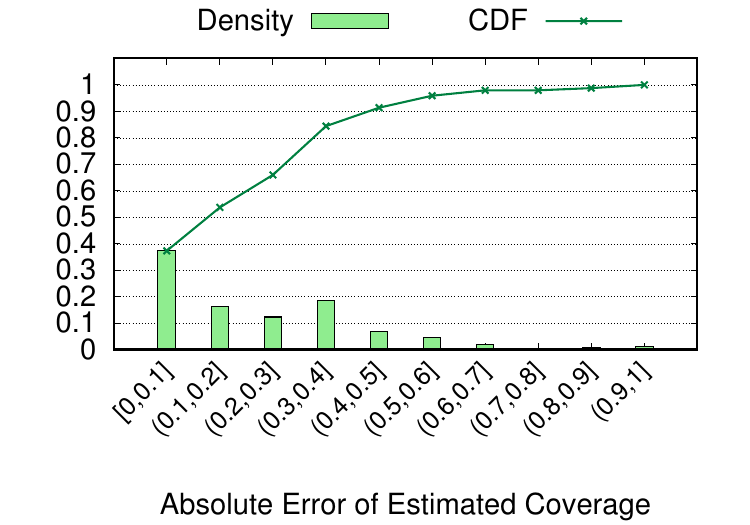}}
\subfigure[\textbf{TPC-DS}]{ \label{fig:coverage-accuracy:tpcds}
    \includegraphics[width=0.23\textwidth]{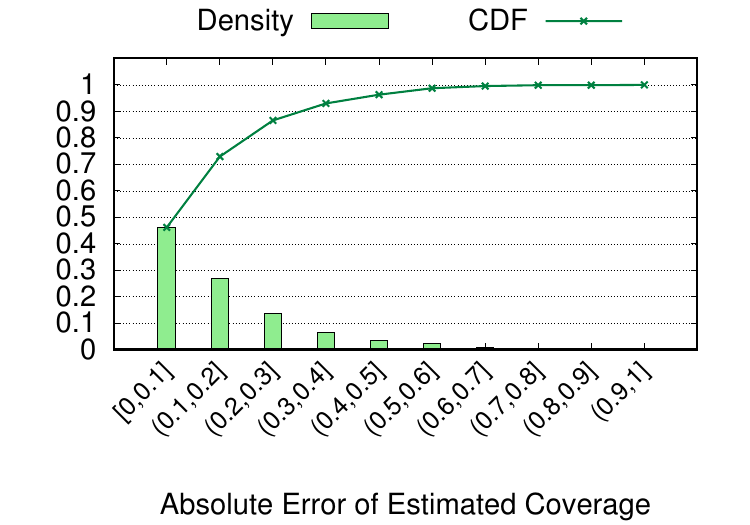}}
%\hspace{0.1\columnwidth}
\subfigure[\textbf{Real-D}]{ \label{fig:coverage-accuracy:real-d}
    \includegraphics[width=0.23\textwidth]{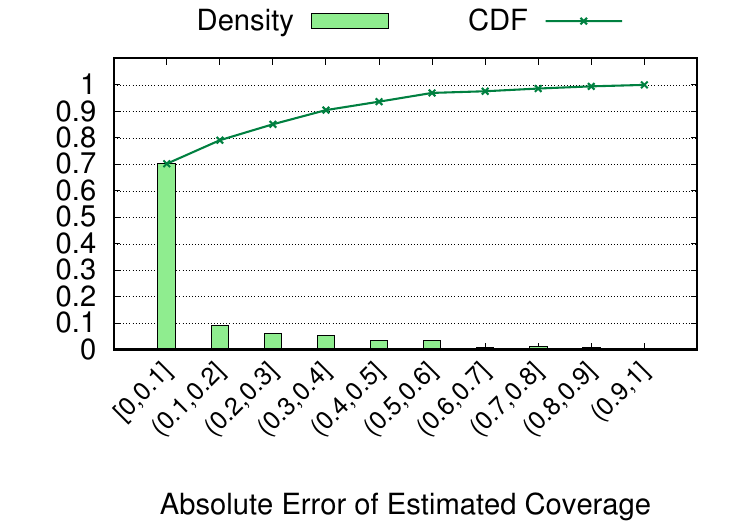}}
%\hspace{0.1\columnwidth}
\subfigure[\textbf{Real-M}]{ \label{fig:coverage-accuracy:real-m}
    \includegraphics[width=0.23\textwidth]{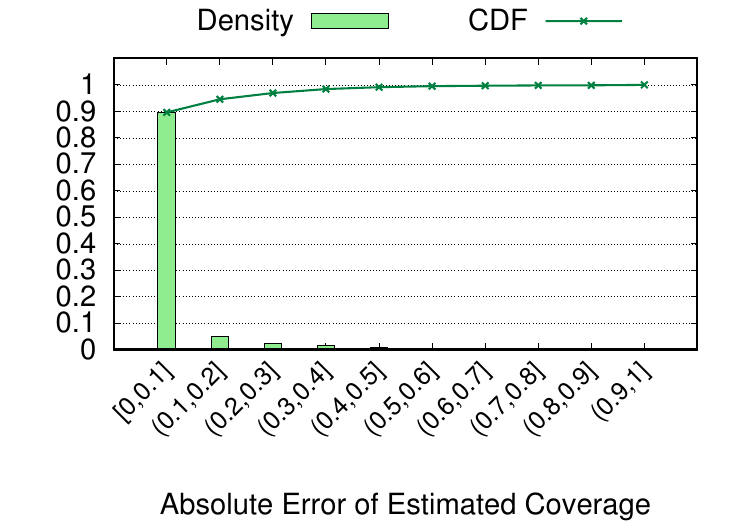}}
\vspace{-1.5em}
\caption{Error distributions of estimated coverage.}
\label{fig:coverage-accuracy}
\vspace{0.5em}
\end{figure*}

Figure~\ref{fig:coverage-accuracy} presents the probability distributions (both the probability \emph{density} and the \emph{cumulative distribution function}, i.e., CDF) for absolute errors on the workloads that we tested.
We observe that, for 66\%, 87\%, 85\%, and 97\% of the query-index pairs collected on \textbf{TPC-H}, \textbf{TPC-DS}, \textbf{Real-D}, and \textbf{Real-M}, their absolute errors of the estimated coverage are below 0.3.
The \emph{mean absolute errors} observed on these workloads are 0.21, 0.16, 0.10, and 0.04, respectively. 
Based on Equation~\ref{eq:lower-bound:call-level:v1}, since we further sum up the MCI's to compute the lower bound, the aggregated error contributed to the lower bound when incorporating coverage-based refinement can be even smaller due to cancellation of the estimation errors made on individual MCI's.

\subsection{Cost Function Properties}
\label{sec:evaluation:cost-func:props}

%\todo{This section can be moved to appendix.}

We validate the \emph{monotonicity} and \emph{submodularity} assumptions of query optimizer cost functions.
%, with the following experiment.

\begin{algorithm}[t]
\small
  \SetAlgoLined
  \KwIn{$W$, the workload.}
  \KwOut{$D_m$, data points for monotonicity validation; $D_s$, data points for submodularity validation.}
  \SetAlgoLined
  \ForEach{query $q\in W$}{
    Run \emph{vanilla greedy} on $\{q\}$ with $K=2$ and $B=\infty$\; %using \emph{actual what-if costs} during configuration enumeration\;
    \ForEach{candidate index $z$ of $q$} {
        Collect all \emph{parent} configurations $P_z=\{x, z\}$ of $z$ with known what-if costs\;
        \ForEach{parent configuration $P_z$} {
            Add three tuples $[c(q,\emptyset), c(q, \{z\})]$, $[c(q,\emptyset),c(q, P_z)]$, and $[c(q, \{z\}), c(q, P_z)]$ into $D_m$ for monotonicity check\;
            Add one tuple $[c(q, \emptyset), c(q, \{z\}), c(q, \{x\}), c(q, P_z)]$ into $D_s$ for submodularity check\;
        }
    }
  }
  \Return{$D_m$ and $D_s$\;}
  \caption{Collect data points for validation of cost function properties (i.e., monotonicity and submodularity).}
\label{alg:validation:cost-func-props}
\end{algorithm}

For each workload, we collect data points using Algorithm~\ref{alg:validation:cost-func-props}.
It runs \emph{vanilla greedy} for each query $q$ (by viewing $q$ as a \emph{singleton workload}) without a budget on the number of what-if calls.
As a result, the actual what-if cost is used for every query-configuration pair.
%encountered in \emph{vanilla greedy}. 
Since this step is costly, we limit the cardinality constraint to $K=2$ (line 2).
After \emph{vanilla greedy} finishes, Algorithm~\ref{alg:validation:cost-func-props} iterates over each candidate index $z$ of the query $q$ to collect corresponding data points for checking monotonicity and submodularity (lines 3 to 8).
For a given candidate index $z$ (i.e., a singleton configuration $\{z\}$), it looks for all of its parent configurations $P_z$, which contain $z$ and one additional candidate index $x$.
At this point, we know $c(q,P_z)$, $c(q,\{z\})$, $c(q,\{x\})$, and $c(q,\emptyset)$:
\begin{itemize}[leftmargin=*]
    \item For monotonicity validation, we check if (1) $c(q,\emptyset) \geq c(q, \{z\})$, (2) $c(q,\emptyset) \geq c(q, \{x, z\})$, and (3) $c(q,\{z\}) \geq c(q, \{x,z\})$ (line 6).
    %That is why we add these three tuples into $D_m$ (line 6).
    Note that we do not need to check whether $c(q, \emptyset)\geq c(q, \{x\})$ and $c(q, \{x\}) \geq c(q, \{x,z\})$ here, since $x$ will also be visited by the iteration at sometime.
    \item For submodularity validation, we check if $c(q,\emptyset)-c(q,\{z\})\geq c(q,\{x\})-c(q, \{x,z\})$ (line 7). 
    %Therefore, we add the corresponding tuple into $D_s$ (line 7).
\end{itemize}

%\todo{Add result figures.}

\begin{table}%[!htb]
\small
\centering
\begin{tabularx}{.8\columnwidth}{|l|X|X|X|X|X|}
\hline
\textbf{Workload} & \textbf{\# Total} & \textbf{\# Yes} & \textbf{\# No} & \textbf{\% Yes} & \textbf{\% No} \\
\hline
\hline
\textbf{TPC-H} & 1,132 & 1,121 & 11 & 99.0\% & 1.0\% \\
\textbf{TPC-DS} & 7,893 & 7,802 & 91 & 98.9\% & 1.1\% \\
\hline
\hline
\textbf{Real-D} & 7,808 & 7,668 & 140 & 98.2\% & 1.8\% \\
\textbf{Real-M} & 120,732 & 115,222 & 5,510 & 95.4\% & 4.6\% \\
\hline
\end{tabularx}
\caption{Validation Results of Monotonicity Assumption.}
\label{tab:validation:monotonicity}
\vspace{-1.5em}
\end{table}

\begin{table}%[!htb]
\small
\centering
\begin{tabularx}{.8\columnwidth}{|l|X|X|X|X|X|}
\hline
\textbf{Workload} & \textbf{\# Total} & \textbf{\# Yes} & \textbf{\# No} & \textbf{\% Yes} & \textbf{\% No} \\
\hline
\hline
\textbf{TPC-H} & 444 & 389 & 55 & 87.6\% & 12.4\% \\
\textbf{TPC-DS} & 3,120 & 2,349 & 771 & 75.3\% & 24.7\% \\
\hline
\hline
\textbf{Real-D} & 3,282 & 2,896 & 386 & 88.2\% & 11.8\% \\
\textbf{Real-M} & 48,166 & 40,976 & 7,190 & 85.1\% & 14.9\% \\
\hline
\end{tabularx}
\caption{Validation Results of Submodularity Assumption.}
\label{tab:validation:submodularity}
\vspace{-1.5em}
\end{table}

\begin{figure*}
\centering
\subfigure[\textbf{TPC-H}]{ \label{fig:delta-negative:histogram:tpch}
    \includegraphics[width=0.23\textwidth]{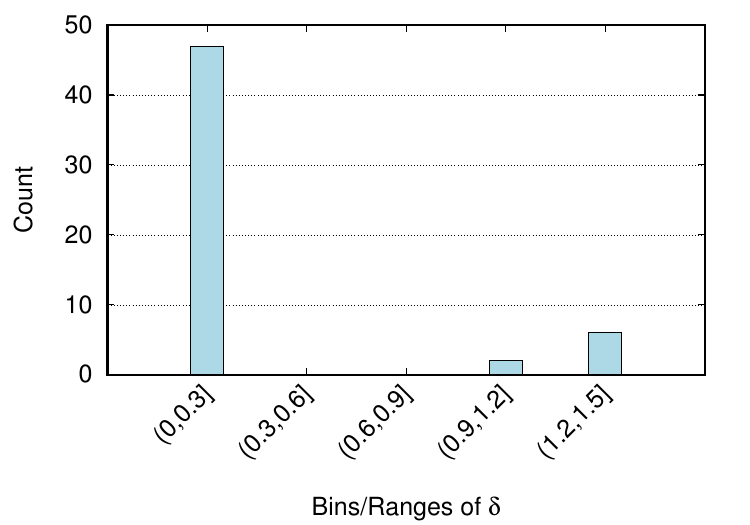}}
\subfigure[\textbf{TPC-DS}]{ \label{fig:delta-negative:histogram:tpcds}
    \includegraphics[width=0.23\textwidth]{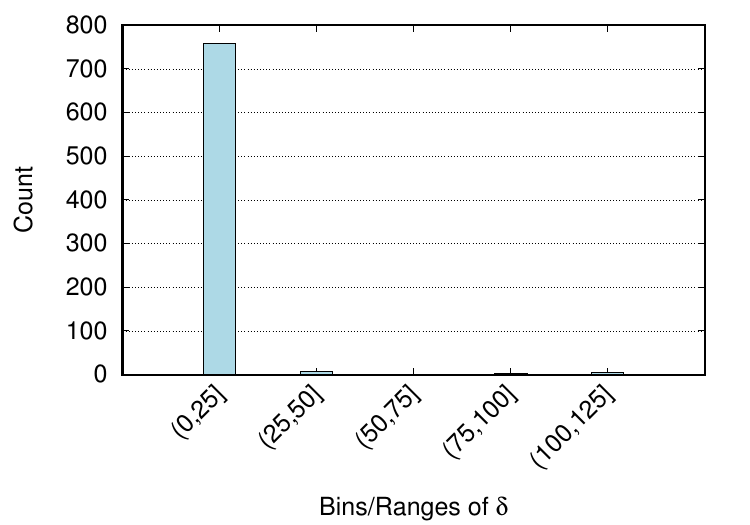}}
%\hspace{0.1\columnwidth}
\subfigure[\textbf{Real-D}]{ \label{fig:delta-negative:histogram:real-d}
    \includegraphics[width=0.23\textwidth]{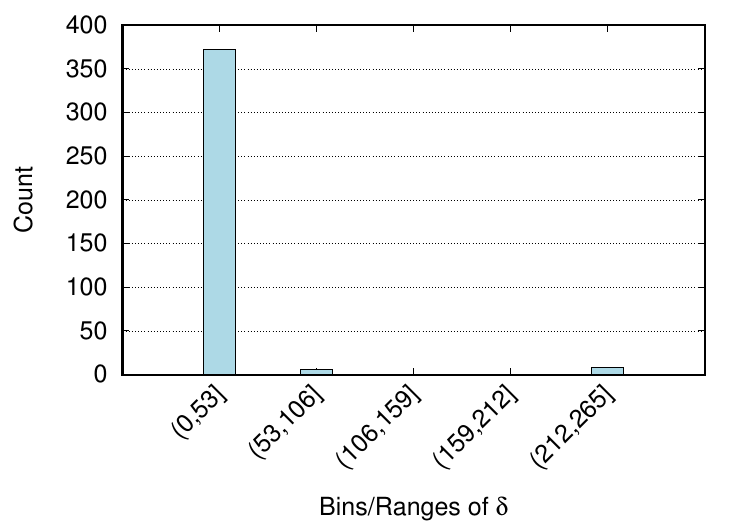}}
%\hspace{0.1\columnwidth}
\subfigure[\textbf{Real-M}]{ \label{fig:delta-negative:histogram:real-m}
    \includegraphics[width=0.23\textwidth]{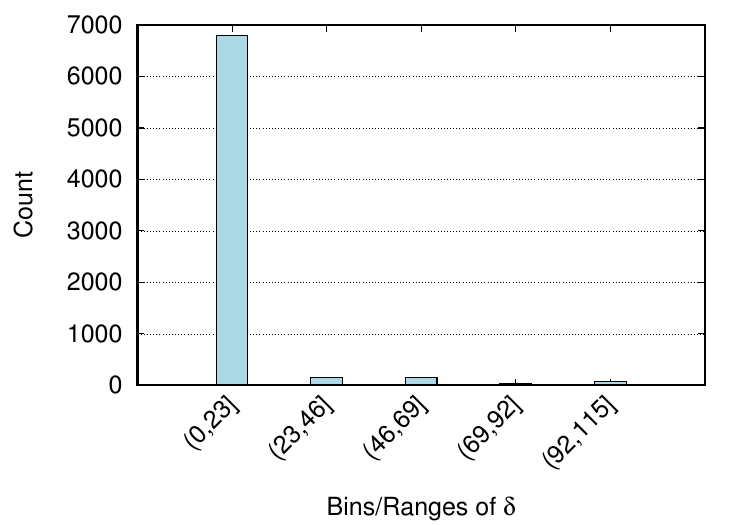}}
\vspace{-1.5em}
\caption{Histograms of $\delta$ when $\delta>0$.}
\label{fig:delta-negative:histogram}
\vspace{0.5em}
\end{figure*}

\iffalse
Given the importance of the submodularity assumption regarding the efficacy of \sysname, we further conducted experiments to validate it by collecting data points using \emph{vanilla greedy} to tune each query in the workload.
Due to space limitation, the details are deferred to~\cite{full-version}.
Table~\ref{tab:validation:submodularity} summarizes the results.
We report the number of data points collected (\textbf{\# Total}), the number (resp. percentage) of data points where submodularity holds (\textbf{\# Yes} resp. \textbf{\% Yes}), and the number (resp. percentage) of data points submodularity does not hold (\textbf{\# No} resp. \textbf{\% No}).
We observe that the probability for submodularity to hold is high on all the workloads that we tested.
\fi

Tables~\ref{tab:validation:monotonicity} and~\ref{tab:validation:submodularity} present validation results of the monotonicity and submodularity assumptions, respectively.
We report the number of data points collected by Algorithm~\ref{alg:validation:cost-func-props} for each validation test (\textbf{\# Total}), the number (resp. percentage) of data points where monotonicity/submodularity holds (\textbf{\# Yes} resp. \textbf{\% Yes}), and the number (resp. percentage) of data points where monotonicity/submodularity does not hold (\textbf{\# No} resp. \textbf{\% No}).
We observe that the probability for monotonicity and submodularity to hold is high on all the workloads that we tested, whereas monotonicity holds with a higher probability ($\geq$95.4\%) than submodularity ($\geq$75.3\%).
%In the following, we show that the violation of the submodularity assumption is typically small.
%\subsection{Violation of Submodularity}

We further looked into the cases where submodularity does not hold, by measuring the difference $\delta$ between $\delta(q,z,\emptyset)=c(q,\emptyset)-c(q,\{z\})$ and $\delta(q,z,\{x\})=c(q,\{x\})-c(q,\{x,z\})$.
That is,
$\delta = \delta(q,z,\{x\})-\delta(q,z,\emptyset).$
Intuitively, a violation of submodularity means $\delta > 0$, and we call $\delta$ the \emph{magnitude of violation}.
Figure~\ref{fig:delta-negative:histogram} presents the histograms for the data points collected on the workloads with $\delta>0$.
We observe the same pattern for all the workloads tested: the distribution of $\delta$ (when $\delta>0$) is \emph{highly skewed}, concentrating on regions where $\delta$ is small.
%It implies that, wherever there is a violation of the submodularity assumption, the violation is typically small.
%\todo{Revise the figure to have distribution for $\delta>0$.}
%Recall that in Algorithm~\ref{alg:get-cost-bound}, when there is a violation of the submodularity assumption, we simply keep the MCI upper bound $u(q, x)$ unchanged.
%Since the violation is usually small, the unchanged upper bound $u(q, x)$ is usually very close to $c(q, C^*)-c(q, C)$, which makes the impact of such violation very small as well.

%This is important, 
%because we rely on the submodularity assumption to refine the MCI upper bounds $u(q, z)$, which will be further summed up to obtain the lower bound %$L(q, C)$.

%\xy{move to experiment?}

%and 158.4 minutes (i.e., 2.4$\times$ speedup), 

%respectively, by using \sysname and \sysname with the coverage-based refinement.
%\todo{Revise the above Real-D numbers.}

\iffalse
For example, without a budget constraint, when tuning the standard \textbf{TPC-H} benchmark with 22 queries, \sysname can reduce index tuning time by 3$\times$ while achieving the same quality on the best configuration found~\cite{full-version}.
\fi

%\balance

%\iffalse
\section{Impact of Coverage on the Confidence}
\label{sec:analysis:impact:coverage}

When we consider using ``coverage'' to estimate the what-if cost $c(q, {z})$ for a singleton configuration $\{z\}$ and thus the corresponding MCI upper-bound $u(q, z)$, the lower bound $L(q, C)$ becomes an estimated value as well.
In the following, we use $\hat{c}(q, {z})$, $\hat{u}(q, z)$, and $\hat{L}(q, C)$ to denote the estimated values based on the estimated coverage $\rho(q, z)$. We present a quantitative analysis regarding the impact of using these estimated values in the confidence-based what-if call skipping mechanism.

%Recall that we have the following relationship between $c(q, \{z\})$ and $\rho(q, z)$ 
By the definition of coverage, we have
\begin{eqnarray*}
c(q, \{z\})= c(q, \emptyset)-\rho(q, z)\cdot\Big(c(q, \emptyset)-c(q, \Omega_q)\Big)
=\Big(1-\rho(q, z)\Big)\cdot c(q, \emptyset) + \rho(q, z)\cdot c(q, \Omega_q).
\end{eqnarray*}
As a result, it follows that
\begin{eqnarray*}
\hat{c}(q, \{z\})= c(q, \emptyset)-\hat{\rho}(q, z)\cdot\Big(c(q, \emptyset)-c(q, \Omega_q)\Big)
=\Big(1-\hat{\rho}(q, z)\Big)\cdot c(q, \emptyset) + \hat{\rho}(q, z)\cdot c(q, \Omega_q).
\end{eqnarray*}
Now, assuming $u(q,z)=c(q,\emptyset)-c(q, \{z\})$, the estimated lower bound becomes
\begin{eqnarray*}
\hat{L}(q, C)&=&c(q,\emptyset)-\sum\nolimits_{z\in C} \hat{u}(q, z)\\
&=& c(q,\emptyset)-\sum\nolimits_{z\in C} \Big(c(q,\emptyset)-\hat{c}(q, \{z\})\Big)\\
&=& c(q,\emptyset)-\sum\nolimits_{z\in C} \hat{\rho}(q, z)\cdot \Big(c(q,\emptyset)-c(q, \Omega_q)\Big)\\
&=& c(q,\emptyset)-\Delta(q, \Omega_q)\cdot \sum\nolimits_{z\in C} \hat{\rho}(q, z).
\end{eqnarray*}
It then follows that the confidence with coverage-based singleton cost estimates is
$$\hat{\alpha}(q, C)=\frac{\hat{L}(q, C)}{U(q, C)}=\frac{c(q,\emptyset)-\Delta(q, \Omega_q)\cdot \sum\nolimits_{z\in C} \hat{\rho}(q, z)}{U(q, C)}.$$
On the other hand, by the definition of confidence we have
$$\alpha(q, C)=\frac{L(q, C)}{U(q, C)}=\frac{c(q,\emptyset)-\Delta(q, \Omega_q)\cdot \sum\nolimits_{z\in C} \rho(q, z)}{U(q, C)}.$$
Combining the above two equations yields
$$U(q,C)\cdot\Big(\hat{\alpha}(q, C)-\alpha(q, C)\Big)=\Delta(q, \Omega_q)\cdot \sum\nolimits_{z\in C} \Big(\rho(q, z)-\hat{\rho}(q, z)\Big).$$
Or equivalently,
\begin{eqnarray*}
\alpha(q, C)&=&\hat{\alpha}(q, C)-\frac{\Delta(q, \Omega_q)}{U(q, C)}\cdot \sum\nolimits_{z\in C} \Big(\rho(q, z)-\hat{\rho}(q, z)\Big)\\
&=&\hat{\alpha}(q, C)+\frac{\Delta(q, \Omega_q)}{U(q, C)}\cdot \sum\nolimits_{z\in C} \Big(\hat{\rho}(q, z)-\rho(q, z)\Big).
\end{eqnarray*}
This implies that the degree of error in the confidence computation using estimated coverage depends on the \emph{sum} of the errors made in estimating coverage for individual indexes (i.e., singleton configurations) within the configuration $C$.
%\fi

\end{document}